\input amstex
\documentstyle{amsppt}
\vsize=47pc
\magnification=\magstep1
\NoBlackBoxes
\def\today{\ifcase\month\or
 January\or February\or March\or April\or May\or June\or
 July\or August\or September\or October\or November\or December\fi
 \space\number\day, \number\year}

\def\strutdepth{\dp\strutbox}
\def\changes{\strut\vadjust{\kern-\strutdepth\specialstar}}
\def\specialstar{\vtop to \strutdepth{\baselineskip\strutdepth
\vss\llap{changes }\null}}

\baselineskip=32pt
\parindent=18pt
\def\nind{\noindent}
\def\RR{\text{{\rm I \hskip -5.75pt R}}}
\def\IN{\text{{\rm I \hskip -5.75pt N}}}

\def\IP{\text{{\rm I \hskip -5.75pt P}}}
\def\CC{\;\text{{\rm \vrule height 6pt width 1pt \hskip -4.5pt C}}}
\def\ZZ{\text{{\rm Z \hskip -8pt Z}}}

\def\As{{\Cal A}}
\def\Bs{{\Cal B}}
\def\Cs{{\Cal C}}
\def\Ds{{\Cal D}}
\def\Es{{\Cal E}}
\def\Fs{{\Cal F}}
\def\Gs{{\Cal G}}
\def\Hs{{\Cal H}}

\def\Js{{\Cal J}}
\def\Ks{{\Cal K}}
\def\Ls{{\Cal L}}
\def\Ms{{\Cal M}}
\def\Ns{{\Cal N}}
\def\Os{{\Cal O}}
\def\Ps{{\Cal P}}

\def\Rs{{\Cal R}}

\def\Ts{{\Cal T}}
\def\Us{{\Cal U}}

\def\Ws{{\Cal W}}

\def\Zs{{\Cal Z}}

\def\wnet{{\{\As(W)\}_{W \in \Ws}}}
\def\wrnet{{\{\Rs(W)\}_{W \in \Ws}}}

\def\inet{{\{\As_i \}_{i \in I}}}
\def\irnet{{\{\Rs_i \}_{i \in I}}}

\def\trnet{{\{\Rs(\Os)\}_{\Os \in \Sr}}}
\def\net{{\{\As(R)\}_{R\in{\Cal R}}}}
\def\cnet{{\{\As(\Os)\}_{\Os\in\Cs}}}
\def\hnet{{\{\hat{\As}(\Os)\}_{\Os\in\Cs}}}

\def\pair{{(W_1,W_2)}}
\def\ad{{\text{ad}}}
\def\sp{{\text{sp}}}
\def\diag{{\text{diag}}}

\def\IL{{\text{InvL}}}
\def\IP{{\text{InvP}}}
\def\ILid{{\text{InvL}_+^{\uparrow}}}
\def\IPid{{\text{InvP}_+^{\uparrow}}}
\def\Pid{{\Ps_+ ^{\uparrow}}}
\def\Lid{{\Ls_+ ^{\uparrow}}}
\def\hal{{\tau}}

\def\Sr{\Im}                             

\def\idty{{\leavevmode\hbox{\rm 1\kern -.3em I}}}

\baselineskip 15pt plus 2pt
\spaceskip=.5em plus .25em minus .20em
\redefine\qed{\hbox{$\boxed{}$}}
\leftheadtext{D. Buchholz, O. Dreyer, M. Florig and S.J. Summers}
\rightheadtext{Geometric Modular Action and Spacetime Symmetry Groups}

\topmatter
\title Geometric Modular Action and Spacetime Symmetry Groups
\endtitle

\author   Detlev Buchholz, Olaf Dreyer, Martin Florig and Stephen J. Summers 
\endauthor     
\address{(Buchholz) Institut f\"ur Theoretische Physik, Universit\"at 
G\"ottingen, Bunsenstr. 9, D-37073 G\"ottingen, Germany \quad , \quad 
(Dreyer) Department of Physics, Pennsylvania State University, University Park,
PA 16802, USA \quad and \quad (Florig and Summers) Department of Mathematics, 
University of Florida, Gainesville, FL  32611, USA} 
\endaddress

\date{May 1998} \enddate

\abstract{A condition of geometric modular action is proposed as a 
selection principle for physically interesting states on general space-times.
This condition is naturally associated with transformation groups 
of partially ordered sets and provides these groups with projective 
representations. Under suitable additional conditions, these groups induce 
groups of point transformations on these space-times, which may be interpreted 
as symmetry groups. The consequences of this condition are studied in 
detail in application to two concrete space-times -- four-dimensional 
Minkowski and three-dimensional de Sitter spaces -- for which it is shown how 
this condition characterizes the states invariant under the respective 
isometry group. An intriguing new algebraic characterization of vacuum states 
is given. In addition, the logical relations between the condition proposed in 
this paper and the condition of modular covariance, widely used in the 
literature, are completely illuminated.}\bigskip

\heading Table of Contents \endheading

\nind I. \ \ Introduction
...........................................................................................
p. \  2 \par 
\nind II. \ Nets of Operator Algebras and Modular Transformation
Groups .... p. \ 5 \par 
\nind III. Geometric Modular Action in Quantum Field Theory ...................... p. \ 9 \par
\nind IV. Geometric Modular Action Associated With Wedges in $\RR^4$ .............. p. 14 \par
   4.1. Wedge Transformations Are Induced By Elements of the 
Poincar\'e Group ......................................................................................................................... p. 16 \par
   4.2. Wedge Transformations Generate the Proper Poincar\'e Group .. p. 31 \par
   4.3. {}From Wedge Transformations Back to the Net: Locality, Covariance and 
Continuity
............................................................................................... p. 40 \par
\nind V. Geometric Action of Modular Groups and the Spectrum Condition
.. p. 51 \par
   5.1. The Modular Spectrum Condition .............................................. p. 52 \par
   5.2. Geometric Action of Modular Groups ......................................... p. 56 \par
   5.3. Modular Involutions Versus Modular Groups .............................. p. 62 \par
\nind VI. Geometric Modular Action and De Sitter Space ................................. p. 66 \par
   6.1. Wedge Transformations in de Sitter Space ................................ p. 67 \par
   6.2. Geometric Modular Action in de Sitter Space and the de Sitter 
Group ......................................................................................................................... p. 72 \par
\nind VII. Summary and Further Remarks ........................................................... p. 74 \par
\nind Appendix. Cohomology and the Poincar\'e Group ....................................... p. 77 \par

\endabstract

\endtopmatter

\document

\bigpagebreak

\heading I. Introduction  \endheading

    In \cite{9}\cite{10}, Bisognano and Wichmann showed that for quantum field 
theories satisfying the Wightman axioms the modular objects associated by 
Tomita-Takesaki theory to the vacuum state and local algebras in wedgelike 
regions in Minkowski space have geometrical interpretation. This fundamental 
insight has opened up a number of fascinating lines of research for algebraic 
quantum field theory. To better appreciate the ramifications of their result, 
it is important to realize that the modular objects of the Tomita-Takesaki 
theory are completely determined by the choice of physical state and algebra
of observables. That these modular objects can also have geometrical and 
dynamical significance thus allows the conceptually important possibility of 
{\it deriving} geometrical and dynamical information from the latter physical 
data. \par
     For example, it has become possible to characterize physically 
distinguished states by the geometric action of the modular objects associated 
with suitably chosen local algebras. This approach was taken in \cite{24}
(cf. also \cite{13}), where it was shown how the vacuum state on 
Minkowski space can be characterized by the action of the modular objects 
associated with wedge algebras and how the dynamics of the theory can be 
derived from the modular involutions. The present paper is in several respects
a refinement and generalization of \cite{24}. \par
     Another program which has grown out of Bisognano and Wichmann's insight 
is the construction of nets of local algebras and representations of a group 
acting covariantly upon the net, starting from a state, a small number of 
algebras, and a suitable ``geometric'' action of the associated modular 
objects upon these algebras. This line was first addressed in \cite{12}, cf.\ 
also \cite{58}. The most complete results in this direction have been,
on the one hand, the construction of conformally covariant nets of local 
algebras in two spacetime dimensions in \cite{70}\cite{72} and, on the other, 
of Poincar\'e covariant nets in three spacetime dimensions in \cite{74} (see 
also \cite{73}\cite{15}). \par
     Yet another closely related research program is the generation of 
unitary representations of spacetime symmetry groups by modular objects which 
are assumed to implement the action of subgroups of these symmetry groups
upon a given net of algebras. This course of study using the unitary modular  
groups was also opened up by Borchers \cite{12} and followed in 
\cite{69}\cite{21}\cite{22}\cite{36}\cite{35}, whereas the derivation of such 
representations from the modular involutions was initiated in \cite{24}. This
aspect we also generalize in this paper. Moreover, we shall clarify 
the relations between these two different approaches to geometric action of
modular objects. For a more detailed review of the prior literature, see 
\cite{14}. \par
     As explained in our first paper on the subject \cite{24}, a further
interesting step is the {\it derivation} of spacetime symmetry groups from 
the underlying algebraic structure and the given state. By ``space-time'' we
here mean some smooth manifold without {\it a priori} given metric or
conformal structure. {}From our point of view, if a given net of observable 
algebras happens to be covariant under the action of a unitary representation 
of some group of point transformations of the underlying manifold, 
then these point transformations should be regarded as the isometries 
of a metric structure to be imposed upon the space-time. We mention 
in this context the papers \cite{43}\cite{76}, in which the causal 
({\it i.e.} conformal) metric structure of the space-time is derived from the 
states and algebras of observables, under certain conditions. \par  
     It is the essential lesson of the present paper that the various goals
mentioned above -- the derivation of spacetime symmetry groups, the generation
of corresponding unitary representations, and the characterization of 
physically distinguished states from the algebraic data -- can all be 
accomplished in physically interesting examples by a Condition of
Geometric Modular Action proposed in \cite{24}. This fact sheds new light on
the results mentioned above and poses some new and intriguing questions. \par 
     We shall present this condition somewhat imprecisely in this introduction 
-- further details will be given in the main text. Let $\Ws$ be a suitable 
collection of open sets on a space-time 
$\Ms$ and $\wnet$ be a net of $C^*$-algebras indexed by $\Ws$, each of which 
is a subalgebra of the $C^*$-algebra $\As$. A state on $\As$ will be denoted 
by $\omega$ and the corresponding GNS representation of $\As$ will be 
signified by $(\Hs,\pi,\Omega)$. For each $W \in \Ws$ the von Neumann 
algebra $\pi(\As(W))''$ will be denoted by $\Rs(W)$. The modular involution
associated to the pair $(\Rs(W),\Omega)$ will be represented by $J_W$, while
the modular group associated to the same pair will be written as 
$\{\Delta_W^{it}\}_{t \in \RR}$.

\proclaim{Condition of Geometric Modular Action} Given the structures 
indicated above, then the pair $(\wrnet,\omega)$ satisfies the Condition of
Geometric Modular Action if the collection of algebras
$\wrnet$ is stable under the adjoint action of the modular 
involution $J_{W}$ associated with the pair $( \Rs (W), \Omega )$, 
for all $W \in \Ws$. In other words, for every pair of regions 
$W_1, W_2 \in \Ws$ there is some region $W_1 \circ W_2 \in \Ws$ such that

$$ J_{W_1} \Rs (W_2 ) J_{W_1} = \Rs (W_1 \circ W_2 ) \quad . \tag{1.1} $$

\endproclaim

     This condition was initially motivated by a number of examples in 
Minkowski space-time in which modular objects have a geometric action  
implying the above condition (see 
\cite{9}\cite{10}\cite{23}\cite{39}\cite{29}). We 
emphasize that this condition does not assume that the adjoint action of
the modular involutions upon the net acts in the detailed manner of the cited 
examples -- indeed, it is not even assumed that this action can be realized as 
a point transformation on the space-time. In fact, we imagine that there will
be situations of physical interest in which this geometric action is {\it not}
implemented by point transformations, but where this condition will still
serve as a useful selection criterion. \par
     Note that this condition can be stated sensibly for arbitrary 
space-time, indeed for arbitrary topological space $\Ms$.
 This enables us to propose this Condition of Geometric Modular Action
as a criterion for selecting physically interesting states on {\it general} 
space-times. We anticipate that in some applications this condition will have 
to be weakened in evident ways. In particular, there are 
circumstances where only (even) {\it products} of modular involutions will act 
``geometrically'' in this manner -- here we think, for example, of the Rindler 
wedge \cite{41}. We expect that also these weakened versions should select 
states of notable physical interest. \par
     We emphasize that our selection criterion is one for a {\it state} and
not an entire folium. In particular, previously suggested criteria, such
as the Hadamard condition \cite{42} and the microlocal spectrum condition
\cite{56}, are valid for an entire folium of states. Though these
criteria are valuable, they beg the question of which state (or states)
of the respective folium is to be regarded as fundamental, {\it i.e.} as a 
reference state.  \par
     In Chapter II we shall state and study our Condition of Geometric
Modular Action in a very general form, which will enable us to explicate
more clearly how it selects an intriguing class of transformation groups on 
the index sets of nets of von Neumann algebras and supplies them with
projective representations. Returning to the original situation of nets indexed
by open subsets of a space-time $\Ms$ in Chapter III, we explain how to 
choose a suitable family $\Ws$ depending only on the space-time itself
and present some results of conceptual importance for our framework. There
we also outline the program opened up by our framework -- a program we carry
out explicitly in two examples in Chapters IV and VI. \par
     In Chapter IV we shall illustrate the power of our condition by choosing
$\Ms$ to be topological $\RR^4$ and $\Ws$ to be the set of wedgelike regions
in $\RR^4$. It will be shown that with a few additional assumptions -- all
expressible in terms of the state, the net of algebras and the associated
modular involutions -- the transformations induced upon the index set $\Ws$
by (1.1) are implemented by point transformations -- in fact, by the
proper Poincar\'e group $\Ps_+$. We obtain after a series of steps a 
representation of $\Ps_+$ which acts covariantly upon the net. Therefore,
we have an algebraic characterization of Poincar\'e invariant states on
nets of algebras indexed by open subsets of $\RR^4$, which induce
Poincar\'e covariant representations of these nets. A more detailed overview of
Chapter IV may be found at its beginning. Yet another example is worked out
in Chapter VI, where it is shown how similar results for the de Sitter group
in three dimensions may be obtained with suitable choices of $\Ms$ and
$\Ws$. \par
     Continuing the development presented in Chapter IV, Chapter V harbors a 
discussion of how also the spectrum condition can 
be characterized in terms of the modular objects, which then leads to how to
derive algebraic PCT- and Spin \& Statistics Theorems in our setting. We 
present a striking new algebraic characterization of vacuum states on
Minkowski space in terms of quantities which have meaning for arbitrary 
space-times. This condition may prove to be useful as a criterion for 
``stability'' for quantum states on general space-times. Moreover, we show 
that if the adjoint action of the modular {\it groups} associated 
to the wedge algebras in $\RR^4$ leaves the set $\wrnet$ invariant, then these 
modular groups satisfy {\it modular covariance}, and all of the results in 
Chapter IV hold once again, along with either the positive or negative 
spectrum 
condition. We provide further details which clarify the relation between our
condition and the widely-used condition of modular covariance. Finally, in 
Chapter VII we collect some further comments and speculations.\par
     An overview of an earlier version of the results of this paper has 
appeared in \cite{62}. In addition to the detailed proofs, most of which were 
suppressed in \cite{62}, the present paper contains somewhat more transparent 
arguments, as well as many additional or strengthened results.

\heading II. Nets of Operator Algebras and Modular Transformation Groups 
\endheading

     We begin the main text of this paper with a more abstract setting of  
our Condition of Geometric Modular Action, since then its connection with 
transformation groups on partially ordered sets and projective representations 
of these groups emerges particularly clearly. We shall return to the original 
situation with further precisions in the next chapter.  \par
     Let $\{ \As_i \}_{i \in I}$ be a collection of $C^*$-algebras labeled 
by the elements of some index set $I$. If $(I,\leq)$ is a directed set and the 
property of isotony holds, {\it i.e.} if for any $i_1,i_2 \in I$ such that 
$i_1 \leq i_2$ one has $\As_{i_1} \subset \As_{i_2}$, then 
$\{ \As_i \}_{i \in I}$ is said to be a net. However, for our purposes it will 
suffice that $(I,\leq)$ be only a partially ordered set and that 
$\{ \As_i \}_{i \in I}$ satisfies isotony. We are therefore working with two 
partially ordered sets, $(I,\leq)$ and $(\{ \As_i \}_{i \in I}, \subseteq)$, 
and we require that the assignment $i \mapsto \As_i$ be an order-preserving 
bijection ({\it i.e.} it is an isomorphism in the structure class of partially 
ordered sets). We note that any such assignment which is not an isomorphism in 
this sense would involve some kind of redundancy in the description. In 
algebraic quantum field theory the index set $I$ is usually a collection of 
open causally closed subsets of an appropriate metric space-time $(\Ms,g)$. In
such a case the algebra $\As_i$ is interpreted as the $C^*$-algebra generated 
by all the observables measurable in the space-time region $i$. Hence, to 
different spacetime regions should correspond different algebras. \par 
     If $\inet$ is a net, then the inductive limit $\As$ of $\inet$ exists
and may be used as a reference algebra. However, even if $\inet$ is not a net, 
it is still possible \cite{32} to naturally embed the algebras 
$\As_i$ in a $C^*$-algebra $\As$ in such a way that the inclusion relations 
are preserved. In the following we need therefore not distinguish these two
cases and refer, somewhat loosely, to any collection $\{ \As_i \}_{i \in I}$ 
of algebras, as specified, as a net. Any  state on $\As$ restricts to
a state on $\As_i$, for each $i \in I$. For that reason, we shall speak of a 
state on $\As$ as being a state on the net $\inet$. \par
     A net automorphism is an automorphism $\alpha$ of the global algebra 
$\As$ such that there 
exists an order-preserving bijection $\hat{\alpha}$ on $I$ satisfying 
$\alpha(\As_i) = \As_{\hat{\alpha}(i)}$. Symmetries, whether dynamical or
otherwise, are generally expressed in terms of net automorphisms (or 
antiautomorphisms) \cite{57}. An internal symmetry of the net is 
represented by an automorphism $\alpha$ such that $\alpha(\As_i) = \As_i$ for 
every $i \in I$, {\it i.e.} the corresponding order-preserving bijection 
$\hat{\alpha}$ is just the identity on $I$. \par
     Given a state $\omega$ on the algebra $\As$, one can consider the 
corresponding GNS representation $(\Hs_{\omega},\pi_{\omega},\Omega)$ and the 
von Neumann algebras $\Rs_i \equiv \pi_{\omega}(\As_i)''$, $i \in I$. We
shall assume that the representation space $\Hs_{\omega}$ is separable. We 
extend the assumption of nonredundancy of indexing to the net 
$\{ \Rs_i \}_{i \in I}$, {\it i.e.} we assume that also the map 
$i \mapsto \Rs_i$ is an order-preserving bijection.\footnote{This is 
automatically the case if the algebras $\As_i$ are von Neumann algebras and
$\omega$ induces a faithful representation of 
$\underset{i \in I}\to{\cup}\As_i$.} If the GNS vector $\Omega$ is cyclic and 
separating for each algebra $\Rs_i$, $i \in I$, then from the modular theory 
of Tomita-Takesaki, we are presented with a collection $\{J_i\}_{i\in I}$ of 
modular involutions (and a collection $\{\Delta_i\}_{i\in I}$ of modular 
operators), directly derivable from the state and the algebras. This 
collection $\{J_i\}_{i\in I}$ of operators on $\Hs_{\omega}$ generates a group 
$\Js$, which becomes a topological group in the strong operator topology on 
$\Bs(\Hs_{\omega})$, the algebra of all bounded operators on $\Hs_{\omega}$. 
Note that $J\Omega = \Omega$ for $J \in \Js$. \par
     In the following we shall denote the adjoint action of $J_i$ upon the 
elements of the net $\irnet$ by $\ad J_i$, {\it i.e.} 
$\ad J_i (\Rs_j) \equiv J_i \Rs_j J_i = \{ J_i A J_i \mid A \in \Rs_j \}$. 
Note that if $\Rs_1 \subset \Rs_2$, then one necessarily has 
$\ad J_i (\Rs_1) \subset \ad J_i (\Rs_2)$, in other words the map $\ad J_i$ is
order-preserving. Hence, the content of the 
Condition of Geometric Modular Action in this abstract setting is that each 
$\ad J_i$ is a net automorphism. Thus, for each $i \in I$, there is an 
order-preserving bijection (an automorphism) $\tau_i$ on $I$ ($(I,\leq)$) such 
that $J_i \Rs_j J_i = \Rs_{\tau_i(j)}$, $j \in I$. The 
group generated by the $\tau_i$, $i \in I$, is denoted by $\Ts$ and forms a
subgroup of the transformations on the index set $I$. For the convenience of 
the reader, we summarize our standing assumptions.  \par

\proclaim{Standing Assumptions} For the net $\inet$ and the state $\omega$
on $\As$ we assume \par
   (i) $i \mapsto \Rs_i$ is an order-preserving bijection; \par
   (ii) $\Omega$ is cyclic and separating for each algebra $\Rs_i$, $i \in I$;
\par
   (iii) each $\ad J_i$ leaves the set $\irnet$ invariant.\footnote{and is
{\it a fortiori} a net automorphism}
\endproclaim

     We collect some basic properties of the group $\Ts$ in the following 
lemma.  \par

\proclaim{Lemma 2.1} The group $\Ts$ defined above has the following
properties. \par 
   (1) For each $i \in I$, $\tau_i ^2 = \iota$, where $\iota$ is the identity 
map on $I$. \par
   (2) For every $\tau \in \Ts$ one has 
$\tau \tau_i \tau^{-1} = \tau_{\tau(i)}$. \par
   (3) If $\tau(k) = k$ for some $\tau \in \Ts$ and some $k \in I$, then 
$\tau\tau_k = \tau_k \tau$.  \par
   (4) One has $\tau_i (i) = i$, for some $i \in I$, if and only 
if the algebra $\Rs_i$ is maximally abelian. If $\Ts$ acts
transitively on $I$, then $\tau_i (i) = i$, for some $i \in I$, if and only if
$\tau_i (i) = i$, for all $i \in I$. Moreover, if $\tau_i(i) = i$ for some 
$i \in I$, then $i$ is an atom in $(I,\leq)$, {\it i.e.} if $j \in I$ and 
$j \leq i$, then $j=i$. \par
   (5) If $i \leq j \leq k \leq l$, then $\tau_i(j) \geq \tau_l(k)$. 
\endproclaim

\demo{Proof} 1. The first assertion is immediate since $J_i ^2 = \idty$, the
identity operator on $\Hs_{\omega}$, hence for each $j \in I$ one has 
$\Rs_j = J_i J_i \Rs_j J_i J_i = J_i \Rs_{\tau_i (j)} J_i = 
\Rs_{\tau_i (\tau_i(j))}$. Standing Assumption (i) then yields 
$\tau_i ^2 = \iota$. \par
   2. Since every element of $\Js$ leaves $\Omega$ invariant, standard
arguments in modular theory show that the basic assumption 
$J_i \Rs_j J_i = \Rs_{\tau_i (j)}$ implies the relation 
$J_i J_j J_i = J_{\tau_i (j)}$. Therefore one has the equalities

$$\Rs_{(\tau_i \tau_j \tau_i) (k)} = J_i J_j J_i \Rs_k J_i J_j J_i = 
J_{\tau_i (j)} \Rs_k J_{\tau_i (j)} = \Rs_{\tau_{\tau_i (j)}(k)}\quad , $$

\nind for every $k \in I$. Once again, the nonredundancy assumption yields the
assertion $\tau_i \tau_j \tau_i = \tau_{\tau_i (j)}$, for each $i,j \in I$. 
Since $\Ts$ is generated by the set $\{ \tau_i \mid i \in I \}$, this entails 
assertion (2). \par
   3. Assume one has 
$J_{i_1}\cdots J_{i_n} \Rs_k J_{i_n}\cdots J_{i_1} = \Rs_k$ for some 
$i_1, \ldots, i_n, k \in I$. Then the (anti)unitary operator 
$J_{i_1}\cdots J_{i_n}$ induces an (anti)automorphism of $\Rs_k$ and leaves 
$\Omega$ invariant. It must therefore commute with the modular objects 
associated with the pair $(\Rs_k,\Omega)$ (see Theorem 3.2.18 in 
\cite{18}). But this implies that
$\tau_{i_1}\cdots \tau_{i_n} \tau_k = \tau_k \tau_{i_1}\cdots \tau_{i_n}$. 
 \par
   4. If $\tau_i (i) = i$ for some $i \in I$, then one has
$\Rs_i ' = J_i \Rs_i J_i = \Rs_{\tau_i (i)} = \Rs_i$, so that $\Rs_i$ is
abelian. Moreover, since $\Omega$ is cyclic for this abelian von Neumann
algebra, it must be maximally abelian. If $\Ts$ acts transitively on $I$,
then since the modular involutions are (anti)unitary, every $\Rs_i$ must
be maximally abelian. On the other hand, if $\Rs_i$ is maximally abelian, one 
has $\Rs_i = \Rs_i ' = J_i \Rs_i J_i = \Rs_{\tau_i (i)}$. Hence, by the 
nonredundancy assumption, one has $\tau_i (i) = i$. It follows that if every 
algebra $\Rs_k$ is maximally abelian, then $\tau_k (k) = k$ for every 
$k \in I$. \par
     As already pointed out, under Standing Assumption (ii), any abelian 
$\Rs_i$ must be 
maximally abelian. Hence, if there exist $i_1 < i_2$ with $\Rs_{i_1}$ and 
$\Rs_{i_2}$ both abelian, then $\Rs_{i_1} \subset \Rs_{i_2}$, which yields 
$\Rs_{i_1} = \Rs_{i_2}$, since both algebras are maximally abelian. This would 
violate Standing Assumption (i). \par
   5. If $i \leq j \leq k \leq l$, then one observes that 
$$J_i \Rs_j J_i \supset J_i \Rs_i J_i = \Rs_i ' \supset \Rs_l '
              = J_l \Rs_l J_l \supset J_l \Rs_k J_l $$
implies $\tau_i(j) \geq \tau_l(k)$. 
\hfill\qed\enddemo

\nind For index sets without atoms, such as the index set $\Ws$ used as an 
example in Chapter IV (however, not the example used in Chapter VI), 
Lemma 2.1 (4) implies that $\Rs_i$ must be nonabelian for every $i \in I$.  
\par
Certain aspects of Lemma 2.1 may be interpreted as follows: given the 
set $I$, we consider functions $\underline{\tau} : I \mapsto \Ts$, where
$\Ts$ is some subgroup of the symmetric group on $I$. There exist two natural
automorphisms on these functions. The first one is given by the 
adjoint action on $\Ts$ - namely, 
$\ad\tau_0 (\underline{\tau})(\cdot) = 
\tau_0 \underline{\tau}(\cdot)\tau_0^{-1}$ for each $\tau_0 \in \Ts$,
and the second one is induced by the action of $\Ts$ on $I$:
$(\underline{\tau} \circ \tau_0)(\cdot) = \underline{\tau}(\tau_0(\cdot))$.
If, for a given function $\underline{\tau}$, these two actions coincide
for all $\tau_0 \in \Ts$, we say that $\underline{\tau}$ is $\Ts$-covariant.
Note that the $\Ts$-covariant functions form a group under pointwise 
multiplication, the identity being the constant function on $I$ with value
$\iota$. A particularly interesting case arises if the range of a function
$\underline{\tau}$ generates $\Ts$; we then say that $\underline{\tau}$ is
a generating function. The preceding proposition thus shows that the condition
of geometric modular action provides us with subgroups $\Ts$ of the symmetric
group on $I$ which admit an idempotent, $\Ts$-covariant generating function.
This is a rather strong consistency condition on $\Ts$. For example, the
full symmetric groups of index sets do not in general admit such functions.
What is of interest here is the fact that the structure is fixed once the
index set $I$ is given. \par
     We feel it is useful to elaborate further the relation between the groups
$\Js$ and $\Ts$. Recall that an operator $Z \in \Js$ is said to be an 
{\it internal symmetry\/} of the net $\irnet$, if $Z\Rs_k Z^{-1} = \Rs_k$ for 
all $k \in I$. 

\proclaim{Proposition 2.2} The surjective map $\xi : \Js \mapsto \Ts$ 
given by
$$\xi(J_{i_1}\cdots J_{i_m}) = \tau_{i_1}\cdots \tau_{i_m}, \qquad 
i_1,\ldots,i_m \in I, \quad m \in \IN,  $$
is a group homomorphism. Its kernel is a subgroup $\Zs $ 
of internal symmetries of the net $\irnet$ which is contained 
in the center of $\Js$.
\endproclaim

\demo{Proof} If $J_{i_1}\cdots J_{i_m} = J_{j_1}\cdots J_{j_n}$, then one has
$$\Rs_{\tau_{i_1}\cdots \tau_{i_m} (k)} = 
J_{i_1}\cdots J_{i_m} \Rs_k J_{i_m}\cdots J_{i_1} =
J_{j_1}\cdots J_{j_n} \Rs_k J_{j_n}\cdots J_{j_1} =
\Rs_{\tau_{j_1}\cdots \tau_{j_n} (k)} \quad , $$

\nind for all $k \in I$. Thus the equality 
$\tau_{i_1}\cdots \tau_{i_m} = \tau_{j_1}\cdots \tau_{j_n}$ follows. It is 
therefore clear that the map $\xi$ is well-defined. Moreover,
$$\xi(J_{i_1}\cdots J_{i_m})\xi(J_{j_1}\cdots J_{j_n}) =
\tau_{i_1}\cdots \tau_{i_m}\tau_{j_1}\cdots \tau_{j_n} =
\xi(J_{i_1}\cdots J_{i_m}J_{j_1}\cdots J_{j_n}) \quad , $$

\nind and by Lemma 2.1 (1), it follows that
$$\xi(J_{i_1}\cdots J_{i_m})^{-1} = \tau_{i_m}^{-1}\cdots \tau_{i_1}^{-1} =
\tau_{i_m}\cdots \tau_{i_1} = \xi(J_{i_m}\cdots J_{i_1}) = 
\xi((J_{i_1}\cdots J_{i_m})^{-1}) \quad . $$

\nind Hence $\xi$ is a group homomorphism. \par
     If $\xi(J_{i_1}\cdots J_{i_m}) = \iota$, then the operator 
$Z = J_{i_1}\cdots J_{i_m}$ is an internal symmetry, by definition. It remains
to be shown that the set $\Zs$ of internal symmetries is contained in the 
center of $\Js$. But as argued before, since $Z\Omega = \Omega$ and 
$Z\Rs_iZ^{-1} = \Rs_i$, for all $i \in I$, it follows from standard arguments 
in modular theory (see Theorem 3.2.18 in \cite{18}) that $Z$ commutes with
the modular involutions $J_i$, $i \in I$. But $\Js$ is generated by these 
operators and $Z$ is an element of $\Js$, so the proof of the statement is
complete. 
\hfill\qed\enddemo

     This proposition may be reformulated as the assertion that there exists a
short exact sequence 

$$ \idty \rightarrow \Zs \overset{\imath}\to{\rightarrow} \Js 
\overset{\xi}\to{\rightarrow} \Ts \rightarrow \iota,$$

\nind where $\imath$ denotes the natural identification map. In other words,
$\Js$ is a central extension of the group $\Ts$ by $\Zs$, a situation 
for which the mathematics has reached a certain maturity. \par
     It is an immediate consequence of the preceding that there exists an
(anti)unitary projective representation of the group $\Ts$ on $\Hs_{\omega}$
by operators in $\Js$. For an arbitrary $\tau \in \Ts$ there may be many ways 
of writing $\tau$ as a product of the elementary $\{\tau_i \mid i \in I \}$. 
For each $\tau \in \Ts$ choose some product 
$\tau = \underset{j = 1}\to{\overset{n(\tau)}\to{\Pi}} \tau_{i_j}$; which 
choice one makes is irrelevant for our immediate purposes. Having made such a 
choice for each $\tau \in \Ts$, define 
$J(\tau) \equiv \underset{j = 1}\to{\overset{n(\tau)}\to{\Pi}} J_{i_j}$.

\proclaim{Corollary 2.3} The above construction provides an (anti)unitary 
projective representation of $\Ts$ on $\Hs_{\omega}$ with coefficients
in an abelian group $\Zs$ of internal symmetries in the center of $\Js$. 
Moreover, one has $J(\tau)\Omega = \Omega$, for all $\tau \in \Ts$, as well as 
$\Zs\Omega = \Omega$.
\endproclaim

\demo{Proof} Consider $\tau,\tau' \in \Ts$ and the corresponding (anti)unitary
operators $J(\tau)$, $J(\tau')$ and $J(\tau\tau')$. If $\xi : \Js \mapsto \Ts$
is the group homomorphism established in Proposition 2.2, one has
$\xi(J(\tau\tau')^{-1}J(\tau)J(\tau')) = \iota$, and the initial assertion 
thus 
follows from that Proposition. The final assertions are trivial, since the 
modular conjugations $J_i$ leave $\Omega$ invariant.
\hfill\qed\enddemo

     It is an interesting mathematical question which groups and corresponding
representations can arise in this manner. As we shall see, both finite 
and continuous groups can be obtained with appropriate choices of net and index
set. Before dealing with infinite groups, let us briefly discuss the finite
case and consider a family $\{ \Rs_i \mid i = 1,\ldots,n \}$ of von 
Neumann algebras with a common cyclic and separating vector $\Omega$ such that 
the corresponding modular conjugations $J_i$ leave this family invariant, 
{\it i.e.} $J_i \Rs_k J_i = \Rs_{\tau_i(k)}$ for $i,k = 1,\ldots,n$. The maps 
$\tau_i$ are in this case permutations on the set $I \equiv \{ 1,\ldots,n \}$ 
which are also involutions. Hence, the group $\Ts$ is a subgroup of the 
symmetric group $S_n$ which is generated by involutions -- a Coxeter group. 
Here we shall only consider the case where $\Ts$ acts transitively upon 
the set $I$. \par
     If the algebras $\Rs_i$ are nonabelian, then it is clear that 
$\Ts$ cannot be primitive. This is because $J_i \Rs_i J_i = \Rs_i{}'$, so that 
if $\Rs_i$ is not maximally abelian, one must have, by hypothesis, 
$\Rs_i{}' = \Rs_{i'}$ for some $i' \neq i$. But, since from 
$J(\tau)\Rs_i J(\tau)^{-1} = \Rs_i{}'$ follows 
$J(\tau)\Rs_i{}' J(\tau)^{-1} = \Rs_i$, the index pair $(i,i')$ is either 
transformed by the elements of $\Ts$ onto itself or onto a disjoint pair. 
In other words, $(i,i')$ is a set of imprimitivity of $\Ts$. \par
     Since $\Ts$ is not primitive, it also is not 2-transitive (Satz II.1.9 in 
\cite{40}). Moreover, since the magnitude of every set of imprimitivity 
in $I$ must be a divisor of the magnitude of $I$, (see, {\it e.g.} Satz II.1.2 
in \cite{40}), the magnitude $n$ of $I$ is then necessarily even. Hence, 
if $n$ is odd, then all the algebras must be maximally abelian (the converse 
is false). \par
     It is easy to compute explicitly the possible groups $\Ts$ which arise
in this manner for small
values of $n$. In the case $n = 2$ one clearly obtains $S_2$; for
$n = 3$ one finds as the only possibility the symmetric group $S_3$. (And one
can give corresponding examples of states and algebras which yield $S_3$.)
The case $n=4$ is not possible for a family of nonabelian algebras, since
then the mentioned sets of imprimitivity are stable under the action of
the group $\Ts$; in other words, $\Ts$ cannot act transitively on $I$ when
$n=4$. This list can be continued without great effort, but a complete
classification of the finite groups $\Ts$ which can be obtained in this 
manner is yet an open problem.  \par

\heading III. Geometric Modular Action in Quantum Field Theory   \endheading

     We turn now to the physically interesting case of nets on a 
space-time manifold $(\Ms,g)$. The index set $I$ appearing in the abstract
formulation of our Condition of Geometric Modular Action in the previous 
chapter will be denoted henceforth by $\Ws$ and will consist of certain open 
subsets $W \subset \Ms$. A natural question is: for the given 
manifold $(\Ms,g)$, how should the index set of the net $\wnet$ of algebras be 
chosen so that any state on that net satisfying the Standing Assumptions of
Chapter II yields a group $\Ts$ which can be identified with a subgroup of
isometries of the space-time? Evidently, not every choice of such regions 
will be appropriate. One purpose of this chapter is to explain which 
considerations should be made when choosing $\Ws$, once the underlying 
space-time has been fixed. After this is done, we specify in detail the 
technical assumptions which constitute our Condition of Geometric Modular 
Action, which was heuristically presented in the introduction. \par  
     We emphasize that in this chapter the starting point is a smooth
manifold $\Ms$ and that some target space-time $(\Ms,g)$ has already been
fixed. In other words, we have in mind a particular metric structure on $\Ms$ 
for which we are looking. If one does not have a specific target, that
is to say if one just has a net $\net$ indexed by open subregions $R$ of the 
manifold without any further clue to the metric structure on the manifold
$\Ms$, then, in principle, one would have to test the Condition of Geometric
Modular Action for various states and for various subnets $\wnet$ of $\net$. 
If the Condition of Geometric Modular Action would hold for one of these, then 
the program outlined below in this section would be applicable to that state 
and subnet. \par
     As the Condition of Geometric Modular Action is to be an 
{\it a priori} criterion for a characterization of elementary states on 
$(\Ms,g)$, the set $\Ws$ should depend only on the space-time manifold 
$(\Ms,g)$. Moreover, it should be sufficiently large to express all desired 
features of nets on $(\Ms,g)$ such as locality, covariance (in the presence of 
spacetime symmetries), {\it etc}. On the other hand, it should be as small as 
possible in order to subsume a large class of theories (on the target
space-time). \par
     In light of these requirements, it is natural to assume that $\Ws$ has
the following properties. \par

\smallpagebreak

     (a) For each $W \in \Ws$ the causal (spacelike) complement $W'$ of
$W$ ({\it i.e.} the interior of the set of all points in $\Ms$ which cannot be 
connected with any point in the closure $\overline{W}$ of $W$ by a causal 
curve) is also contained in $\Ws$. It is convenient to require each $W \in \Ws$
to be causally closed, that is to say $W = (W')' \equiv W''$. Moreover, the 
collection $\Ws$ should be large enough to separate spacelike separated points 
in $\Ms$. \par
     (b) The set $\Ws$ is stable under the action of the group of isometries
(spacetime symmetries) of $(\Ms,g)$. \par

\smallpagebreak

\nind The latter constraint is consistent with the idea that the Condition
of Geometric Modular Action should characterize the most elementary states
on $(\Ms,g)$ with the highest symmetry properties. \par
     We append to the preceding conditions another constraint of a topological
nature. In order to motivate it, let us assume for a moment that the 
transformations $\tau_W$, $W \in \Ws$, on the index set $\Ws$ arising from a 
given net and state satisfying the Condition of Geometric Modular Action are 
induced by diffeomorphisms (or even just homeomorphisms) of $\Ms$ and together 
act transitively on $\Ws$. This is only possible if all regions in 
$\Ws$ belong to the same homotopy class. We therefore assume the following 
additional condition. \par

\smallpagebreak

     (c) All regions $W \in \Ws$ are contractible. \par

\smallpagebreak

     Condition (c) excludes, for example, the appearance of double cones in
$\Ws$ when $(\Ms,g)$ is asymptotically flat (such as Minkowski space), since
their causal complements, which are to be elements of $\Ws$ by condition (a),
are not contractible. But double cones {\it would} be admissible in
space-times such as the Einstein universe. We shall call families $\Ws$
of open regions $W \subset \Ms$ satisfying (a)-(c) {\it admissible}. \par
     Given an admissible family $\Ws$ of regions, it may contain proper 
subfamilies $\Ws_0 \subset \Ws$ which are also admissible. One could then base 
the Condition of Geometric Modular Action on the subnet indexed by $\Ws_0$, 
instead. It should be noticed that there may exist nets
which satisfy our condition with respect to $\Ws$ but not for $\Ws_0$. In
other words, the subgroup $\Ts_0 \subset \Ts$ induced by the underlying
modular involutions corresponding to $W_0 \in \Ws_0$ may not be a stability
group of $\Ws_0$ in certain cases. However, it seems plausible that there 
exists a larger class of theories (nets and states) satisfying the condition
based on $\Ws_0$ than that based on $\Ws$, since there are fewer constraints
imposed on the nets in the former case. So from this point of view, it appears
to be natural to select sets $\Ws$ which, heuristically speaking, are small.
\par
     It is of interest in this context that for certain space-times $(\Ms,g)$
with large isometry groups, there exist distinguished families $\Ws$ which
are generated by applying the isometry group to a single region $W$, which 
itself has a maximal stability group, ({\it i.e.} a group which cannot be 
extended to the stability group of some other region which is still a member 
of the admissible family). Identifying $\Ws$ with the collection of 
corresponding coset spaces, it is then meaningful to say that these families 
are minimal and thus very natural candidates for a concrete formulation of the 
Condition of Geometric Modular Action. We shall consider certain examples of 
this type in the subsequent chapters. \par
     As was explained in the introduction, it is one of the aims of the 
Condition of Geometric Modular Action to distinguish, for any given net $\wnet$
of $C^*$-algebras on the manifold $\Ms$, states $\omega$ on the net which can 
be attributed to the most symmetric physical systems in the space-time 
$(\Ms,g)$. Fix an admissible family $\Ws$ of regions in $(\Ms,g)$ and consider 
the von Neumann algebras $\wrnet$ associated to $(\wnet,\omega)$ as before. We
state our Condition of Geometric Modular Action (henceforth, CGMA) for this 
structure. \par

\proclaim{Condition of Geometric Modular Action}
Let $\Ws$ be an admissible family of open regions in the space-time $(\Ms,g)$, 
let $\wnet$ be a net of $C^*$-algebras indexed by $\Ws$, and let $\omega$ be 
a state on $\wnet$. The CGMA is fulfilled if the corresponding net $\wrnet$ 
satisfies \par
   (i) $W \mapsto \Rs(W)$ is an order-preserving bijection, \par
   (ii) for $W_1,W_2 \in \Ws$, if $W_1 \cap W_2 \neq \emptyset$, then $\Omega$ 
is cyclic and separating for $\Rs(W_1) \cap \Rs(W_2)$, \par
   (iii) for $W_1,W_2 \in \Ws$, if $\Omega$ is cyclic and separating for 
$\Rs(W_1) \cap \Rs(W_2)$, then 
$\overline{W_1} \cap \overline{W_2} \neq \emptyset$, \par
\nind and \par
   (iv) for each $W \in \Ws$, the adjoint action of $J_W$ leaves the set
$\wrnet$ invariant.
\endproclaim

     The somewhat curious lack of symmetry in conditions (ii) and (iii)
is introduced in order to admit theories for which 
$W_1 \cap W_2 = \emptyset$, but nonetheless the vector $\Omega$ is cyclic and 
separating for the intersection $\Rs(W_1) \cap \Rs(W_2)$. This can occur, for 
example, in certain massless models in Minkowski space, when $W_1$ and $W_2$ 
are disjoint wedgelike regions but where $\overline{W_1} \cap \overline{W_2}$
contains an unbounded lower-dimensional set. \par
     We would like to emphasize that this condition is to be viewed as a 
selection criterion for states of {\it particular} physical interest. We
do not assert that {\it every} state of physical interest will satisfy this
condition. We observe that its formulation does not require any specific 
structure of the net $\wnet$ such as local commutativity, existence of 
spacetime symmetries, and so forth. As a matter of fact, $\wnet$ could be a 
free net on the manifold $\Ms$ satisfying no other relations but isotony. 
The above assumptions (i)-(iv) imply the Standing Assumptions of Chapter II, 
so that all the results from that chapter will be available to us. 
In particular, we have a group $\Ts$ of bijections acting on $\Ws$. The 
corresponding maps $\tau_W$ on $\Ws$ have additional convenient properties. 
\par

\proclaim{Proposition 3.1} Let $\Ws$ be an admissible family of open regions
in the space-time $(\Ms,g)$, and let $\wnet$ be a net of $C^*$-algebras
indexed by $\Ws$. If $\omega$ is a state on $\wnet$ such that the CGMA
is satisfied, then the involutions 
$\tau_W : \Ws \mapsto \Ws$, $W \in \Ws$, satisfy the following conditions:
$$\overline{W_1} \cap \overline{W_2} = \emptyset \quad \text{implies} \quad  
\tau_W(W_1) \cap \tau_W(W_2) = \emptyset \quad , \tag{3.1} $$
and 
$$W_1 \subset W_2 \quad \text{if and only if} \quad \tau_W(W_1) \subset 
\tau_W(W_2) \quad , \tag{3.2}$$ 
with $W_1, W_2 \in \Ws$.
\endproclaim

\demo{Proof} Since each $J_W$ is antiunitary and leaves $\Omega$ invariant,
it is evident that the set $(\Rs(W_1) \cap \Rs(W_2))\Omega$ is dense if and
only if the set 
$$\align
J_W(\Rs(W_1) \cap \Rs(W_2))\Omega &= 
(J_W\Rs(W_1)J_W \cap J_W\Rs(W_2)J_W)\Omega \\
&= (\Rs(\tau_W(W_1)) \cap \Rs(\tau_W(W_2)))\Omega
\endalign $$
is dense. Hence (3.1) follows from (ii) and (iii). The assertion (3.2) 
is a consequence of (i). 
\hfill\qed\enddemo

     The lack of symmetry in conditions (ii) and (iii) above entails the
lack of symmetry in (3.1). If the map $\tau_W$ were continuous in the 
obvious sense, then (3.1) would imply 
$$\tau_W(W_1) \cap \tau_W(W_2) = \emptyset \quad \text{if and only if} \quad
W_1 \cap W_2 = \emptyset \quad . \tag{3.3} $$
For the two examples worked out in the present paper, it will be seen 
that in Minkowski space the maps on the index sets $\Ws$ do indeed satisfy 
(3.3). In de Sitter space, condition (iii) is trivial and will be supplemented
by an algebraic condition yielding (3.3).  \par
     Having thus fixed the framework in detail, there arises the interesting
question: which transformation groups $\Ts$ are associated with states 
fulfilling this criterion and how do they act on the corresponding nets? In 
particular, are they implemented by point transformations on the manifold
$\Ms$, and are these isometries of the space-time $(\Ms,g)$? A
comprehensive answer to this question does not seem to be an easy problem, but
there are some engaging facts of a quite general nature which we wish to 
explain.  \par
     Let us first consider the question of whether the elements of $\Ts$
could be implemented by point transformations on $\Ms$. {\it If} we knew from 
the outset that the maps $\tau_W$ also leave stable a larger net $\trnet$ 
containing $\wrnet$ and indexed by a {\it base} for the topology on $\Ms$, we 
could rely upon an approach initiated by Araki \cite{4} (building upon 
\cite{7}\cite{6}) and further developed by Keyl \cite{43} in order to prove 
under certain conditions that the maps $\tau_W$ are induced by point 
transformations of $\Ms$ which generate a group $\Gs$. If these maps also 
preserved the causal structure on $(\Ms,g)$ in the sense of 
$$\tau_W(W_0)' = \tau_W(W_0{}') \quad , \quad  \text{for all} \quad W_0 \in \Ws \quad , \tag{3.4}$$
for each $W \in \Ws$,\footnote{It is of interest to note that we shall
{\it derive}, not assume, (3.4) in our examples, hence deduce, not 
postulate, locality and Haag duality for wedge algebras.} then since the 
regions in $\Ws$ separate spacelike separated points, we could, for a 
significant class of spacetimes, appeal to the well-known result of 
Alexandrov \cite{2}\cite{3}, (see also Zeeman \cite{77}, Borchers and 
Hegerfeldt \cite{11}, Lester \cite{48}, Benz \cite{8}) and conclude 
that the group $\Gs$ is a subgroup of the conformal group of $(\Ms,g)$.
\footnote{Some of the details of the argument which would be involved here may 
be gleaned from the proofs presented in Section 4.1. The basic ideas are 
sketched in Section 3 of \cite{62}.} Moreover, as was shown in 
the preceding chapter, there exists an (anti)unitary projective representation 
of $\Ts$ (and thus of $\Gs$) on the Hilbert space $\Hs_{\omega}$. Well-known 
examples which nicely illustrate this scenario are conformal quantum field 
theories on compactified Minkowski space (see \cite{21}). \par
     However, in order to cover a larger class of spacetimes, we would like
to avoid the initial strong assumption that the adjoint action of the
modular involutions $\{ J_W \mid W \in \Ws \}$ leaves the net $\trnet$
invariant. In particular, the CGMA
can obtain without the maps $\tau_W$ being induced by point
transformations of $\Ms$. In order to indicate what can occur, let us 
consider any decreasing net $\{ \underset{i}\to{\cap} W_{i,n} \}_{n \in \IN}$
which converges to some point $x \in \Ms$. Because $\tau_W$ is 
order-preserving, the images 
$\{ \underset{i}\to{\cap} \tau_W(W_{i,n}) \}_{n \in \IN}$ also form a 
decreasing net, and if the limit set is nonempty, it is straightforward to
show that it consists of a single point (see \cite{4}). But the net may have
no limit for certain points $x \in \Ms$. Hence, loosely speaking, our 
CGMA admits the possibility of singular point
transformations which are {\it not} contained in the conformal group of
$(\Ms,g)$ but which nonetheless preserve the causal structure.\footnote{See
also the example discussed at the end of Section 4.1.} This 
flexibility is actually very advantageous for our purposes, 
since the conformal group is rather small for certain space-times and thus not 
suitable for the characterization of elementary physical states. Hence, the
CGMA may still be a useful selection 
criterion for physically interesting states even in these cases, where the
point transformation group $\Gs$ has very little indeed to say about the
underlying space-time. \par     
     We conclude this chapter with a list of mathematical problems which
naturally arise if one wants to use our principle of geometric modular action 
for the determination of the possible symmetry groups $\Ts$ and their action on
nets for a given space-time $(\Ms,g)$. The first step is to pick an admissible
family $\Ws$ of regions $W \subset \Ms$. We do not have a general algorithm 
for the choice of $\Ws$, but, as previously mentioned, there do exist 
space-times for which the family $\Ws$ is uniquely fixed by our general 
requirements. One then has to solve, step by step, each of the following 
problems. \par

\smallpagebreak

   1) Are the transformations on $\Ws$ satisfying the conditions (3.1) and
(3.2) induced by (singular) point transformations on $(\Ms,g)$ (forming a group
$\Gs$)? \par
   2) Which subgroups $\Ts$ of the symmetric group on $\Ws$ can appear?
More precisely, which groups are generated by
families $\{ \tau_W \}_{W \in \Ws}$ of such automorphisms for which

$$\tau_{W_1}\tau_{W_2}\tau_{W_1} = \tau_{\tau_{W_1}(W_2)} \quad , \quad 
\text{for} \quad W_1,W_2 \in \Ws \quad ? $$

\nind Of special 
interest are cases where $\Ts$ is large and acts transitively on $\Ws$. \par
   3) Do $\Ws$ and $\Ts$ (as an abstract group) determine the action of the
automorphisms $\{\tau_W \}_{W \in \Ws}$? \par
   4) If the group $\Gs$ of point transformations is a continuous group or 
contains a continuous subgroup, (when) do the underlying modular involutions 
induce a continuous unitary projective 
representation of $\Gs$, respectively of its continuous subgroup? \par
   5) Can this projective representation be lifted to a continuous unitary 
representation of $\Gs$? \par
   6) If there exists a one-parameter subgroup in $\Gs$ which can be 
interpreted as time evolution on $(\Ms,g)$, what are the spectral properties
of the generator of the corresponding unitary representation? In particular,
when is the spectrum bounded from below (as one would expect in the case of
elementary physical states such as the vacuum)? \par

\smallpagebreak

     Whereas the latter three problems are standard in the representation
theory of groups, the first three are problems in the theory of transformation
groups of subsets of topological spaces, which apparently have not received the
attention they seem to deserve. We discuss in the subsequent chapters the
physically interesting examples of Minkowski space and de Sitter space,
for which the preceding program can be completely carried out. Our proofs
are largely based on explicit calculations which do not yet provide the basis
for a more general argument. But as our results are promising, we believe that 
a more systematic study of these mathematical problems would be worthwhile. 
\par

\heading IV. Geometric Modular Action Associated With Wedges in $\RR^4$  
\endheading

    We now carry out the program outlined at the end of the preceding chapter
for the case of four-dimensional Minkowski space with the standard metric

$$g = \diag(1,-1,-1,-1) \equiv 
\left( \matrix 1 & 0 & 0 & 0 \\ 
               0 & -1 & 0 & 0 \\
               0 & 0 & -1 & 0 \\
               0 & 0 & 0 & -1  \endmatrix \right) 
\quad . \tag{4.1}$$

\nind in proper coordinates as the target space. The isometry group of this 
space is the Poincar\'e group $\Ps$ and an admissible family $\Ws$ of regions 
is obtained by applying the elements of $\Ps$ to a single wedge-shaped region 
of the form
$$W_R \equiv \{ x \in \RR^4 \mid x_1 > \vert x_0 \vert \} \quad , \tag{4.2}$$
{\it i.e.} $\Ws = \{ \lambda W_R \mid \lambda \in \Ps \}$, where
$\lambda W_R = \{ \lambda(x) \mid x \in W_R \}$. It is
easy to show that $\Ws$ is an admissible family in four-dimensional
Minkowski space. Because of the requirement that the admissible family be 
mapped onto itself by the isometry group of the space-time, an admissible 
family $\Ws$ in the case of Minkowski space must contain the orbit of each of 
its elements under the action of the Poincar\'e group. Recall that an 
admissible family $\Ws$ is called minimal if it coincides with the orbit under
the action of the isometry group of a single region with a maximal
stability group. As the only open, causally closed regions which are invariant
under the stability group $\IP(W_R)$ of $W_R$ are $W_R$ itself, its
causal complement $W_R '$ and the entire space $\RR^4$, one concludes that
$\RR^4$ is the only open, causally closed region which is stable under the 
action of any proper extension of $\IP(W_R)$. Hence, $\Ws$ is a minimal
admissible family for four-dimensional Minkowski space. We therefore base the 
analysis in this chapter on this canonical choice of regions. We remark that,
in fact, one has $\Ws = \{ \lambda W_R \mid \lambda \in \Pid \}$, where
$\Pid$ is the identity component of the Poincar\'e group. \par
    Note that the metric is introduced because a specific target space is
envisioned. The wedges in the smooth manifold $\RR^4$ can be 
defined without reference to the Minkowski metric by introducing coordinates.
Then the set $\Ws$ of wedges is determined only up to diffeomorphism,
which is all we shall require. Nonetheless, it is clear that there is nothing 
intrinsic about such a definition of wedges. For a discussion of a possible 
means to determine an {\it intrinsic} algebraic characterization of ``wedges'' 
for our purpose, see Chapter VII. \par
     We commence with a state on an initial net $\wnet$ which satisfies the 
CGMA discussed in the previous chapter. In
Section 4.1 we consider the elements of the transformation group $\Ts$ 
associated with any such state and establish a considerable extension
of the Alexandrov-Zeeman-Borchers-Hegerfeldt theorems by showing that these
maps are induced by point transformations which form a subgroup $\Gs$ of the
Poincar\'e group. This section also contains a simple example of a space-time 
manifold and well-behaved transformations of a corresponding family of regions 
which are {\it not} induced by point transformations. \par
     In Section 4.2 and the subsequent 
sections, we restrict attention to those cases where the transformation group 
$\Ts$ is large enough to act transitively upon the set $\Ws$. It turns out 
that $\Gs$ then contains the full identity component $\Pid$ of the Poincar\'e 
group $\Ps$. The specific form of the Poincar\'e elements corresponding to
the generating involutions in $\Ts$, which themselves arise from the adjoint
action of the initial modular conjugations upon the net $\wrnet$, is also
identified in Section 4.2, and it is found that this form is uniquely fixed
and agrees with the one first determined by Bisognano and Wichmann for the
case of the vacuum state on Minkowski space and any net of von Neumann algebras
locally associated with a quantum field satisfying Wightman's axioms 
\cite{9}\cite{10}. It then follows from this explicit knowledge of the form
of the implementing Poincar\'e elements that $\Gs$ is exactly equal to the
proper Poincar\'e group $\Ps_+$. Thus, starting with the CGMA, we find a 
unique and familiar solution for the possible 
symmetry groups and their respective actions. \par
     In the remaining portion of Chapter IV we discuss the properties of the
representations of $\Ts$ -- and hence of $\Gs = \Ps_+$ -- which are induced by 
the modular conjugations. In Section 4.3 we shall identify a natural continuity
condition on the net $\wrnet$ which implies that there exists a strongly
continuous (anti)unitary projective representation of $\Gs = \Ps_+$. This
requires a certain choice of product decomposition in the definition of the
projective representation (cf. the discussion before Corollary 2.3). These 
results are used in Section 4.3 for the proof, first of all, 
that one can always lift this projective representation to a continuous 
unitary representation of the covering group of $\Pid$. Our analysis, which is 
based on results in Borel measurable group cohomology theory and is carried 
out in the Appendix, parallels to some extent the discussion in \cite{22}; 
but our more global point of view and our explicit construction of the
projective representation provide certain simplifications. In particular, we 
shall not need to argue {\it via} the Lie algebra, since the results of 
Section 4.2 and modular theory give us sufficient control over our explicit 
representation. And then we show that, after all, this representation of the 
covering group provides a strongly continuous representation of $\Pid$ and
{\it coincides} with the initially and explicitly constructed projective 
representation. \par
     It is worth emphasizing that we explicitly construct a strongly continuous
unitary representation of the translation subgroup (using ideas of \cite{24}), 
thereby determining the generator of the timelike translations, which has the 
physical interpretation of the Hamiltonian, or total energy operator, of the 
theory. In other words, we derive the dynamics of the theory from the physical 
data of the state and net of observable algebras. \par
     We recall that it is the main purpose of this chapter to illustrate the
steps which are necessary to apply the CGMA in our program.
As already mentioned at the end of Chapter III, the mathematics relevant to
the first three group theoretical problems does not seem to be sufficiently
well developed for our purposes, and we must therefore rely on explicit and
sometimes tedious computations to carry out our program. But our results
demonstrate that the CGMA, which at first
glance appears very general and diaphanous, actually imposes strong constraints
on the admissible states and allows one to characterize the vacuum states
in the case of Minkowski space. \par  

\bigpagebreak
{\bf 4.1. Wedge Transformations Are Induced By Elements of the Poincar\'e 
Group} 
\bigpagebreak

     The aim of this section is to show that the elements of the transformation
group $\Ts$ acting upon the wedges $\Ws$, which arises when one assumes the
CGMA discussed in the
previous chapter, are induced by point transformations on Minkowski space,
indeed, by elements of the Poincar\'e group. In other words, we wish to show
that $\Ts$ can be identified with a subgroup of the Poincar\'e group. Since 
one can define points as intersections of edges of suitable wedges, it is an 
intuitively appealing possibility that transformations of wedges could lead to 
point transformations. The assumptions made in this section are slightly more 
general than actually needed for our primary purpose, but these somewhat more 
general results have interest going beyond the immediate problem we are 
addressing. In particular, we shall also employ these 
results in Chapter V, where we consider the consequences of the geometric 
action of modular groups. \par

     In the remainder of this section, we shall assume that we have a 
bijective map $\hal : \Ws \mapsto \Ws$ with the following properties:  \par

\smallpagebreak

   (A) If $W_1, W_2 \in \Ws$ satisfy 
$\overline{W_1} \cap \overline{W_2} = \emptyset$, then  
$\hal(W_1) \cap \hal(W_2) = \emptyset$ and 
$\hal^{-1}(W_1) \cap \hal^{-1}(W_2) = \emptyset$; \par
   (B) $W_1, W_2 \in \Ws$ satisfy $W_1 \subset W_2$ if and only 
if $\hal(W_1) \subset \hal(W_2)$. \par

\smallpagebreak

     By Prop. 3.1, these are properties shared by the maps $\tau_W$, 
$W \in \Ws$, arising from states complying with the CGMA. We do not assume in 
this section that the map $\hal$ is an involution or that (3.3) holds. 
We shall show that conditions (A) and (B) {\it imply} (3.3). \par
     We introduce the following notation: $\ell \in \RR^4$ denotes a 
future-directed lightlike vector and $p \in \RR$ a real parameter. For given 
$\ell, p$ we define the characteristic half-spaces
$$H_p[\ell]^{\pm} \equiv \{ x \in \RR^4 \mid \pm (x \cdot \ell - p) > 0 \} 
\quad . \tag{4.1.1} $$
Note that the boundary of such a half-space,
$H_p[\ell] = \partial H_p[\ell]^{\pm} = $ \newline
$ = \{ x \in\RR^4\mid x \cdot\ell = p \}$,
is a characteristic hyperplane with the properties that all lightlike vectors
parallel to this hyperplane are parallel to $\ell$ and all other vectors 
parallel to $H_p[\ell]$ are spacelike.
Given two such pairs, $\{\ell_i, p_i\}$, $i = 1,2$, where $\ell_1$ and
$\ell_2$ are not parallel, then $W = H_{p_1}[\ell_1]^+ \cap H_{p_2}[\ell_2]^-$ 
is a wedge. All wedges can be obtained in this manner. In particular, for any 
wedge $W \in \Ws$ there exist two future-directed lightlike vectors 
$\ell_{\pm}$ such that $W \pm \ell_{\pm} \subset W$. These vectors are unique 
up to a positive scaling factor. The half-spaces $H^{\pm}$ generating $W$ as 
above are given by
$$H^{\pm} = \underset{\lambda \in \RR}\to{\cup} (W + \lambda \ell_{\mp}) \quad
. \tag{4.1.2}$$
In the sequel, we shall denote by $\Fs^{\pm}$ the following family
of wedges:

$$\Fs^{\pm} \equiv \{ W + \lambda \ell_{\mp} \mid \lambda \in \RR \} \quad . $$

\nind We shall say that $\Fs^{\pm}$ generates $H^{\pm}$ via (4.1.2). Note that 
every such family $\Fs^{\pm}$ has the following properties:
\par
    (i) $\Fs^{\pm}$ is linearly ordered, {\it i.e.} if 
$W_1, W_2 \in \Fs^{\pm}$, then either $W_1 \subset W_2$ or $W_2 \subset W_1$.
\par
    (ii) $\Fs^{\pm}$ is maximal in the sense that if $W_1, W_2 \in \Fs^{\pm}$
satisfy $W_1 \subset W_2$ and there exists a wedge $W \in \Ws$ such that
$W_1 \subset W \subset W_2$, then $W \in \Fs^{\pm}$. \par 
    (iii) $\Fs$ has no upper or lower bound in $(\Ws,\subset)$, {\it i.e.} 
there exists no element $W_< \in \Ws$ such that $W_< \subset W$ for all
$W \in \Fs$ and also no element $W_> \in \Ws$ such that $W_> \supset W$ for 
all $W \in \Fs$. \par

     We shall call a collection of wedges $\Fs \subset \Ws$ with the 
properties (i)-(iii) a characteristic family of wedges. Every characteristic family of
wedges is, in fact, of the form of $\Fs^{\pm}$. The proof of this assertion
rests upon the following well-known properties of wedges. For
wedges $W, W_0 \in \Ws$ with $W_0 \subset W$ and $W_0 \neq W$, there exists
a space- or lightlike translation $a \in \RR^4$ such that

$$W_0 = W+a \subset W + \lambda a \subset W \qquad \text{for all} \quad
0 \leq \lambda \leq 1 \quad . $$

\nind If the edge of $W_0$ lies on the boundary of $W$, then the translation
$a$ can be chosen to be lightlike (and is therefore a multiple of one of the
lightlike vectors $\ell_{\pm}$ determining $W$). On the other hand, if the
edge of $W_0$ lies in the interior of $W$, then there exists an open 
set $\Ns \subset \RR^4$ such that $W_0 \subset W + a \subset W$, for 
all $a \in \Ns$. As in \cite{24}, we shall say that two wedges 
$W_1, W_2 \in \Ws$ are coherent if one is obtained from the other by a 
translation, or, equivalently, if there exists another wedge $W_3$ such that 
$W_1 \subset W_3$ and $W_2 \subset W_3$. Hence, all wedges in a characteristic 
family are mutually coherent. We now prove the initial assertion.   \par

\proclaim{Lemma 4.1.1} Every characteristic family of wedges $\Fs$ has the 
form \newline
$\Fs = \{ W + \lambda\ell \mid \lambda \in \RR \}$, for some wedge $W \in \Ws$
and some future-directed lightlike vector $\ell$ with the property that
$W + \ell \subset W$ or $W - \ell \subset W$.
\endproclaim

\demo{Proof} Let $W_0,W \in \Fs$. By the linear
ordering of $\Fs$, one may assume without loss of generality that
$W_0 \subset W$. If the edge of $W_0$ would lie in the interior of $W$,
then, as mentioned above, there exists an open set $\Ns$ in $\RR^4$ such that 
$W_0 \subset W + a \subset W$, for all $a \in \Ns$. By the maximality of 
$\Fs$ in $\Ws$, this would entail that $W + a \in \Fs$, for all $a \in \Ns$. 
However, the elements of $\{ W + a \mid a \in \Ns \}$
clearly violate the linear ordering of $\Fs$. Hence, the edge of $W_0$ must
lie on the boundary of $W$, so there exists a lightlike translation 
$a \in \RR^4$ such that $W_0 = W + a \subset W$. \par
     Let now $W, W+a, W+b \in \Fs$ be chosen such that $a$ and $b$ are
lightlike and $W + a \subset W \subset W + b$. As in the preceding paragraph
one shows that the edge of $W+a$ lies on the boundary of $W+b$. The assumed
inclusion then implies that the edge of $W$ lies on the same characteristic
hyperplane. This entails that $a$ and $b$ are proportional, {\it i.e.} the 
elements of $\Fs$ are all of the form $W + \lambda\ell$ with real $\lambda$ and
future-directed lightlike vector $\ell \in \RR^4$. That every 
$\lambda \in \RR$ must occur follows at once from properties (ii) and (iii)
of characteristic families. 
\hfill\qed\enddemo

     In the next lemma we show that order-preserving bijections 
$\hal : \Ws \mapsto \Ws$ map characteristic families onto characteristic 
families.

\proclaim{Lemma 4.1.2} Let $\hal : \Ws \mapsto \Ws$ be a bijective map with
the property (B). Then $\hal$ maps every characteristic family $\Fs$ of
wedges onto a characteristic family \newline
$\hal(\Fs) \equiv \{ \hal(W) \mid W \in \Fs \}$. In fact, if
$\Fs_1 = \{ W_1 + \lambda\ell_1 \mid \lambda \in \RR \}$, for some wedge 
$W_1 \in \Ws$ and some future-directed lightlike vector $\ell_1$ with the 
property that $W_1 + \ell_1 \subset W_1$ or $W_1 - \ell_1 \subset W_1$, and if
$\hal(W_1) = W_2$, then $\hal(W_1 + \lambda\ell_1) = W_2 + f(\lambda)\ell_2$,
where $f : \RR \mapsto \RR$ is a continuous monotonic bijection,
$f(0) = 0$, and $\ell_2$ 
is a future-directed lightlike vector with the property that 
$W_2 + \ell_2 \subset W_2$ or $W_2 - \ell_2 \subset W_2$.
\endproclaim

\demo{Proof} Since $\hal$ is an order isomorphism, the linear ordering
of $\hal(\Fs)$, property (i), follows at once. If one has for some 
$W \in \Ws$ and $W_1,W_2 \in \Fs$ the inclusions 
$\hal(W_1) \subset W \subset \hal(W_2)$, one must also have the inclusions 
$W_1 \subset \hal^{-1}(W) \subset W_2$, since $\hal^{-1}$ is also an order 
isomorphism. Hence, by the maximality of $\Fs$ it follows that 
$\hal^{-1}(W) \in \Fs$, so that $W \in \hal(\Fs)$, establishing the
maximality of $\hal(\Fs)$. \par
     Finally, if there were to exist a lower bound $W_< \in \Ws$ to
$\hal(\Fs)$, then since $\hal$ is an order isomorphism, the wedge
$\hal^{-1}(W_<)$ would be a lower bound for $\Fs$, a contradiction. 
Similarly, one can exclude the existence of an upper bound in $\Ws$ for
$\hal(\Fs)$. \par
     Let $\Fs_1$, $W_1$, $\ell_1$, and $W_2$ be as indicated in the hypothesis.
Since it has just been established that inclusion-preserving bijections on 
$\Ws$ map characteristic families of wedges onto characteristic families, one 
sees from Lemma 4.1.1 that there exist future-directed
lightlike vectors $k_1,k_2$, such that $W_2 + k_1 \subset W_2$ and
$W_2 - k_2 \subset W_2$, and a function $f : \RR \mapsto \RR$ such that for
all $\lambda \in \RR$ either

$$\tau(W_1 + \lambda\ell_1) = W_2 + f(\lambda)k_1 \quad \text{or} \quad
\tau(W_1 + \lambda\ell_1) = W_2 - f(\lambda)k_2 \quad . $$

\nind Since $\tau$ is an inclusion-preserving bijection, $f$ is bijective
and monotone; hence $f$ is continuous.
\hfill\qed\enddemo

     We wish now to show that the apparent asymmetry in condition (A) can be 
removed without loss of generality; in other words, condition (3.3) holds for
the mappings considered in this section.  \par

\proclaim{Corollary 4.1.3} Let $\hal : \Ws \mapsto \Ws$ be a bijection which 
satisfies conditions (A) and (B). Then $\hal$ also satisfies 
$$W_1 \cap W_2 = \emptyset \quad \text{if and only if} \quad 
\hal(W_1) \cap \hal(W_2) = \emptyset \quad .  \tag{4.1.3}$$

\nind Relation (4.1.3) is also true for the mapping $\tau^{-1}$.
\endproclaim

\demo{Proof} Let $W_1,W_2 \in \Ws$ such that $W_1 \cap W_2 = \emptyset$ but
$\overline{W_1} \cap \overline{W_2} \neq \emptyset$. It suffices to show
that in this case one has $\tau(W_1) \cap \tau(W_2) = \emptyset$. \par
     First note that if $N$ is a convex subset of the boundary 
$\overline{W}\setminus W$ of the wedge $W$, it is contained in one 
of the two characteristic hyperplanes $H_p[\ell_{\pm}]$ determined by $W$, and 
thus it is easy to see that either

$$N \cap \overline{W + \lambda\ell_{+}} = \emptyset \quad \text{or} \quad 
N \cap \overline{W - \lambda\ell_{-}} = \emptyset \quad , $$

\nind for all $\lambda > 0$. Since both $\overline{W_1}$ and $\overline{W_2}$ 
are convex, so is their intersection 
$\overline{W_1} \cap \overline{W_2} \subset \overline{W_1}\setminus W_1$; 
hence, with $\ell_1$, $\ell_2$ future-directed lightlike vectors with 
$W_1 + \ell_1 \subset W_1$ and $W_1 - \ell_2 \subset W_1$, it follows that

$$\emptyset = \overline{W_1 + \lambda\ell_1} \cap 
   (\overline{W_1} \cap \overline{W_2}) =  
   \overline{W_1 + \lambda\ell_1} \cap \overline{W_2} $$

\nind or

$$\emptyset = \overline{W_1 - \lambda\ell_2} \cap 
   (\overline{W_1} \cap \overline{W_2}) =  
   \overline{W_1 - \lambda\ell_2} \cap \overline{W_2} \quad , $$

\nind for all $\lambda > 0$. 
     Consider the first case and note that Lemma 4.1.2 entails that
$\hal(W_1 + \lambda\ell_1) = \hal(W_1) + f(\lambda)\ell$, with 
$\hal(W_1) + \ell \subset \tau(W_1)$ or $\hal(W_1) - \ell \subset \tau(W_1)$
and $f : \RR \mapsto \RR$ a continuous bijection which is either monotone
increasing or monotone decreasing. Consider the subcase where $f$ is monotone
increasing and $\hal(W_1) + \ell \subset \tau(W_1)$. Then by the continuity
of $f$, one has 
$$\align
\tau(W_1) \cap \tau(W_2) &= (\hal(W_1) + f(0)\ell) \cap \tau(W_2) \\
 &= (\underset{\lambda > 0}\to{\cup}(\hal(W_1) +f(\lambda)\ell)) \cap \tau(W_2) \\
 &= \underset{\lambda > 0}\to{\cup}(\tau(W_1 + \lambda\ell_1) \cap \tau(W_2)) \\
  &= \emptyset \quad , 
\endalign $$
using assumption (A). On the other hand, the subcase $f$ monotone 
decreasing and $\hal(W_1) + \ell \subset \tau(W_1)$ cannot arise, since $\hal$ 
is inclusion-preserving. Similarly, the subcase $f$ monotone increasing and 
$\hal(W_1) - \ell \subset \tau(W_1)$ cannot occur. Finally, in the subcase $f$
monotone decreasing and $\hal(W_1) - \ell \subset \tau(W_1)$ one finds
the same chain of equalities as above. \par
     In the second case, namely
$\emptyset = \overline{W_1 - \lambda\ell_2} \cap \overline{W_2}$, for 
all $\lambda > 0$, one similarly sees that the subcases 
$\hal(W_1) + \ell \subset \hal(W_1)$
with $f$ increasing, and $\hal(W_1) - \ell \subset \hal(W_1)$ with $f$
decreasing are excluded by the inclusion-preserving property of $\hal$. In
the other two subcases, one has from Lemma 4.1.2 in a like manner
$$\align
\tau(W_1) \cap \tau(W_2) &= (\hal(W_1) + f(0)\ell) \cap \tau(W_2) \\
 &= (\underset{\lambda > 0}\to{\cup}(\hal(W_1) +f(-\lambda)\ell)) \cap \tau(W_2) \\
 &= \underset{\lambda > 0}\to{\cup}(\tau(W_1 - \lambda\ell_2) \cap \tau(W_2)) \\
  &= \emptyset \quad , 
\endalign $$
by assumption (A). Thus, one has proven that
$W_1 \cap W_2 = \emptyset$ implies $\tau(W_1) \cap \tau(W_2) = \emptyset$.
The argument for $\hal^{-1}$ is identical, completing the proof of the lemma.
\hfill\qed\enddemo

    To proceed further, it is convenient to use the following notation for 
wedges. For any linearly independent future-directed lightlike vectors 
$\ell_1, \ell_2 \in \RR^4$  and any $a \in \RR^4$, we define the wedge
$$\align
W[\ell_1,\ell_2,a] &\equiv \{ \alpha\ell_1 + \beta\ell_2 + \ell^{\bot} + a 
\mid \alpha > 0, \beta < 0, \ell^{\bot} \in \RR^4, 
\ell^{\bot}\cdot\ell_1 = \ell^{\bot}\cdot\ell_2 = 0 \} \\
 &= W[\ell_1,\ell_2,0] + a \quad , 
\endalign $$  

\nind where the dot product here represents the Minkowski scalar product. 
Then with 
$$\ell_{1\pm} = (1,\pm 1, 0, 0) \, , \, \ell_{2\pm} = (1,0,\pm 1, 0) \, , \,
\ell_{3\pm} = (1,0,0,\pm 1) \, , $$

\nind one sees that $W_R = W[\ell_{1+},\ell_{1-},0]$. Note that with this 
notation, one has $W[\ell_1,\ell_2,a] + \ell_1 \subset W[\ell_1,\ell_2,a]$
and $W[\ell_1,\ell_2,a] - \ell_2 \subset W[\ell_1,\ell_2,a]$, {\it i.e.}
for this wedge $\ell_+$ is a positive multiple of $\ell_1$ and $\ell_-$ is a
positive multiple of $\ell_2$. Moreover, the half-spaces $H^{\pm}$ generating 
$W[\ell_1,\ell_2,a]$ as above are given by $H^+ = H_{a\cdot\ell_2}[\ell_2]^+$ 
and $H^- = H_{a\cdot\ell_1}[\ell_1]^-$, and the associated characteristic 
families are given by
$$\Fs^+ = \{ W[\ell_1,\ell_2,a+\lambda\ell_2] \mid \lambda \in \RR \} \quad
\text{and} \quad 
\Fs^- = \{ W[\ell_1,\ell_2,a+\lambda\ell_1] \mid \lambda \in \RR \} \quad . $$

     We next show a useful characterization of pairs of spacelike separated
wedges.

\proclaim{Lemma 4.1.4} Let $W_1,W_2$ be wedges. $W_1 \subset W_2'$ if and only
if the two characteristic families $\Fs_2^+$ and $\Fs_2^-$ containing $W_2$ 
satisfy $W_1 \cap W = \emptyset$ for every $W \in \Fs_2^+ \cup \Fs_2^-$.
\endproclaim

\demo{Proof} Let $H_2^{\pm}$ be the characteristic half-spaces generated by
the families $\Fs_2^{\pm}$, so that one has $W_2 = H_2^+ \cap H_2^-$ and
$W_2 ' = H_2^{+c} \cap H_2^{-c}$, where the superscript $c$ signifies that
one takes the complementary half-space. {}From $W_1 \subset W_2 '$ follows
therefore the containment $W_1 \subset H_2^{\pm c}$ and hence also
$W_1 \cap H_2^{\pm} = \emptyset$. Conversely, the last equality follows
from the disjointness of $W_1$ from each member of the set
$\Fs_2^+ \cup \Fs_2^-$, so that one must have 
$W_1 \subset H_2^{+c} \cap H_2^{-c} = W_2 '$.
\hfill\qed\enddemo

     It is next established that bijections on $\Ws$ satisfying conditions 
(A) and (B) preserve causal complements and thus causal structure.

\proclaim{Corollary 4.1.5} A bijection $\tau : \Ws \mapsto \Ws$ which fulfills
conditions (A) and (B) also satisfies the following condition:
$$\tau(W') = \tau(W)' \quad , \quad \text{for any} \quad
W \in \Ws \quad . \tag{4.1.4}$$
\endproclaim

\demo{Proof} Consider an arbitrary wedge $W \in \Ws$, and let $\Fs^+$ and
$\Fs^-$ be the characteristic families of wedges containing $W'$. By Lemma 
4.1.2, $\tau$ maps $\Fs^+$ and $\Fs^-$ onto two characteristic families
$\tau(\Fs^+)$ and $\tau(\Fs^-)$ containing $\tau(W')$. Lemma 4.1.4
entails that $W$ is disjoint from every element of $\Fs^+ \cup \Fs^-$,
and hence Corollary 4.1.3 implies that $\tau(W)$ is disjoint from every element
of $\tau(\Fs^+) \cup \tau(\Fs^-)$. Thus, Lemma 4.1.4 yields the containment
$\tau(W') \subset \tau(W)'$. The reverse containment follows by applying
the same argument to $\tau^{-1}$.
\hfill\qed\enddemo

    We continue now with our development of point transformations. A pair 
$(W_1,W_2)$ of disjoint wedges will be called maximal if there
is no wedge $W$ properly containing $W_1$, resp. $W_2$, such that
$W \cap W_2 = \emptyset$, resp. $W \cap W_1 = \emptyset$. Note that a
bijection $\hal : \Ws \mapsto \Ws$ fulfilling conditions (A) and (B) maps
maximal pairs of wedges onto maximal pairs of wedges. We need a computational 
characterization of a maximal pair of wedges. To this end, we remark that
given a pair $(W_1,W_2)$ such that $W_2$ is not a translate of $W_1$ or 
$W_1'$, there exists a Poincar\'e transformation $(\Lambda,x)$ mapping $W_1$ 
onto $W_R$ and $W_2$ onto either the wedge $W[\ell_{2+},\ell,d]$ or its
causal complement $W[\ell_{2+},\ell,d]'$, where $\ell$ is some positive
lightlike vector which is not parallel to $\ell_{2+}$ and $d \in \RR^4$.
This follows from the observations that there always exists a Lorentz
transformation $\Lambda_1$ such that $\Lambda_1 W_1 = W_R$ and that every
positive lightlike vector not parallel to $\ell_{1\pm}$ is mapped by some 
element of the invariance group of $W_R$ to $\ell_{2+}$. We shall therefore 
consider the pair $(W_R,W[\ell_{2+},\ell,d])$ -- indeed, without loss of 
generality, the pair $(W_R,W[\ell_{2+},\ell,d])$, for suitable 
$\ell = (1,a,b,c)$ with $a^2 + b^2 + c^2 = 1$, 
$b \neq 1$, and $d \in \RR^4$ -- 
and determine under which conditions this pair is maximal. In preparation, we 
prove the following simple lemma.  

\proclaim{Lemma 4.1.6} Let $P : \RR^4 \mapsto \RR^2$ be given by
$P(x_0,x_1,x_2,x_3) = (x_0,x_1)$ and let $W =
W[\ell_{2+}, \ell, d]$ with $\ell = (1,a,b,c)$, where 
$a,b,c \in \RR$ satisfy $a^2 + b^2 + c^2 = 1$, $b \neq 1$, and 
$d \in \RR^4$. Then $PW = \RR^2$ for $b < 0$ or $c \neq 0$. On the other
hand, if $0 \leq b < 1$ and $c = 0$, one has

$$PW = \{ x \in \RR^2 \mid (x - Pd) \cdot (1-b,-a) > 0 \} \quad , $$

\nind where here the dot product represents the Euclidean scalar product on
$\RR^2$.
\endproclaim

\demo{Proof} Without loss of generality, one may assume $d = 0$. One has
$$\align
PW &= P\{\alpha\ell_{2+} + \beta(1,a,b,c) + s(c,0,c,1-b) + t(a,1-b,a,0) \mid \\
   &\qquad\qquad \alpha > 0, \beta < 0, s,t \in \RR \} \\
&= \{ \alpha(1,0) + \beta(1,a) + s(c,0) + t(a,1-b) \mid \alpha > 0, \beta < 0, s,t \in \RR \} \quad .
\endalign $$
And since $1-b \neq 0$, this shows that $PW = \RR^2$ for $c \neq 0$. 
Hence, one may restrict one's attention to $c = 0$. Since $(1-b,-a)$ is a 
normal vector for the line $\{ t(a,1-b) \mid t \in \RR \}$, the remaining 
assertions readily follow from
$$\alpha(1,0) \cdot (1-b,-a) = \alpha(1-b) > 0 \quad , $$
for $\alpha > 0$, and 
$$\beta(1,a) \cdot (1-b,-a) = \beta(1 - b - a^2) = \beta(b^2 - b) \quad  , $$
which is nonnegative for $0 \leq b < 1$ and negative for $b < 0$.
\hfill\qed\enddemo

     This straightforward observation leads to the following characterization 
of maximal pairs of wedges.

\proclaim{Lemma 4.1.7} The wedges $W_R = W[\ell_{1+},\ell_{1-},0]$ and 
$W = W[\ell_{2+},\ell,d]$, where $\ell = (1,a,b,c)$ and $a,b,c \in \RR$ 
satisfy $a^2 + b^2 + c^2 = 1$, $b \neq 1$, and $d \in \RR^4$, form a maximal 
pair of wedges if and only if $0 < a < 1$, $0 < b < 1$, $c=0$, and the vector
$d$ is a linear combination of vectors whose associated translations leave
either $W_R$ or $W$ fixed. The statement is true if $W$ is replaced by
$W'$ and the condition $0 < a < 1$ is replaced by $-1 < a < 0$ or also if
$\ell_{2+}$ is replaced by $\ell_{2-}$ and $0 < b < 1$ by $-1 < b < 0$.
\endproclaim

\demo{Proof} Using the projection $P$ from Lemma 4.1.6, note that
$x \in PW_R$ if and only if 
$$x = \alpha(1,1) - \beta(1,-1) \quad \text{for suitable} \quad
\alpha,\beta > 0 \quad . \tag{4.1.5} $$

\nind $W_R$ is invariant with respect to translations by vectors in the
subspace generated by $(0,0,1,0)$ and $(0,0,0,1)$, so one has
$W_R \cap W = \emptyset$ if and only if $PW_R \cap PW = \emptyset$.
By Lemma 4.1.6, the condition $PW_R \cap PW = \emptyset$ is equivalent to
$c = 0$, $0 \leq b < 1$ and (by (4.1.5))
$$\align
0 &\geq (\alpha(1,1) - \beta(1,-1) - Pd) \cdot (1-b,-a) \\
&= (\alpha - \beta)(1-b) - (\alpha + \beta)a - Pd\cdot(1-b,-a) \quad ,
\endalign $$
for all $\alpha,\beta > 0$. This clearly entails that $a \geq 0$.
Note also that $a,b \geq 0$ and $c=0$ imply $a>0$, since $b \neq 1$.
It is then easy to check that this implies 
$$1-b-a = (1,1)\cdot(1-b,-a) \leq 0 \tag{4.1.6}$$
and
$$1-b+a = (1,-1)\cdot(1-b,-a) > 0 \quad . \tag{4.1.7} $$
Hence, $W_R \cap W = \emptyset$ is equivalent to the conditions $c=0$, 
$0 \leq b < 1$, $a>0$, and $-Pd\cdot(1-b,-a) \leq 0$. \par
     Assume first the maximality of the pair $(W_R,W)$. Then 
$-Pd\cdot(1-b,-a) \leq 0$ and the conditions just established entail
$$(x-Pd)\cdot(1-b,-a) \leq -Pd\cdot(1-b,-a) \leq 0 \quad , \tag{4.1.8}$$
for all $x \in PW_R$. The maximality then implies the equality
$$-Pd\cdot(1-b,-a) = 0 \quad , \tag{4.1.9}$$
since, if not, one could obtain a wedge which properly contains
$W$ and yet is still disjoint from $W_R$ by choosing a different $d$ such that
(4.1.8) is still satisfied. Thus, one concludes that $Pd$ is a multiple
of $(a,1-b) = P(a,1-b,a,0)$. Therefore, $d$ is a linear combination of the
vectors $(a,1-b,a,0)$, $(0,0,0,1)$ and $(0,0,1,0)$, where translations by the
former two leave $W$ invariant and translations by the latter two leave $W_R$
fixed. \par
     The possibility that $b=0$ still remains to be excluded. But $b=0$ entails
$a=1$, so $W$, resp. $PW$, is invariant with respect to translations by
multiples of $(1,1,1,0)$, resp. $(1,1)$. Translating the disjoint pair 
$(W,W_R)$ by $d = -(1,1,1,0)$, one would therefore obtain another disjoint 
pair $(W,W_2)$ such that $W_2 = W[\ell_{1+},\ell_{1-},d]$ properly 
contains $W_R$, contradicting the assumed maximality of $(W,W_R)$. \par
     For the converse, assume that $W$ has the stated form. By the first
part of this proof, one already knows that $W$ and $W_R$ are then disjoint.
Only the proof of maximality remains. By hypothesis, (4.1.9) holds in this
direction, as well. Furthermore, (4.1.6) and (4.1.7) are fulfilled. Note that
if $b \neq 0$, then (4.1.6) holds with strict inequality. A wedge 
$W_3$ which contains $W_R$ must be coherent with $W_R$ and is thus obtained 
by translating $W_R$ by a vector of the form 
$-\alpha_0\ell_{1+} + \beta_0\ell_{1-}$, with $\alpha_0,\beta_0 \geq 0$.
For $W_3 \neq W_R$, {\it i.e.} for $\alpha_0 \neq 0$ or $\beta_0 \neq 0$,
(4.1.6) and (4.1.7) imply

$$P(-\alpha_0\ell_{1+} + \beta_0\ell_{1-})\cdot(1-b,-a) > 0 \quad . $$

\nind The vertex of $PW_3$ lies in $PW$ (Lemma 4.1.6 and (4.1.9)), hence
$PW_3 \cap PW \neq \emptyset$ and so $W_3 \cap W \neq \emptyset$. One can
argue similarly to eliminate the possibility that there does not exist a
wedge properly containing $W$ and yet being disjoint from $W_R$. \par
     To establish the final assertions of the lemma, one need but consider the
wedges transformed by suitable reflections.
\hfill\qed\enddemo

    Since the union of the elements of a characteristic family of wedges 
yields a characteristic half-space, it is natural to use Lemma 4.1.2 
to extend the map $\hal$ to the set $\Hs$ of all characteristic half-spaces in
$\RR^4$. In order to establish that this extension is well-defined, it is 
necessary to consider the possibility that two characteristic families 
generate the same half-space. \par
    According to Lemma 4.1.1, every characteristic family $\Fs$ can be 
represented in the form $\Fs = \{ W + \lambda\ell \mid \lambda \in \RR \}$. We 
define the complementary characteristic family 
$\Fs^c \equiv \{ (W + \lambda\ell)' \mid \lambda \in \RR \}$. The families
$\Fs$ and $\Fs^c$ generate complementary characteristic half-spaces $H$ and
$H^c$, respectively, {\it i.e.} $H^c = \RR^4 \setminus \overline{H}$. In order
to simplify notation, we shall write $\Fs_1 \cap \Fs_2 = \emptyset$ for two
characteristic families to mean $W_1 \cap W_2 = \emptyset$ for all 
$W_1 \in \Fs_1$ and all $W_2 \in \Fs_2$. Hence, one has 
$\Fs \cap \Fs^c = \emptyset$, for any characteristic family $\Fs$. \par

\proclaim{Lemma 4.1.8} Let $\hal : \Ws \mapsto \Ws$ be a bijection with 
properties (A) and (B). Moreover, let $\Fs_1$ and $\Fs_2$ be two 
characteristic families of wedges generating the same half-space, {\it i.e.} 
$\underset{W_1 \in \Fs_1}\to{\cup}W_1 = \underset{W_2 \in \Fs_2}\to{\cup}W_2$.
Then one has $\underset{W_1 \in \Fs_1}\to{\cup}\hal(W_1) = 
\underset{W_2 \in \Fs_2}\to{\cup}\hal(W_2)$.
\endproclaim

\demo{Proof} Since $\Fs_1$ and $\Fs_2$ generate the same half-space, one must 
have $\Fs_1 \cap \Fs_2^c = \emptyset$. Hence, Corollary 4.1.3 entails 
$\hal(\Fs_1) \cap \hal(\Fs_2^c) = \emptyset$. Similarly, one 
derives $\hal(\Fs_1^c) \cap \hal(\Fs_2) = \emptyset$. {}From (4.1.4) it also
follows that $\hal(\Fs^c) = \hal(\Fs)^c$, so that one finds
$\hal(\Fs_1)\cap\hal(\Fs_2)^c = \emptyset$ and 
$\hal(\Fs_1)^c\cap\hal(\Fs_2) = \emptyset$. By Lemma 4.1.2, $\hal(\Fs_1)$
and $\hal(\Fs_2)$ generate half-spaces $H_1$ and $H_2$, respectively,
for which the following relations must therefore hold: 
$H_1 \cap H_2^c = \emptyset$ and $H_1^c \cap H_2 = \emptyset$. It follows that
$H_1 = H_2$.
\hfill\qed\enddemo

     Lemmas 4.1.2 and 4.1.8 ensure that the following map is well-defined: \par

\proclaim{Definition} Let $\hal : \Ws \mapsto \Ws$ be a bijection satisfying
the properties (A) and (B). Then an associated map $\hal : \Hs \mapsto \Hs$ 
is obtained by setting for $H \in \Hs$

$$\hal(H) \equiv \underset{W \in \Fs}\to{\cup} \hal(W) \quad , $$

\nind where $\Fs$ is any characteristic family generating $H$. 
\endproclaim

     We permit ourselves this abuse of notation in order to keep the notation
as simple as possible, and because there will be no possibility of confusion
of context. We next collect some useful properties of this map. We let 
$\Hs^{\pm} \subset \Hs$ denote the set of all future-directed (resp. 
past-directed) characteristic half-spaces $H^{\pm}$. \par

\proclaim{Lemma 4.1.9} Let $\hal : \Ws \mapsto \Ws$ be a bijection satisfying
the properties (A) and (B), and let $\hal : \Hs \mapsto \Hs$ be the 
associated mapping of characteristic half-spaces. \par
   (1) $\hal$ is bijective on $\Hs$; \par
   (2) $\hal(H^c) = \hal(H)^c$, for all $H \in \Hs$; \par
   (3) for $H_1,H_2 \in \Hs$, $H_1 \cap H_2 = \emptyset$ if and only if
$\hal(H_1) \cap \hal(H_2) = \emptyset$; moreover, $H_1 \subset H_2$ if and
only if $\hal(H_1) \subset \hal(H_2)$; \par
   (4) for given $H \in \Hs$ and every element $a \in \RR^4$ there exists an 
element $b \in \RR^4$ (and {\it vice versa}) such that 
$\hal(H+a) = \hal(H)+b$; \par
   (5) for any $W \in \Ws$, $W = H_+ \cap H_-$ if and only if 
$\hal(W) = \hal(H_+) \cap \hal(H_-)$; \par
   (6) either $\hal(\Hs^{\pm}) = \Hs^{\pm}$ or $\hal(\Hs^{\pm}) = \Hs^{\mp}$.
\endproclaim

\demo{Proof} 1. Let $\Fs_1,\Fs_2$ be characteristic families such that
$\underset{W_1 \in \Fs_1}\to{\cup}\hal(W_1) = 
\underset{W_2 \in \Fs_2}\to{\cup}\hal(W_2)$.
Since $\hal^{-1}$ has the same properties as $\hal$ does, Lemma
4.1.8 entails that $\underset{W_1 \in \Fs_1}\to{\cup}W_1 = 
\underset{W_2 \in \Fs_2}\to{\cup}W_2$, {\it i.e.} $\hal$ is injective on $\Hs$.
Let now $H \in \Hs$ be generated by a characteristic family $\Fs$:
$H = \underset{W \in \Fs}\to{\cup}W$. Then defining 
$H_0 = \underset{W \in \Fs}\to{\cup}\hal^{-1}(W)$, one has $\hal(H_0) = H$.
{\it i.e.} $\hal$ is surjective on $\Hs$. \par
     2. Assertion (2) is an immediate consequence of the property (4.1.4) of 
the map $\hal$ on $\Ws$. \par
     3. Let $\Fs_1,\Fs_2$ be characteristic families which generate the 
characteristic half-spaces $H_1,H_2$, respectively. If 
$H_1 \cap H_2 = \emptyset$, then $\Fs_1 \cap \Fs_2 = \emptyset$, which implies
$\hal(\Fs_1) \cap \hal(\Fs_2) = \emptyset$, by property (4.1.3) of the 
transformation $\hal$. Hence one has 
$\hal(H_1) \cap \hal(H_2) = \emptyset$. The converse is proven using the
fact that the map $\hal^{-1}$ also has the stated properties. \par
     If one has instead the inclusion $H_1 \subset H_2$, then by Lemma 4.1.1 
there exist wedges $W_1,W_2$ such that 
$H_i = \cup \{ W_i + \lambda\ell \mid \lambda \in \RR \}$, $i = 1,2$, for a 
fixed future-directed lightlike vector $\ell$ (one characteristic half-space is
contained in another only if their boundaries are parallel hyperplanes).
One can choose $W_1,W_2$ such that $W_1 \subset W_2$. {}From condition (B) it 
then follows that 
$\hal(W_1 + \lambda\ell) \subset \hal(W_2 + \lambda\ell)$ for all
$\lambda \in \RR$, so that one must have the inclusion 
$\hal(H_1) \subset \hal(H_2)$.  \par
     4. One first notes some general properties of characteristic
half-spaces: if $H_1,H_2$ are half-spaces with $H_1 \subset H_2$, then
there exists a translation $c \in \RR^4$ such that $H_2 = H_1 + c$. If, on
the other hand, the latter relation holds, then one must have either
$H_1 \subset H_2$ or $H_2 \subset H_1$. \par
     Let now $H \in \Hs$ and $a \in \RR^4$ be given. Then either
$H \subset H + a$ or $H + a \subset H$. In the former case, part (3) of this
lemma entails the inclusion $\hal(H) \subset \hal(H+a)$, so that
$\hal(H+a) = \hal(H)+b$ for some $b \in \RR^4$. The second case is handled
analogously. Since the map $\hal^{-1}$ on $\Hs$ satisfies assertions (1)-(3)
of this lemma, the assertion (4) also follows when the roles of $a$ and
$b$ are exchanged.  \par
     5. Given a wedge $W \in \Ws$ there exist unique characteristic 
half-spaces $H^{\pm}$ such that $W = H^+ \cap H^-$. They are determined by
the characteristic families 
$\Fs^{\pm} = \{ W + \lambda\ell_{\mp} \mid \lambda \in \RR \}$, where 
$\ell^{\pm}$ are future-directed lightlike vectors such that 
$W \pm \ell_{\pm} \subset W$. Clearly one has 
$\hal(W) \in \hal(\Fs^{\pm})$. Since $\Fs^{\pm}$ are characteristic
families, by Lemma 4.1.1 there exist future-directed lightlike vectors
$\ell_{\hal}^{\pm}$ such that 
$\hal(\Fs^{\pm}) = \{ \hal(W) + \lambda\ell_{\hal}^{\pm} \mid 
\lambda\in\RR\}$. Since the set $\Fs^+ \cup \Fs^-$ is not linearly ordered,
condition (B) entails that also the set $\hal(\Fs^+) \cup \hal(\Fs^-)$
is not linearly ordered, in other words, $\hal(\Fs^+) \neq \hal(\Fs^-)$.
Hence the vectors $\ell_{\hal}^{+}$ and $\ell_{\hal}^{-}$ are not 
parallel. Therefore, the intersection of the half-spaces
$\hal(H^{\pm})$ generated by $\hal(\Fs^{\pm})$ must coincide with
$\hal(W)$. \par
     6. Let $H^{\pm} \in \Hs^{\pm}$. If the hyperplanes which form the 
boundaries of $H^{\pm}$ are parallel, then one must have either
$H^+ \cap H^- = \emptyset$ or $H^{+{}c} \cap H^{-{}c} = \emptyset$. Parts
(2) and (3) of this lemma then entail that either
$\hal(H^+) \cap \hal(H^-) = \emptyset$ or 
$\hal(H^{+})^c \cap \hal(H^{-})^c = \emptyset$ must hold. Hence the 
boundary hyperplanes of the characteristic half-spaces $\hal(H^{\pm})$
are parallel, and the time-like orientations of these half-spaces are
oppositely directed. On the other hand, if the boundary hyperplanes of 
$H^{\pm}$ are not parallel, then their intersection $H^+ \cap H^- = W$ is a 
wedge, and it follows from part (4) that 
$\hal(H^+) \cap \hal(H^-) = \hal(W) \in \Ws$. Hence, also in this situation 
the time-like orientations of the half-spaces $\hal(H^{\pm})$ are oppositely 
directed. \par
     Fixing $H^-$ and letting $H^+$ range through $\Hs^+$, one concludes
that either $\hal(\Hs^+) \subset \Hs^+$ and $\hal(H^-) \in \Hs^-$ or 
$\hal(\Hs^+) \subset \Hs^-$ and $\hal(H^-) \in \Hs^+$. Varying $H^-$ while 
holding $H^+$ fixed completes the proof of assertion (6), when one recalls the 
result of part (1).
\hfill\qed\enddemo

     Each characteristic half-space $H_p[\ell]^{\pm}$ determines uniquely
a characteristic hyperplane 
$H_p[\ell] = \overline{H_p[\ell]^+} \cap \overline{H_p[\ell]^-}$, and so
the map $\hal$ on $\Hs$ naturally induces a map on the set of characteristic
hyperplanes.

\proclaim{Definition} Let $\hal : \Ws \mapsto \Ws$ be a bijection satisfying
properties (A) and (B) and $\hal : \Hs \mapsto \Hs$ the associated mapping
of characteristic half-spaces. Then 

$$\hal(H_p[\ell]) \equiv 
\overline{\hal(H_p[\ell]^+)} \cap \overline{\hal(H_p[\ell]^-)}$$

\nind defines a mapping of characteristic hyperplanes onto characteristic
hyperplanes.
\endproclaim

     The following properties of this mapping of characteristic hyperplanes
are an immediate consequence of Lemma 4.1.9.

\proclaim{Corollary 4.1.10} Let $\hal : \Ws \mapsto \Ws$ be a bijection 
satisfying properties (A) and (B) and $\hal$ be the associated mapping of 
characteristic hyperplanes. \par
   (1) $\hal$ is bijective on the set of characteristic hyperplanes in 
$\RR^4$; \par
   (2) for a given hyperplane $H_p[\ell]$ and every element $a \in \RR^4$
there exists an element $b \in \RR^4$ (and {\it vice versa}) such that
$\hal(H_p[\ell] + a) = \hal(H_p[\ell]) + b$; \par
   (3) $\hal$ maps distinct parallel characteristic hyperplanes onto
distinct parallel characteristic hyperplanes.
\endproclaim

     We next prove some further properties of this mapping $\hal$ which
are not quite so obvious.

\proclaim{Lemma 4.1.11} Let $\hal : \Ws \mapsto \Ws$ be a bijection 
satisfying properties (A) and (B) and $\hal$ be the associated mapping of 
characteristic hyperplanes. If $\ell_1,\ell_2,\ell_3,\ell_4$ are
linearly dependent future-directed lightlike vectors such that any two of
them are linearly independent, then

$$\underset{i=1}\to{\overset{4}\to{\cap}} \hal(H_0[\ell_i]) =
\underset{i \neq k}\to{\cap} \hal(H_0[\ell_i]) \quad \text{for} \quad
k = 1,2,3,4 \quad . $$

\endproclaim

\demo{Proof} \! \! As pointed out earlier, an arbitrary maximal pair \! 
$(W[\widetilde{\ell}_1,\widetilde{\ell}_2,d_1],
W[\widetilde{\ell}_3,\widetilde{\ell}_4,d_2])$ with 
$\{\widetilde{\ell}_1, \widetilde{\ell}_2 \} \neq 
\{ \widetilde{\ell}_3, \widetilde{\ell}_4 \}$ 
can be brought into the form
$(W_R,W[\ell_{2+},\widetilde{\ell},d])$ 
(or $(W_R,W[\ell_{2+},\widetilde{\ell},d]')$ by a suitable
Poincar\'e transformation, and by Lemma 4.1.7 it is no loss of generality
to take $d = 0$. Hence, $H_0[\ell_{1+}]$, $H_0[\ell_{1-}]$, $H_0[\ell_{2+}]$
and $H_0[\widetilde{\ell}]$ are 
the characteristic hyperplanes determined by these
wedges. Since Lemma 4.1.7 entails that $ \widetilde{\ell}= (1,a,b,0)$, with
$0 < a < 1$ and $0 < b < 1$, one observes that any three of the four vectors
$\ell_{1+}$, $\ell_{1-}$, $\ell_{2+}$ and 
$\widetilde{\ell}$ are linearly independent.
Hence, the intersection of any three of the hyperplanes 
$H_0[\ell_{1+}]$, $H_0[\ell_{1-}]$, $H_0[\ell_{2+}]$, 
$H_0[\widetilde{\ell}]$ is 
one-dimensional. But, on the other hand, one evidently has
$$\{ c(0,0,0,1) \mid c \in \RR \} \subset 
H_0[\ell_{1+}] \cap H_0[\ell_{1-}] \cap H_0[\ell_{2+}] 
\cap H_0[\widetilde{\ell}]
\quad . \tag{4.1.10}$$
Therefore, one may conclude that the right-hand side of (4.1.10) is
equal to the one-dimensional intersection of any three of the hyperplanes
in that expression. Employing the suitable Poincar\'e transformation, one
sees that
$$\underset{i=1}\to{\overset{4}\to{\cap}} H_{c_i}[\widetilde{\ell}_i] = 
\underset{i \neq j}\to{\cap} 
H_{c_i}[\widetilde{\ell}_i] \quad \text{for} \quad
j = 1,2,3,4 \quad , \tag{4.1.11}$$
where 
$\{ H_{c_i}[\widetilde{\ell}_i]\}_{i=1}^4$ 
are the hyperplanes determined by the
maximal pair \newline 
$(W[\widetilde{\ell}_1,\widetilde{\ell}_2,d],
W[\widetilde{\ell}_3,\widetilde{\ell}_4,d'])$. \par
     Returning to the vectors $\{\ell_1, \ldots,\ell_4\}$ of the hypothesis,
there exists a Lorentz transformation $\Lambda$ with 
$\Lambda\ell_1 = a_1 \ell_{1+}$, $\Lambda\ell_2 = a_2 \ell_{1-}$, 
$\Lambda\ell_3 = a_3 \ell_{2+}$, and $\Lambda\ell_4 = a_4 {\ell}$, where 
$${\ell} = (1,a,b,0), \quad  a,b \in \RR \quad ,\quad  a^2 + b^2 = 1 \quad ,$$ 
and $a_i$, $i = 1,\ldots,4$, are positive constants. Hence, one may once again 
consider the pair
$(\Lambda W[\ell_1,\ell_2,0],\Lambda W[\ell_3,\ell_4,0]) = 
(W_R,W[\ell_{2+},{\ell},0])$
without loss of generality, since $\tau \circ \Lambda^{-1}$ maps 
maximal pairs onto maximal pairs.
If this pair is maximal, then (4.1.11) yields the desired assertion. If this 
pair and $(W_R{}',W[\ell_{2+},{\ell},0])$ are not maximal, then 
Lemma 4.1.7 entails
$b \leq 0$. But, in fact, $b=0$ is excluded by the linear independence
assumption. Set $\ell_0 = (1,\frac{1}{\sqrt{2}},\frac{1}{\sqrt{2}},0)$. Then
using (4.1.11) for the maximal pairs $(\tau(W_R),\tau(W[\ell_{2+},\ell_0,0]))$
and $(\tau(W[\ell_{2+},\ell_{2-},0]),\tau(W[\ell_{1+},\ell_0,0]))$ 
as well as Lemma 4.1.7 and the fact that $\tau$ preserves the maximality of 
pairs of wedges, one finds 
$$\align
\underset{i=1}\to{\overset{3}\to{\cap}}\hal(\Lambda H_0[\ell_i]) &=
  \hal(H_0[\ell_{1+}]) \cap \hal(H_0[\ell_{1-}]) \cap \hal(H_0[\ell_{2+}]) \\
&= \hal(H_0[\ell_{1+}]) \cap \hal(H_0[\ell_{1-}]) \cap
\hal(H_0[\ell_{2+}]) 
\cap \hal(H_0[\ell]) \\  
&= \hal(H_0[\ell_{1+}]) \cap \hal(H_0[\ell_0]) \cap \hal(H_0[\ell_{2+}]) \\
&= \hal(H_0[\ell_{1+}]) \cap \hal(H_0[\ell_0]) \cap \hal(H_0[\ell_{2+}])
\cap \hal(H_0[\ell_{2-}]) \\
&\subset \hal(H_0[\ell_{2-}]) \quad . \tag{4.1.12}
\endalign   $$ 
If $a\neq 0$, then either $(W_R,W[\ell_{2-},{\ell},0])$ or
$(W_R,W[\ell_{2-},{\ell},0]')$ 
is maximal, by Lemma 4.1.7. Hence, (4.1.11)
and (4.1.12) yield 
$$\align
\underset{i=1}\to{\overset{3}\to{\cap}}\hal(\Lambda H_0[\ell_i]) &\subset
  \hal(H_0[\ell_{1+}]) \cap \hal(H_0[\ell_{1-}]) \cap \hal(H_0[\ell_{2-}]) \\
&= \hal(H_0[\ell_{1+}]) \cap \hal(H_0[\ell_{1-}]) \cap 
\hal(H_0[\ell_{2-}]) \cap \hal(H_0[{\ell}]) \\
&\subset \hal(H_0[{\ell}]) \quad ,
\endalign  $$ 
implying the desired assertion. If, on the other hand, $a=0$, then
${\ell}$ is a positive multiple of $\ell_{2-}$, so that 
$H_0[\ell] = H_0[\ell_{2-}]$ and use of (4.1.12) completes the proof.
\hfill\qed\enddemo

     It is evident that the intersection of four hyperplanes $H_{c_i}[\ell_i]$
corresponding to a linearly independent set of four future-directed lightlike
vectors $\ell_i$ and four real numbers $c_i$ is a set containing a single 
point. We now have established sufficient background to prove that the map
$\hal$ preserves this property.

\proclaim{Lemma 4.1.12} Let $\hal : \Ws \mapsto \Ws$ be a bijection 
satisfying properties (A) and (B) and $\hal$ be the associated mapping of 
characteristic hyperplanes. Then the intersection 
$\cap_{\ell} \, \hal(H_0[\ell])$ taken over all future-directed lightlike
vectors is a singleton set (a point) in $\RR^4$.
\endproclaim

\demo{Proof} Since, from Corollary 4.1.10, $\tau$ maps parallel characteristic
hyperplanes onto parallel characteristic hyperplanes, there exist suitable 
pairwise linearly independent lightlike vectors 
$\widetilde{\ell}_1,\widetilde{\ell}_2,\widetilde{\ell}_3,\widetilde{\ell}_4$
and 
$c_1,c_2,c_3,c_4 \in \RR$ such that 
$\hal(H_0[\ell_{1+}]) = H_{c_1}[\widetilde{\ell}_1]$, 
$\hal(H_0[\ell_{1-}]) = H_{c_2}[\widetilde{\ell}_2]$,
$\hal(H_0[\ell_{2+}]) = H_{c_3}[\widetilde{\ell}_3]$, and
$\hal(H_0[\ell_{3+}]) = H_{c_4}[\widetilde{\ell}_4]$. 
By part (2) of Corollary 4.1.10, there
exist real numbers $b_1,b_2,b_3,b_4 \in \RR$ such that 
$\hal(H_{b_1}[\ell_{1+}]) = H_0[\widetilde{\ell}_1]$, 
$\hal(H_{b_2}[\ell_{1-}]) = H_0[\widetilde{\ell}_2]$,
$\hal(H_{b_3}[\ell_{2+}]) = H_0[\widetilde{\ell}_3]$, and
$\hal(H_{b_4}[\ell_{3+}]) = H_0[\widetilde{\ell}_4]$. 
If $\{\widetilde{\ell}_i\}_{i=1,\ldots,4}$ is a
linearly dependent set, then Lemma 4.1.11 applied to $\hal^{-1}$ as a mapping
on the set of characteristic hyperplanes would entail
that $\{\ell_{1+},\ell_{1-},\ell_{2+},\ell_{3+}\}$ is linearly dependent,
a contradiction. Hence, 
$\{\widetilde{\ell}_i\}_{i=1,\ldots,4}$ is a linearly independent set
and so the intersection 
$\cap_{i=1}^4 H_{c_i}[\widetilde{\ell}_i]$ is a singleton set. \par
     An arbitrary lightlike vector $\ell \neq 0$ is a linear combination of
$\ell_{3+}$ and two linearly independent lightlike vectors $\ell_1$, $\ell_2$ 
with zero $x_3$-component. By Lemma 4.1.11, it follows that

$$\hal(H_0[\ell_1]) \cap \hal(H_0[\ell_2]) \cap \hal(H_0[\ell_{3+}])
\subset \hal(H_0[\ell])$$

\nind (note that if $\ell_1,\ell_2,\ell_{3+},\ell$ are not pairwise linearly 
independent, then $\ell$ is a positive multiple of one of the others and 
determines the same hyperplane as the latter) and also
$$\hal(H_0[\ell_{1+}]) \cap \hal(H_0[\ell_{1-}]) \cap
\hal(H_0[\ell_{2+}]) \subset \hal(H_0[\ell_j]) \quad \text{for} \quad
j = 1,2 \quad . $$
This proves the claim, since 
$$\underset{i=1}\to{\overset{4}\to{\cap}}H_{c_i}[\widetilde{\ell}_i] =
\hal(H_0[\ell_{1+}]) \cap \hal(H_0[\ell_{1-}]) \cap
\hal(H_0[\ell_{2+}]) \cap \hal(H_0[\ell_{3+}]) \subset \hal(H_0[\ell]) 
\quad , $$
for arbitrary future-directed lightlike $\ell \neq 0$. 
\hfill\qed\enddemo

     This result entails that $\tau$ induces a point transformation on $\RR^4$.

\proclaim{Definition} For each $x \in \RR^4$, $W \in \Ws$, and each 
characteristic hyperplane $H$, let $T_x(H) \equiv H + x$ and 
$T_x(W) \equiv W + x$. Let $\hal : \Ws \mapsto \Ws$ be a bijection satisfying 
properties (A) and (B) and $\hal$ be the associated mapping of characteristic 
hyperplanes. Then define $\delta : \RR^4 \mapsto \RR^4$ by

$$\{ \delta(x)\} \equiv \underset{\ell}\to{\cap} \hal(T_x H_0[\ell])
\quad \text{for} \quad x \in \RR^4 \quad , $$

\nind where the intersection is taken over all non-zero future-directed 
lightlike vectors $\ell \in \RR^4$.
\endproclaim

    Note that the mapping $\tau \circ T_x$ has the same properties as $\tau$;
applying Lemma 4.1.12 to this mapping implies that
$\delta$ is well-defined. We next need to show that this point transformation
is consistent with the mapping $\tau$.

\proclaim{Proposition 4.1.13} Let $\hal : \Ws \mapsto \Ws$ be a bijection 
satisfying properties (A) and (B) and $\delta$ be the associated point
transformation. Then $\delta$ is a bijection and 

$$
\tau(W) = \{ \delta(x) \mid x \in W \} \quad \text{for all} \quad W \in \Ws
\quad . $$
\endproclaim

\demo{Proof} Define a mapping $\gamma : \RR^4 \mapsto \RR^4$ by

$$\{\gamma(y)\} \equiv \underset{\ell}\to{\cap} \tau^{-1}(T_yH_0[\ell]) \quad . $$

\nind For a fixed $x \in \RR^4$, consider $y \equiv \delta(x)$, so that
$y \in \tau(T_xH_0[\ell])$ for all non-zero positive lightlike vectors $\ell$. 
But, by Corollary 4.1.10, for each such $\ell$ there exists a non-zero 
positive lightlike vector $\ell'$ such that 
$\tau(T_xH_0[\ell]) = T_yH_0[\ell']$. Since $\hal$ is bijective on the set 
of characteristic hyperplanes in $\RR^4$, it follows that

$$
\{\gamma(\delta(x))\} = \underset{\ell'}\to{\cap} \tau^{-1}(T_yH_0[\ell'])
= \underset{\ell}\to{\cap} \tau^{-1}(\tau(T_xH_0[\ell]) = \{ x \} \quad ; $$

\nind hence, one has $\gamma = \delta^{-1}$ and $\delta$ is a bijection. \par
     For arbitrary $W_0 \in \Ws$ and $y \in W_0$, there exists a wedge
$W_1 \subset W_0$ such that $y$ lies in the edge of $W_1$ and such that the
characteristic hyperplanes determined by $W_0$ are different from (though 
parallel to) those determined by $W_1$. By Corollary 4.1.10, the same must be 
true of the hyperplanes determined by the wedges $\tau(W_1) \subset \tau(W_0)$.
Thus, one has $\overline{\tau(W_1)} \subset \tau(W_0)$. Let $H_1$ and $H_2$ be
the characteristic hyperplanes determined by $W_1$. There are two 
characteristic families $\Fs_1$ and $\Fs_2$, containing $W_1$, with
$H_1 = \partial(\cup_{W \in \Fs_1}W)$ and 
$H_2 = \partial(\cup_{W \in \Fs_2}W)$. The wedge $\tau(W_1)$ is contained
in both $\hal(\Fs_1)$ and $\hal(\Fs_2)$, so that 
$\hal(H_1) = \partial(\cup_{W \in \tau(\Fs_1)}W)$ and
$\hal(H_2) = \partial(\cup_{W \in \tau(\Fs_2)}W)$ are the characteristic
hyperplanes determined by $\tau(W_1)$. The characteristic hyperplanes 
containing the point $y$ ($H_1$ and $H_2$ belong to this set) are mapped by
$\hal$ into the set of characteristic hyperplanes containing $\delta(y)$,
{\it i.e.} $\delta(y) \in \tau(H_1)$ and $\delta(y) \in \tau(H_2)$. This
shows that $\delta(y)$ lies in the two characteristic hyperplanes determined
by $\tau(W_1)$. But this entails 
$\delta(y) \in \overline{\tau(W_1)} \subset \tau(W_0)$, which yields
$$\{\delta(x) \mid x \in W \} \subset \tau(W) \quad \text{for every} \quad
W \in \Ws \quad . \tag{4.1.13}$$
Since, by Corollary 4.1.3, $\hal^{-1}$ has the same properties as
$\hal$, one has similarly

$$\{\delta^{-1}(x) \mid x \in W \} \subset \tau^{-1}(W) \quad \text{for every} \quad
W \in \Ws \quad . $$

     Now let $y \in \tau(W)$. Then one has 
$x \equiv \delta^{-1}(y) \in \hal^{-1}(\hal(W)) = W$, and since
$\delta(x) = y$, it follows that $\tau(W) \subset \{\delta(x) \mid x \in W\}$. 
The containment (4.1.13) completes the proof.
\hfill\qed\enddemo

    We recall the well-known result of Alexandrov \cite{2}\cite{3} (see also 
Zeeman \cite{77}, Borchers and Hegerfeldt \cite{11}) to the 
effect that bijections on $\RR^4$ mapping light cones to light cones must be 
elements of the extended Poincar\'e group, $\Ds\Ps$, generated by the 
Poincar\'e group and the dilatation group. The above-established results can
be used to show that the bijection $\delta : \RR^4 \mapsto \RR^4$ constructed
above does indeed map light cones onto light cones. However, a more concise
argument can be obtained by appealing to a related result of Alexandrov 
\cite{3}, to wit: a bijection on $\RR^4$, who along with its inverse maps
spacelike separated points onto spacelike separated points, is an element
of the extended Poincar\'e group.

\proclaim{Lemma 4.1.14} Let $\tau : \Ws \mapsto \Ws$ be a bijection satisfying
properties (A) and (B) and $\delta : \RR^4 \mapsto \RR^4$ be the associated
point transformation. Then $\delta$ is an element of $\Ds\Ps$. 
\endproclaim
\demo{Proof} Note that two points $x,y \in \RR^4$ are spacelike separated
if and only if there exists a wedge $W \in \Ws$ such that $x \in W$ and
$y \in W'$. But by Prop. 4.1.13 and Corollary 4.1.5, one sees that $x \in \Ws$ 
and $y \in W'$ if and only if $\delta(x) \in \tau(W)$ and 
$\delta(y) \in \tau(W') = \tau(W)'$, {\it i.e.} $\delta(x)$ and $\delta(y)$
are spacelike separated. It is therefore evident that both $\delta$ and
$\delta^{-1}$ preserve spacelike separation. The desired assertion then 
follows from Theorem 1 of \cite{3}. 
\hfill\qed\enddemo

     We have therefore established the following result, which we regard as a 
considerable extension of the theorems of Alexandrov {\it et alia} just cited. 

\proclaim{Theorem 4.1.15} Let $\hal : \Ws \mapsto \Ws$ be a bijection with
the properties (A) and (B). Then there exists an element $\delta$ of the
extended Poincar\'e group $\Ds\Ps$ such that for all $W \in \Ws$ one has

$$\hal(W) = \{ \delta(x) \mid x \in W \} \quad . $$
\endproclaim

\demo{Proof} This is an immediate consequence of Corollary 4.1.10, Prop. 
4.1.13 and Lemma 4.1.14.
\hfill\qed\enddemo

     Turning to the more special case of the transformations $\tau_W$ on $\Ws$ 
which arise when the CGMA holds, we know from Prop. 3.1 that they satisfy 
conditions (A) and (B) and are involutions. Each of those transformations will 
therefore fulfill the hypotheses of the next Corollary.

\proclaim{Corollary 4.1.16} Let $\hal : \Ws \mapsto \Ws$ be an involutive
bijection with the properties (A) and (B). Then there exists an element
$\delta$ of the Poincar\'e group such that for all $W \in \Ws$

$$\hal(W) = \{ \delta(x) \mid x \in W \} \quad . $$
\endproclaim

\demo{Proof} It follows from the preceding proposition that there exists an
element $\delta$ of the extended Poincar\'e group such that the stated 
equality of sets holds. Since $\hal$ is an involution, one sees that
$W = \hal^2(W) = \{ \delta^2(x) \mid x \in W \}$, for each $W \in \Ws$.
Hence, by taking suitable intersections one may conclude that 
$\delta^2(x) = x$, for all $x \in \RR^4$. Since $\delta$ is an affine map, it 
is then clear that it cannot contain a nontrivial dilatation. 
\hfill\qed\enddemo  

     Since the group $\Ts$ is generated by elements satisfying the hypothesis
of Corollary 4.1.16, and since Poincar\'e transformations are completely
fixed by their action on the wedges $\Ws$, we conclude that $\Ts$ is
isomorphic to a subgroup $\Gs$ of the Poincar\'e group. In order to indicate
the strength of this result, we shall outline a closely related example,
where the respective transformations are {\it not} induced by point 
transformations, even though properties (A) and (B) obtain. \par
     We consider the manifold $\Ms \equiv \RR^4 \setminus \overline{V_+}$,
which is the complement in Minkowski space of the closure of 
the forward light cone with apex at the origin, with the conformal structure
inherited from Minkowski space, and we take as an admissible family
$\Ws_+$ the set of all regions $W_+ = W \setminus \overline{V_+}$, where
$W$ ranges through the wedges in Minkowski space considered above. Note that
$W$ is uniquely determined once $W_+$ is given and that 
$(W_1)_+ \cap (W_2)_+ = \emptyset$ if and only if $W_1 \cap W_2 = \emptyset$.
However, this latter implication fails to be true in general for the 
intersection of more than two regions. Moreover, we also note that the
equality $(W_+)' = (W')_+$ holds for all wedges $W$. We pick now any Lorentz
transformation which interchanges the forward and backward light cones in
$\RR^4$, such as time reversal $T$, and define on $\Ws_+$ the mapping

$$\tau(W_+) \equiv (TW)_+ \quad , \quad W_+ \in \Ws_+ \quad . $$

\nind It follows from the preceding remarks that $\tau : \Ws_+ \mapsto \Ws_+$
is well-defined and has properties (A) and (B). But if the intersection
of three (or more) partial wedges $W_+$ is contained in the backward light
cone $V_-$, their images under the map $\tau$ have empty intersection. This
shows that $\tau$ cannot be induced by a point transformation on $\Ms$. \par

\bigpagebreak
 {\bf 4.2. Wedge Transformations Generate the Proper Poincar\'e Group}
\bigpagebreak

     In the preceding section we have seen that for any theory on $\RR^4$
satisfying the CGMA for the wedge regions $\Ws$, the corresponding 
transformation group $\Ts$ is isomorphic
to a subgroup $\Gs$ of the Poincar\'e group $\Ps$. So the next question in
our program is: which subgroups of $\Ps$ can appear in this way? We do not aim
here at a complete answer to this question and restrict attention to those
cases where the group $\Ts$ is ``large''. A natural way of expressing this
mathematically is to assume that the group $\Ts$ acts transitively upon the 
set $\Ws$. It would be interesting to consider situations where this 
transitive action fails\footnote{We shall return to this point in a subsequent
publication.}. However, as
our intention in this paper is to illustrate the application of our
approach to just a few, albeit physically important cases, we make this
additional asssumption and leave the other possibilities uninvestigated for
the present. We remark that the condition that $\Ts$ acts 
transitively upon the set $\Ws$ is implied by the algebraic postulate that
the adjoint action of the modular conjugations $\{ J_W \mid W \in \Ws \}$
acts transitively upon the set $\wrnet$. Hence, this condition also is 
expressible in terms of algebraically determined quantities. \par  
     We have constructed a subgroup $\Gs$ of the Poincar\'e group, which is 
isomorphic to $\Ts$ and related to the group $\Ts$ as follows:
For each $\tau \in \Ts$ there exists an element $g_{\tau} \in \Gs$ such that 
$\tau(W) = g_{\tau}W \equiv \{ g_{\tau}(x) \mid x \in W \}$. To each of the 
defining involutions
$\tau_W \in \Ts$, $W \in \Ws$, there exists a unique corresponding involution
$g_W \in \Gs \subset \Ps$. The Poincar\'e group has four connected components,
and the transitivity of the action of $\Gs$ upon the set $\Ws$, which 
implies that for every $W_1,W_2 \in \Ws$, there exists an element $g \in \Gs$ 
such that $gg_{W_1}g^{-1} = g_{gW_1} = g_{W_2}$, entails the relation 

$$g_{W_1}g_{W_2} = g_{W_1}gg_{W_1}g^{-1} = g_{W_1}gg_{W_1}^{-1}g^{-1} \quad ,$$

\nind since $g_{W_1}$ is an involution. But the right-hand side is 
a group commutator, and in the Poincar\'e group such commutators are always 
contained in the identity component $\Pid$. Hence, for any wedges 
$W_1,W_2 \in \Ws$ the product of the corresponding group elements 
$g_{W_1}g_{W_2}$ must be contained in $\Pid$, and the same is true for 
products of an even number of the generating involutions of $\Gs$. Now
pick a wedge $W \in \Ws$ and consider the corresponding involution
$g_W \in \Gs$, which must lie in one of the four components of $\Ps$. 
One then notes that if $n \in \IN$ is odd, then it follows from 
$g_{W_1} \cdots g_{W_n} = g_W (g_W g_{W_1} \cdots g_{W_n})$ that
$g_{W_1} \cdots g_{W_n}$ must lie in the same component of $\Ps$ as
$g_W$. But this implies the following lemma.

\proclaim{Lemma 4.2.1} The group $\Gs$ has nonempty intersection with at most 
one connected component of the Poincar\'e group $\Ps$ other than $\Pid$.
\endproclaim

     Thus we are dealing with a subgroup $\Gs$ of $\Ps$ which is generated
by involutions, intersects at most two of the four connected components of
$\Ps$ and acts transitively on $\Ws$ in the obvious sense. Which subgroups
can such $\Gs$ be? Answering this question turned out to be a somewhat
laborious task. We begin by discussing an analogous problem for the Lorentz
group. \par 
     Consider again the reference wedge 
$W_R = \{ x \in \RR^4 \mid x_1 > \vert x_0 \vert\}$, whose edge contains the
origin, and let 
$\IL(W_R) \equiv \{ \Lambda \in \Ls \mid \Lambda W_R = W_R \}$ be its
invariance subgroup in the full Lorentz group $\Ls$. The involutions in
$\IL(W_R)$ given by the identity $\diag(1,1,1,1) \in \Ls^{\uparrow}_+$, the 
temporal reflection $T = \diag(-1,1,1,1) \in \Ls^{\downarrow}_-$, the 
reflection through the 3-axis (in other words, about the 
$x_0x_1x_2$-hyperplane) $P_3 = \diag(1,1,1,-1) \in \Ls^{\uparrow}_-$, 
and their product $P_3 T = \diag(-1,1,1,-1) \in \Ls^{\downarrow}_+$ are
distinguished, because all elements of $\IL(W_R)$ can be obtained by 
multiplying elements of $\ILid(W_R) \equiv \IL(W_R) \cap \Ls^{\uparrow}_+$ by 
these involutions. It is important in what is to come that $\ILid(W_R)$ is an 
abelian group, since it is generated by rotations about the 1-axis and velocity
transformations (boosts) in the 0-1 direction, whereas $\IL(W_R)$ is not 
abelian, precisely because of the mentioned involutions. The fact that
$\ILid(W_R)$ is abelian is heavily used in our arguments, and for that reason
our proof does not function in higher-dimensional Minkowski spaces. We wish to 
prove the following proposition. \par

\proclaim{Proposition 4.2.2} Any subgroup $G$ of the identity component $\Lid$
of the Lorentz group, which acts transitively upon the set $\Ws_0$ of wedges 
whose edges contain the origin of $\RR^4$, must equal $\Lid$. Furthermore, any 
subgroup $G$ of the Lorentz group $\Ls$, which is generated by a collection of 
involutions, has nontrivial intersection with at most two connected components 
of $\Ls$ and acts transitively upon the set $\Ws_0$, must contain $\Lid$.
\endproclaim

\nind The proof will proceed in a number of steps, since we find it convenient 
to consider the following alternatives: 
(i) $G \cap \IL(W_R)$ is trivial, {\it i.e} consists only of the identity $1$, 
(ii) $G \cap \IL(W_R)$ is nontrivial but $G \cap \ILid(W_R)$ is trivial, 
or (iii) $G \cap \ILid(W_R)$ is nontrivial. We shall show that cases (i) and 
(ii) cannot obtain under our assumptions and that case (iii) implies the 
desired conclusion. \par

     We shall exclude case (i) by proving the following claim.

\proclaim{Lemma 4.2.3} Let $G$ be a subgroup of $\Ls$ which acts transitively
upon the set $\Ws_0$. Then one must have $G \cap \IL(W_R) \neq \{ 1 \}$.
\endproclaim

      If we knew from the outset that the group $G$ in the 
statement of Prop. 4.2.2 has the property that also $G_+ \equiv G \cap \Lid$
acts transitively on $\Ws_0$, its proof would follow directly from a 
simplified version of this lemma and the fact that $\Lid$ is a simple group.
For then the adjoint action of $G_+$ applied to the nontrivial element in
$G_+ \cap \IL(W_R)$ would generate all of $\Lid$. In particular, if
$G_+$ acts transitively upon $\Ws_0$, then there exists for each 
$\Lambda \in \Lid$ a $g_{\Lambda} \in G_+$ and some
$\widetilde{\Lambda} \in \ILid(W_R)$ such that
$g_{\Lambda} = \Lambda\widetilde{\Lambda}$. Moreover, Lemma 4.2.3 would yield
the existence of some nontrivial element $h_0 \in G_+ \cap \IL(W_R)$. Since
$G_+ \cap \IL(W_R) \subset \ILid(W_R)$ and the latter group is abelian, we
conclude
$\Lambda h_0 \Lambda^{-1} = \Lambda \widetilde{\Lambda}h_0 \widetilde{\Lambda}^{-1}\Lambda^{-1} = g_{\Lambda}h_0 g_{\Lambda}^{-1} \in G_+$, for all
$\Lambda \in \Lid$. But $\Lid$ is simple (see, {\it e.g.} Sect. I.2.8 in
\cite{37}), so it follows in this case that $G_+ = \Lid$. 
What makes the proofs somewhat cumbersome is the {\it a priori} possibility
that for the transitivity of the action of $G$ on $\Ws_0$ elements in
$G \setminus G_+$ are essential. Note, however, given Lemma 4.2.3, the
argument just given establishes the first assertion in Prop. 4.2.2. \par
     As the proof of Lemma 4.2.3 is itself quite lengthy, we shall break it up 
into a series of sublemmas. The assumption that the intersection 
$G \cap \IL(W_R)$ is trivial and that $G$ acts transitively on $\Ws_0$ entail 
that for every $\Lambda \in \Ls_+^{\uparrow}$ there exists {\it exactly one} 
$g_{\Lambda} \in G$ and a {\it unique} $\widetilde{\Lambda} \in \IL(W_R)$ such 
that $g_{\Lambda} = \Lambda \widetilde{\Lambda}$ (otherwise, one would have
$\Lambda = g_1 \widetilde{\Lambda}_1^{-1} = g_2 \widetilde{\Lambda}_2^{-1}$, 
for $g_1,g_2 \in G$ and 
$\widetilde{\Lambda}_1, \widetilde{\Lambda}_2 \in \IL(W_R)$, which entails 
$g_2^{-1}g_1 = \widetilde{\Lambda}_2^{-1} \widetilde{\Lambda}_1$, yielding a 
contradiction unless both sides are equal to the identity in $\Ls$). Thus, 
under the given assumption we have a map $m: \Lid \mapsto \IL(W_R)$ with 
$m(\Lambda) = \widetilde{\Lambda} = \Lambda^{-1}g_{\Lambda}$. Note that, in 
view of the assumption $G \cap \IL(W_R) = \{ 1 \}$, the map 
$m:\Lid\mapsto\IL(W_R)$ is the identity map when restricted to $\ILid(W_R)$. 
Moreover, for any $\Lambda \in \Lid$ the elements $m(\Lambda)$ and 
$g_{\Lambda}$ lie in the same component of the Lorentz group. \par
     Utilizing the fact that $G$ is a group yields a strong condition on the
map $m$. Consider any two elements $\Lambda_1,\Lambda_2 \in \Lid$ and the 
corresponding $g_{\Lambda_1}, g_{\Lambda_2} \in G$. Then since $G$ is a group,
we must have

$$g_{\Lambda_1}g_{\Lambda_2} = 
\Lambda_1\widetilde{\Lambda_1}\Lambda_2\widetilde{\Lambda_2} = 
\Lambda_1(\widetilde{\Lambda_1}\Lambda_2\widetilde{\Lambda_1}^{-1})\widetilde{\Lambda_1}
\widetilde{\Lambda_2}\in G \quad .$$

\nind Setting 
$\Lambda = \Lambda_1(\widetilde{\Lambda_1}\Lambda_2\widetilde{\Lambda_1}^{-1})$
we have on the other hand $g_{\Lambda} = \Lambda\widetilde{\Lambda}$ with
$g_{\Lambda} \in G$ and consequently 
$g_{\Lambda}^{-1}g_{\Lambda_1}g_{\Lambda_2} = 
\widetilde{\Lambda}^{-1}\widetilde{\Lambda_1}\widetilde{\Lambda_2} \in 
G \cap \IL(W_R) = \{ 1 \}$. This yields the equation
$$m(\Lambda_1)m(\Lambda_2) = 
m(\Lambda_1 m(\Lambda_1) \Lambda_2 m(\Lambda_1)^{-1}) \quad , \tag{4.2.1} $$
for all $\Lambda_1,\Lambda_2 \in \Lid$. \par
     For the solution of this equation it is convenient to proceed to the
covering group $SL(2,\CC)$ of $\Ls^{\uparrow}_+$. One then has to consider the
action of space and time reflections on $SL(2,\CC)$. Adopting standard
conventions (see, {\it e.g.} \cite{60}), one obtains by a straightforward 
computation the following result, which we state without proof. 

\proclaim{Lemma 4.2.4} Space and time reflections ($P$ and $T$) acting on 
four-dimensional Minkowski spacetime induce the same automorphic action upon 
$SL(2,\CC)$, given by $\pi(A) = A^{*-1}$, whereas the reflection of the 3-axis
$P_3$ induces the action $\pi_3(A) = -R\overline{A}R^*$, where
$R = \left( \matrix i & 0 \\ 0 & -i \endmatrix \right)$ and the bar denotes
complex conjugation.
\endproclaim

     With $\rho : SL(2,\CC) \mapsto \Lid$ the canonical homomorphism from the
covering group, we proceed from $m$ to the map $M: SL(2,\CC) \mapsto \IL(W_0)$ 
given by $M \equiv m \circ \rho$. Note that according to our assumptions on
$G$, the set $M(SL(2,\CC))$ is contained in at most two connected components
of the Lorentz group. With $\Lambda_1 = \rho(A)$ and
$\Lambda_2 = \rho(B)$, $A,B \in SL(2,\CC)$, and the fact that $\rho$ is a 
homomorphism, equation (4.2.1) yields the following functional equation for 
$M$:
$$M(A)M(B) = M(A\gamma_A(B)) \quad , \quad A,B \in SL(2,\CC) \quad .
\tag{4.2.2}$$

\nind $\gamma_A$ is the unique automorphism of $SL(2,\CC)$ satisfying
$\rho \circ \gamma_A (\cdot) \! = M(A) \! 
\rho(\cdot) M(A)^{-1}$. More concretely,
for each $A \in SL(2,\CC)$, $M(A)$ can be written uniquely as a product of one 
of the reflections $1$, $T$, $P_3$ or $TP_3$ and an element of $\ILid(W_R)$. 
The subgroup of $SL(2,\CC)$ corresponding to $\ILid(W_R)$ (with the 
appropriate choice of coordinates) is the maximally abelian subgroup $\Ds$ of 
matrices in $SL(2,\CC)$ of the form
$$\left( \matrix
\lambda & 0 \\
0 & \lambda^{-1} 
\endmatrix \right) \qquad , \qquad \lambda \in \CC \setminus \{ 0 \} \quad . $$

\nind  Hence, any choice of $A \in SL(2,\CC)$ determines by the above
decomposition of $M(A)$ such an element $D_{\lambda} \in \Ds$ (up to a sign). 
With this in mind, the action of $\gamma_A$ on $SL(2,\CC)$ can be determined
with the help of Lemma 4.2.4 and is given by
$$\gamma_A(\left( \matrix \alpha & \beta \\ 
\gamma & \delta \endmatrix \right)) =
\cases
\left( \matrix \alpha & \lambda^2 \beta \\ 
\lambda^{-2} \gamma & \delta \endmatrix \right), 
& \text{if $M(A) \in \Lid$} \quad , \qquad (a) \\ 
\left( \matrix \overline{\delta} & -\lambda^2 \overline{\gamma} \\ 
-\lambda^{-2} \overline{\beta} & \overline{\alpha} \endmatrix \right),
& \text{if $M(A) \in \Ls^{\downarrow}_-$} \quad , \qquad (b) \\ 
\left( \matrix \overline{\alpha} & -\lambda^2 \overline{\beta} \\ 
-\lambda^{-2} \overline{\gamma} & \overline{\delta} \endmatrix \right), 
& \text{if $M(A) \in \Ls^{\uparrow}_-$} \quad , \qquad (c) \\ 
\left( \matrix \delta & \lambda^2 \gamma \\ 
\lambda^{-2} \beta & \alpha \endmatrix \right), 
& \text{if $M(A) \in \Ls^{\downarrow}_+$} \quad , \qquad (d) \\ 
\endcases $$

\nind where $\alpha,\beta,\delta,\gamma \in \CC$ with 
$\alpha\delta - \beta\gamma = 1$. We shall refer to these four possibilities
in the following as cases (a), (b), (c) and (d). \par 

     After these preparations, we now turn to the solution of equation (4.2.2)
and hence of equation (4.2.1). Let 
$\Us_{C} \equiv \{ \left( \matrix 1 & z \\ 0 & 1 \endmatrix \right) 
\mid z \in \CC \}$ be the subgroup of upper triangular matrices  and 
$\Ls_{C} \equiv \{ \left( \matrix 1 & 0 \\ z & 1 \endmatrix \right) \mid
z \in \CC \}$ be the subgroup of lower triangular matrices in $SL(2,\CC)$. 
Note that in cases (a) and (c) $\gamma_A$ leaves the sets $\Us_C$ and $\Ls_C$ 
invariant, while in the other cases $\gamma_A$ interchanges the two. Moreover, 
as long as $A$ is in case (a) and $\gamma_A$ is not the identity, one has for
some $\lambda^2 \neq 1$
$$\gamma_A(\left( \matrix 1 & 0 \\ z & 1 \endmatrix \right))
\left( \matrix 1 & 0 \\ z & 1 \endmatrix \right)^{-1} =
\left( \matrix 1 & 0 \\ (\lambda^{-2}-1)z & 1 \endmatrix \right)$$
and
$$\gamma_A(\left( \matrix 1 & z \\ 0 & 1 \endmatrix \right))
\left( \matrix 1 & z \\ 0 & 1 \endmatrix \right)^{-1} =
\left( \matrix 1 & (\lambda^{2}-1)z \\ 0 & 1 \endmatrix \right) \quad , $$
which entail
$\{ \gamma_A(X)X^{-1} \mid X \in \Ls_C \} = \Ls_C$, as well
as $\{ \gamma_A(X)X^{-1} \mid X \in \Us_C \} = \Us_C$. The following result
is a simple consequence of the latter observation. 

\proclaim{Lemma 4.2.5} For any triangular matrix $A$ in $SL(2,\CC)$ such that
$M(A) \in \Lid$, one has $M(A) = 1$.
\endproclaim

\demo{Proof} Let $A$ be contained in $\Us_C$ or $\Ls_C$ and satisfy
$M(A) \in \Lid$. If $\gamma_A$ is not trivial, then from the above remarks 
there exists a matrix $X \in SL(2,\CC)$ such that 
$\gamma_A(X)X^{-1} = A^{-1}$. Therewith one has the equality 
$A\gamma_A(X) = X$, and equation (4.2.2) implies $M(A) = 1$. This is a 
contradiction, since then $\gamma_A$ is trivial. Therefore $\gamma_A$ must act 
as the identity map on $SL(2,\CC)$, and $M(A)$ has to lie in the center of
$\Lid$, {\it i.e.} $M(A) = 1$.
\hfill\qed\enddemo

     Some elementary properties of the elements of $SL(2,\CC)$ which are 
mapped by $M$ to the identity are collected in the following lemma.

\proclaim{Lemma 4.2.6} Let $\Es$ consist of all $A \in SL(2,\CC)$ such that
$M(A) = 1$. Then \par
   (1) $\Es$ is a subgroup of $SL(2,\CC)$, and \par
   (2) one has $M(AB) = M(B)$ for all $A \in \Es$ and $B \in SL(2,\CC)$.
\endproclaim

\demo{Proof} If $A \in \Es$ and $B \in SL(2,\CC)$, then equation (4.2.2)
and the triviality of $\gamma_A$ entail that 
$M(AB) = M(A\gamma_A(B)) = M(A)M(B) = M(B)$, proving assertion
(2). Clearly the identity element of $SL(2,\CC)$ is contained in $\Es$, and if
$A,B \in \Es$, one has $M(AB) = M(B) = 1$. Thus, $\Es$ is closed under 
products and taking inverses, hence assertion (1) follows.
\hfill\qed\enddemo

     We exploit these results to show that, in fact, the image of any 
triangular matrix in $SL(2,\CC)$ under $M$ is the identity. 

\proclaim{Lemma 4.2.7} For any triangular matrix $A$ in $SL(2,\CC)$, one has 
$M(A) = 1$, {\it i.e.} $\Us_C \cup \Ls_C \subset \Es$.
\endproclaim

\demo{Proof} Since Lemma 4.2.5 has already established the claim for any 
triangular $A$ in case (a) and since $M(A) \neq 1$ in the remaining cases, it 
is necessary to show that case (b), (c) and (d) cannot occur. Note that the 
set $\{M(A) \mid A \not\in \Es \}$ of Lorentz transformations lies in a single 
component of the Lorentz group $\Ls$ (unless, of course, it is empty), as a 
consequence of the assumption that the given group $G$ intersects at most two 
components of $\Ls$ and here $M(A) \not\in \Lid$. Hence if $A,B \not\in \Es$, 
it follows that $M(A)M(B) \in \Lid$ and consequently (4.2.2) yields 
$M(A\gamma_A(B)) \in \Lid$. The details of the exclusion of cases (b)-(d) will 
be given for such $A \in \Ls_C$ - the argument for $A \in \Us_C$ is similar. 
\par
   With $A \in \Ls_C$, one has $A\gamma_A(B) \in \Ls_C$ whenever (if $A$
is in case (c)) $B \in \Ls_C$, respectively (if $A$ is in case (b) or (d))
$B \in \Us_C$. Moreover, the equation $A\gamma_A(X) = 1$ has, for any 
$A \in \Ls_C$, the solution $X = \gamma_A^{-1}(A^{-1})$ in $\Ls_C$ (in case
(c)), respectively in $\Us_C$ (in cases (b) and (d)), since $\gamma_A^{-1}$ is 
an isomorphism between the respective groups. Given an element $A_0 \in \Ls_C$ 
which is not contained in $\Es$, one can
choose $X_0$ such that $A_0\gamma_{A_0}(X_0) = 1$ holds and get
$M(A_0)M(X_0) = M(A_0\gamma_{A_0}(X_0)) = 1$. This shows that also $X_0$ does
not lie in $\Es$, hence, by the first paragraph, one finds that 
$M(A\gamma_A(X_0)) \in \Lid$ whenever $A \not\in \Es$. If, in addition, 
$A \in \Ls_C$, then by the second paragraph one has $A\gamma_A(X_0) \in \Ls_C$,
and Lemma 4.2.5 implies that  
$$M(A)M(X_0) = M(A\gamma_{A}(X_0)) = 1 \quad . $$
Thus one concludes that the Lorentz element $M(A)$ does not depend upon 
the choice of the element $A$ contained in $\Ls_C$ but not contained in $\Es$. 
The same is therefore true for the corresponding automorphisms $\gamma_A$. \par
     Let $\Es_0$ denote the subgroup $\Es \cap \Ls_C$ of $\Es$ and choose now 
$A,B \in \Ls_C \setminus \Es_0$. By the preceding paragraph one also has
$\gamma_B^{-1}(B^{-1}) \in \Ls_C \setminus \Es_0$ (respectively 
$\gamma_B^{-1}(B^{-1}) \in \Us_C \setminus \Es_0$). Thus, taking into account
the first paragraph and the fact that $\gamma_A = \gamma_B$ and $M(A) =M(B)$,
one finds, using (4.2.2), 
$$\align
M(AB^{-1}) &= M(A\gamma_A(\gamma_B^{-1}(B^{-1}))) = M(A)M(\gamma_B^{-1}(B^{-1})) \\
&= M(B)M(\gamma_B^{-1}(B^{-1})) = M(B\gamma_B(\gamma_B^{-1}(B^{-1}))) =
M(1) = 1 \quad . 
\endalign $$
Therefore, $AB^{-1} \in \Es_0$. \par
     It has therefore been established that (1) $\Es_0 \subset \Ls_C$ is a
group, (2) if $A \in \Ls_C \setminus \Es_0$, then 
$\Es_0 \cdot A \subset \Ls_C \setminus \Es_0$ (this is the content of Lemma
4.2.6 (2)), and (3) if 
$A,B \in \Ls_C \setminus \Es_0$, then $AB^{-1} \in \Es_0$. Hence, for each
$A \in \Ls_C \setminus \Es_0$ one has the disjoint decomposition
$\Ls_C = \Es_0 \cup (\Es_0 \cdot A)$. But for each $A \in \Ls_C$ there exists
an element $X \in \Ls_C$ such that $X^2 = A$. If $X \in \Es_0$, then so is
$A$, since $\Es_0$ is a group. Thus for $A \in \Ls_C \setminus \Es_0$
one must have $X \not\in \Es_0$. But on the other hand, if 
$X \in \Es_0 \cdot A$, then $XA^{-1} \in \Es_0$, so that 
$1 = X \cdot XA^{-1} \in X \cdot \Es_0$, which implies $X^{-1} \in \Es_0$. 
Then again one has $X \in \Es_0$. This is a 
contradiction unless the set $\Ls_C \setminus \Es_0$ is empty.   
\hfill\qed\enddemo 

     We are now in the position to complete the proof of Lemma 4.2.3. Since 
$\Us_C \cup \Ls_C$ generates all of $SL(2,\CC)$, we may conclude from Lemmas 
4.2.6 and 4.2.7 that $M$ maps $SL(2,\CC)$ onto $\{ 1 \}$, and consequently
$m$ maps $\Lid$ onto $\{ 1 \}$. But this contradicts the fact that $m$ must 
be the identity map on $\ILid(W_R)$, so the assertion in Lemma 4.2.3 follows.
Next we turn to case (ii), which is easily eliminated by pointing out the 
following simple consequence of Lemma 4.2.3.

\proclaim{Lemma 4.2.8} If the group $G \subset \Ls$ intersects
at most one connected component of $\Ls$ other than $\Lid$ and acts 
transitively upon the set $\Ws_0$, then the group $G \cap \ILid(W_R)$ is 
nontrivial.
\endproclaim

\demo{Proof} {}From Lemma 4.2.3 there exists a nontrivial element $g_0$ in the 
group $G \cap \IL(W_R)$. If one such $g_0$ happens to lie in $\Lid$, the proof 
is over. So assume that all such $g_0$ are not contained in $\Lid$. Since $G$ 
intersects at most one other component of $\Ls$ besides $\Lid$, one must have 
$G = G_+ \cup G_+ g_0$, where $G_+ = G \cap \Lid$. Thus, the transitivity of 
the action of $G$ upon $\Ws_0$ implies

$$\Ws_0 = G \cdot W_R = G_+ \cdot W_R \, \cup \, G_+ \cdot g_0W_R = 
G_+ \cdot W_R \quad . $$

\nind In other words, also the group $G_+$ acts transitively upon the set
$\Ws_0$, even though $G_+ \cap \IL(W_R) = \{ 1 \}$. But this possibility has 
been excluded by Lemma 4.2.3.
\hfill\qed\enddemo

    We are ready to show that in the only remaining case, case (iii),
the identity component of the Lorentz group must be contained in $G$, which is
the statement of Prop. 4.2.2. We begin by noting that for any involutive 
element $j \in \Ls$, there exists some wedge $W \in \Ws_0$ which is mapped by
$j$ either onto itself or onto its causal complement $W' = -W$. This 
follows from the fact that either $j$ maps every lightlike vector $\ell$ onto
$\ell$, respectively $-\ell$, or there exists a lightlike vector $\ell_1$ such 
that its (lightlike) image $\ell_2 = j \ell_1$ is not parallel to $\ell_1$. In 
the latter case, the pair $(\ell_1,\ell_2)$ is mapped onto itself by $j$, 
since $j$ is an involution. As every wedge is determined by two lightlike 
vectors, the statement then follows after a moment's reflection. \par
     Now, as above, let $G_+ = G \cap \Lid$ and let $G_- = G \setminus G_+$.
We first consider the case where $G_-$ is empty. Then $G_+$ acts transitively
upon $\Ws_0$ and we can conclude from the simplicity of $\Lid$ that 
$G_+ = \Lid$ in this case (cf. the argument directly following the statement 
of Lemma 4.2.3). \par
     Note that the assumption that $G$ is generated by involutions
has not been used in the preceding paragraph. This will be exploited now in
the case where $G_-$ is nonempty. For then there must be some involution
$j \in G_-$ and a wedge $W \in \Ws_0$ such that either $jW = W$ or $jW = -W$.
Without loss of generality, we may assume that $W = W_R$. Since 
$G = G_+ \cup G_+ j$, the relation $jW_R = W_R$ lets us conclude, as in 
the proof of Lemma 4.2.8, that $G_+$ acts transitively on $\Ws_0$ and hence
$G_+ = \Lid$ by the preceding argument. \par
     In the remaining case, where $jW_R = -W_R$, we have

$$\Ws_0 = G \cdot W_R = G_+ \cdot W_R \, \cup \, - G_+ W_R \quad . $$

\nind In other words, for each $W \in \Ws_0$ there exists an element
$g \in G_+$ such that either $gW_R = W$ or $gW_R = - W$. \par
     Now consider the element $R_0 \in \Lid$ which implements the
rotation of angle $\pi$ about the $x_2$-axis: $R_0 = \text{diag}(1,-1,1,-1)$.
This element maps $W_R$ to its causal complement: $R_0 W_R = -W_R$.
Moreover, conjugation by $R_0$ takes the elements of $\ILid(W_R)$ into their 
inverses, {\it i.e.} for each $\widetilde{\Lambda} \in \ILid(W_R)$ we find
$$R_0 \widetilde{\Lambda} R_0^{-1} = \widetilde{\Lambda}^{-1} \quad . 
\tag{4.2.3} $$

    Since $\Ws_0 = \Lid W_R$, we may conclude from the above arguments
that for every $\Lambda \in \Lid$ there exists an element $g_{\Lambda} \in G_+$
such that $\Lambda W_R = g_{\Lambda} W_R$ or 
$\Lambda W_R = - g_{\Lambda} W_R = g_{\Lambda}R_0 W_R$. Hence, for every
$\Lambda \in \Lid$ there exist elements $g_{\Lambda} \in G_+$ and 
$\widetilde{\Lambda} \in \ILid(W_R)$ so that either (1) 
$g_{\Lambda} = \Lambda\widetilde{\Lambda}$ or (2)
$g_{\Lambda} = \Lambda R_0 \widetilde{\Lambda}$. Define therefore the subset 
$\Lid^{(1)}$, resp. $\Lid^{(2)}$, consisting of those elements $\Lambda$ of 
$\Lid$ in case (1), resp. case (2). We have 
$\Lid = \Lid^{(1)} \cup \Lid^{(2)}$. \par
     According to Lemma 4.2.8 there exists a nontrivial element 
$h_0 \in G \cap \ILid(W_R)$. Since $\ILid(W_R)$ is abelian, we find as before
for any $\Lambda \in \Lid^{(1)}$ the relation $\Lambda h_0 \Lambda^{-1} = 
g_{\Lambda} h_0 g_{\Lambda}^{-1} \in G_+$. Moreover, since $G \cap \ILid(W_R)$ 
is a group, it also contains the element $h_0^{-1}$. It follows that for any 
$\Lambda \in \Lid^{(2)}$, we have
$$\Lambda h_0 \Lambda^{-1} = 
\Lambda R_0 \widetilde{\Lambda} h_0^{-1} \widetilde{\Lambda}^{-1} R_0^{-1}\Lambda^{-1} =
g_{\Lambda}h_0^{-1}g_{\Lambda}^{-1} \in G_+ \quad , $$

\nind using (4.2.3). It has therefore been established that 
$\Lambda h_0 \Lambda^{-1} \in G_+$ for any element $\Lambda \in \Lid$. Once
again, it then follows from the simplicity of $\Lid$ that $G_+ = \Lid$.  \par
     The proof of Proposition 4.2.2 is therewith completed. The next step is 
to show that a similar statement holds also for the Poincar\'e group.  \par

\proclaim{Proposition 4.2.9} Any subgroup $G$ of the identity component $\Pid$
of the Poincar\'e group, which acts transitively upon the set $\Ws$ of 
wedges in $\RR^4$, must equal $\Pid$. Moreover, 
any subgroup $G$ of the Poincar\'e group $\Ps$, which is generated by 
involutions, intersects at most two of the four connected components of $\Ps$ 
and which acts transitively upon the set $\Ws$ of wedges in $\RR^4$, must 
contain $\Pid$.
\endproclaim  

\demo{Proof} As a first step, consider the canonical homomorphism 
$\sigma : G \mapsto \Ls$ which acts as $\sigma(\Lambda,a) = \Lambda$
for $(\Lambda,a) \in G$. Since $G$ acts transitively on the set of wedges,
it follows that $\sigma(G)$ acts transitively on the subset $\Ws_0$ of
wedges whose edges contain the origin. For if $W \in \Ws_0$ there exists
an element $(\Lambda,a) \in G$ such that $W = \Lambda W_R + a$, and since
$\Lambda W_R \in \Ws_0$, it follows that $W = \Lambda W_R$. Since
$\sigma(G) \subset \Ls$ is also generated by its involutions and intersects
with at most two components of $\Ls$, one may apply Prop. 4.2.2 and conclude 
that $\Lid \subset \sigma(G)$. Consider now the following alternatives. \par 
   (1) There exist an element $\Lambda \in \Ls$ and 
$a,b \in \RR^4$ with $a \neq b$ such that {\it both} elements $(\Lambda,a)$ 
and $(\Lambda,b)$ are contained in $G$. Since $G$ is a group, it follows that
$(\Lambda,a)(\Lambda,b)^{-1} = (1,a-b) \in G$. As has already been seen, for 
every element $\Lambda \in \Ls^{\uparrow}_+$ there exists some element 
$(\Lambda,c) \in G$; hence it follows that

$$(\Lambda,c)(1,a-b)(\Lambda,c)^{-1} = (1,\Lambda(a-b)) \in G \qquad ,
  \qquad \Lambda \in \Ls^{\uparrow}_+ \quad . $$

\nind Since $(1,c) \in G$ implies that $(1,-c) \in G$, one may conclude 
that $G$ contains all translations $(1,x)$ with $x \cdot x$ equal to some
fixed constant $\kappa$. Since $(1,x),(1,x') \in G$ imply that 
$(1,x + x') \in G$, and since every $y \in \RR^4$ can be written in the form
$y = \sum^4_1 x_i$ with $x_i \cdot x_i = \kappa$, $i=1,\ldots,4$, it 
also follows that $G$ contains {\it all} translations. \par
     Consider now for given $\Lambda \in \Ls^{\uparrow}_+$ an element 
$c \in \RR^4$ for which $(\Lambda,c) \in G$. Then one has by the preceding
result

$$(\Lambda,c)(1,-\Lambda^{-1}c) = (\Lambda,0) \in G \quad ;$$

\nind in other words, $G$ also contains all the pure Lorentz transformations,
as well. Thus, in this case one has $\Ps^{\uparrow}_+ \subset G$. \par
   (2) For every element $\Lambda \in \sigma(G)$ there exists {\it 
exactly one} $a(\Lambda) \in \RR^4$ such that $(\Lambda,a(\Lambda)) \in G$. 
Since $G$ is a group, this entails the following cocycle relation for the 
translations: 
$$a(\Lambda\Lambda') = a(\Lambda) + \Lambda a(\Lambda') \quad ,
  \quad \Lambda,\Lambda' \in \sigma(G) \quad . \tag{4.2.4} $$
Consider the subgroup $G_0 \subset G$ whose elements 
translate the wedge $W_R$ without rotating it. The elements of $G_0$ have the 
form  $(\Lambda,a(\Lambda))$ with $\Lambda \in \IL(W_R)$. So it follows from
the first paragraph of this proof that for $G_0^+ \equiv G_0 \cap \Pid$ the 
equality $\sigma(G_0^+) =\IL(W_R) \cap \Lid$ holds. Since
$\IL(W_R)\cap\Lid$ is abelian, the cocycle equation (4.2.4) implies that

$$a(\Lambda) + \Lambda a(\Lambda') = a(\Lambda \Lambda') =
a(\Lambda' \Lambda) = a(\Lambda') + \Lambda' a(\Lambda) \quad ,$$

\nind for every $\Lambda, \Lambda' \in \IL(W_R)\cap\Lid$, which itself entails 
that

$$(1 - \Lambda')a(\Lambda) = (1 - \Lambda)a(\Lambda') \quad .$$
 
\nind Fixing an element $\Lambda' \in \IL(W_R)\cap\Lid$ such that the matrix 
$(1 - \Lambda')$ is invertible and setting 
$a \equiv (1 - \Lambda')^{-1}a(\Lambda')$, one obtains
$$a(\Lambda) = (1 - \Lambda)a \qquad , \qquad \Lambda \in 
\IL(W_R)\cap\Lid \quad . \tag{4.2.5} $$
Hence $G_0^+$ is comprised of the elements
$\{ (\Lambda,(1-\Lambda)a) \mid \Lambda \in \IL(W_R)\cap\Lid \}$ for some fixed
$a \in \RR^4$. \par
     Now, if $G_0^- \equiv G_0 \setminus G_0^+$ is nonempty, there exists some
$g_0 = (\Lambda_0,a_0) \in G_0^-$ such that $G_0^- = G_0^+ \cdot g_0$ (recall
that $G$ intersects at most two of the connected components of $\Ps$).
Hence, without loss of generality, one may assume that $(1 + \Lambda_0)$ is 
invertible. Since $g_0^2 = (\Lambda_0^2 , a_0 + \Lambda_0 a_0) \in G_0^+$, it
follows from equations (4.2.4) and (4.2.5) that
$(1 - \Lambda_0^2)a = a(\Lambda_0^2) = a(\Lambda_0) + \Lambda_0 a(\Lambda_0) =
(1 + \Lambda_0)a(\Lambda_0)$ and consequently 
$a(\Lambda_0) = (1 - \Lambda_0)a$. Applying equation (4.2.4) another time
yields
$$a(\Lambda\Lambda_0) = a(\Lambda) + \Lambda a(\Lambda_0) = (1 - \Lambda\Lambda_0)a$$
for arbitrary $\Lambda \in \IL(W_R) \cap \Lid$, which finally shows
that \newline
$G_0 = \{ (\Lambda, (1 - \Lambda)a) \mid \Lambda \in \sigma(G_0) \}$.
Hence, $G_0$ induces solely translations of the edge of the wedge $W_R$ along
some (two-sheeted) hyperbola or light ray, contradicting the assumption
that $G$ acts transitively on $\Ws$. Therefore, only case (1) can arise and 
the proof of the proposition is complete.
\hfill\qed\enddemo

     Summing up the results obtained so far in this section, we see that the 
symmetry groups $\Gs$ which arise by the CGMA in Minkowski
space theories must contain the proper orthochronous Poincar\'e group
$\Pid$ if they act transitively on the set of wedges $\Ws$. This result will
enable us in the next step to determine $\Gs$ exactly, as well as the 
action of its generating involutions on Minkowski space.  \par 
     
\proclaim{Proposition 4.2.10} Let the group $\Ts$ act transitively upon the
set $\Ws$ of wedges in $\RR^4$, and let $\Gs$ be the corresponding subgroup
of the Poincar\'e group. Moreover, let  $g_{W_R} = (\Lambda_{W_R},a_{W_R})$ be 
the involutive element of the Poincar\'e group corresponding to the involution 
$\tau_{W_R} \in \Ts$. Then $a_{W_R} = 0$ and 
$\Lambda_{W_R} = P_1 T = \diag(-1,-1,1,1)$, where $P_1$ is the reflection
through the 1-axis and $T$ is the time reflection. Since all wedges are 
transforms of $W_R$ under $\Pid$, these assertions are also true, with the 
obvious modifications, for the involution $g_W$ corresponding to any wedge 
$W \in \Ws$. In particular, one has $g_W W = W'$, for every $W \in \Ws$. In 
addition, $\Gs$ exactly equals the proper Poincar\'e group $\Ps_+$, and every 
element of $\Pid$ can be obtained as a product of an even number of 
involutions, $g_W$, $W \in \Ws$.
\endproclaim

\demo{Proof} If $\tau_0 \in \Ts$ leaves a given wedge $W \in \Ws_0$ fixed, then
Lemma 2.1 (3) entails that $\tau_W \tau_0 = \tau_0 \tau_W$. Hence, if
$g_W$ and $g_0$ are the corresponding elements in the Poincar\'e group,
one must have $g_0 g_W g_0^{-1} = g_W$. In light of Prop. 4.2.9, this
implies that $g_W$ must commute with every element of the invariance group 
$\IPid(W)$. With $g_W = (\Lambda_W,a_W)$, it follows that one must have

$$(\Lambda_0 \Lambda_W \Lambda_0^{-1}, a_0 + \Lambda_0 a_W - 
\Lambda_0 \Lambda_W \Lambda_0^{-1}a_0) = (\Lambda_W,a_W) \quad , $$

\nind for arbitrary $(\Lambda_0,a_0) \in \IPid(W)$. By setting $a_0 = 0$
and letting $\Lambda_0$ vary freely through $\ILid(W)$, this equation
implies $a_W = 0$, and therefore 
$\Lambda_0 \Lambda_W \Lambda_0^{-1} = \Lambda_W$ { \it and} 
$$(1 - \Lambda_W)a_0 = 0 \quad , \tag{4.2.6}$$ 
for all $(\Lambda_0,a_0) \in \IPid(W)$. Furthermore, one has
$\Lambda_W^2 = 1$, since $g_W$ is an involution. \par
     Choosing $W = W_R$, one concludes from (4.2.6) that $\Lambda_W$ must 
have the form 
$$\Lambda_W = \left( \matrix X & 0 \\ Y & 1 \endmatrix \right) \quad , $$
for suitable $2 \times 2$-matrices $X,Y$. Since $\Lambda_W$ is a Lorentz 
transformation, it is easy to see that $Y = 0$. The facts that $\Lambda_W$
must commute with the Lorentz boosts in the 1-direction (leaving $W_R$
invariant) and that $\Lambda_W^2 = 1$ lead then, after some elementary
computation, to $X = \pm 1$. But in the case where the positive sign is taken, 
one would have $\tau_{W_R}(W_R) = W_R$, which is excluded by Lemma 2.1 (4) and 
the fact that there are no atoms in $\Ws$. \par
     The remaining assertions are now easy to verify.
\hfill\qed\enddemo

\bigpagebreak
 {\bf 4.3 {}From Wedge Transformations Back to the Net: 
Locality, Covariance and 
Continuity}
\bigpagebreak

     Having established the geometrical features of the elements of the group
$\Ts$, we turn now to the discussion of its representations induced by the
modular conjugations. Proposition 4.2.10 implies that there exists a projective
representation $J(\Ps_+)$ of the proper Poincar\'e group with coefficients in 
the internal symmetry group of the net $\wrnet$. The next step is to verify 
that this projective representation acts geometrically correctly upon the net, 
in other words that the net is Poincar\'e covariant under this projective 
representation.

\proclaim{Proposition 4.3.1} Let the CGMA obtain with the choices 
$\Ms = \RR^4$ and $\Ws$ equal to the set of wedgelike regions in $\RR^4$, and 
let the adjoint action of $\Js$ upon the set $\wrnet$ be transitive.
Then the projective representation $J(\Ps_+)$ of the proper Poincar\'e
group whose existence is entailed by Corollary 2.3 and Prop.\ 4.2.10
acts geometrically correctly upon the net $\wrnet$, i.e.\ for
each $\Lambda \in \Ps_+$ and each $W \in \Ws$ one has

$$J(\Lambda)\Rs(W)J(\Lambda)^{-1} = \Rs(\Lambda W) \quad . $$

\nind Furthermore, Haag duality holds for $\wrnet$, hence the net
$\wrnet$ satisfies Einstein locality.
\endproclaim

\demo{Proof} By construction, for each $g \in \Gs = \Ps_+$ there exists an
element $\hal_g \in \Ts$ such that $\hal_g(W) = gW$ for all $W \in \Ws$. 
Hence, for all $W \in \Ws$, one has 

$$ J(g)\Rs(W)J(g)^{-1} = 
\Rs((\underset{j = 1}\to{\overset{n(\hal_g)}\to{\Pi}} \hal_{i_j})(W)) = 
\Rs(\hal_g(W)) = \Rs(gW) \quad , 
$$

\nind where the product indicated is taken over the chosen product for
the element $\hal_g \in \Ts$ implicit in the definition of the projective 
representation $J(\Ts)$.  \par
     By Lemma 4.2.10, one has

$$\Rs(W)' = J_W \Rs(W) J_W = \Rs(g_W W) = \Rs(W') \quad , $$

\nind for each $W \in \Ws$. So Haag duality holds; thus, for each 
$W_1 \subset W'$, one has $\Rs(W_1) \subset \Rs(W') = \Rs(W)'$. 
\hfill\qed\enddemo

     Note that these results do not depend upon the choice of projective
representation $J(\Ps_+)$. We next provide conditions on the net $\wrnet$ 
which imply that there exists a strongly continuous projective representation 
of $\Pid$. These conditions essentially involve a
continuity property of the map $W \mapsto \Rs(W)$. First note that since
$\Ws = \Pid W_R$, $\Ws$ is in 1-1 correspondence with the quotient space
$\Pid / \IPid(W_R)$; the latter's topology induces thereby a topology on $\Ws$.
Consider then a continuous collection $\{W_{\epsilon}\}_{\epsilon > 0}$ of 
wedges in $\Ws$ such that $W_{\epsilon} \rightarrow W$ as 
$\epsilon \rightarrow 0$, for some fixed $W \in \Ws$. For $\delta > 0$, let
$A_{\delta} \equiv \underset{0 \leq \epsilon < \delta}\to{\cup} W_{\epsilon}$
and
$I_{\delta} \equiv \underset{0 \leq \epsilon < \delta}\to{\cap} W_{\epsilon}$,
where $W_0 \equiv W$. Define
$\Rs(I_{\delta}) \equiv \underset{0 \leq \epsilon < \delta}\to{\cap} \Rs(W_{\epsilon})$ and 
$\Rs(A_{\delta}) \equiv (\underset{0 \leq \epsilon < \delta}\to{\cup} \Rs(W_{\epsilon}))''$ 
to be the indicated intersection and union of wedge algebras. Note that
$\{ \Rs(A_{\delta})\}_{\delta > 0}$, resp. $\{ \Rs(I_{\delta})\}_{\delta > 0}$,
is a monotone decreasing, resp. increasing, family of von Neumann algebras.
Our net continuity assumption is given next.

\proclaim{Net Continuity Condition} For any $W \in \Ws$ and any continuous 
collection $\{W_{\epsilon}\}_{\epsilon > 0} \subset \Ws$ converging to $W$, 
the net $\wrnet$ satisfies
$\Rs(W) = (\underset{\delta > 0}\to{\cup} \Rs(I_{\delta}))'' =
\underset{\delta > 0}\to{\cap} \Rs(A_{\delta})$. Moreover, there exists
a $\delta_0 > 0$ such that $\Omega$ is cyclic for the algebras 
$\Rs(I_{\delta})$, with $0 < \delta < \delta_0$.
\endproclaim

     We proceed with the following result, which establishes that the 
mentioned net continuity condition implies a certain continuity in nets of 
associated modular objects in the context of the CGMA. Condition (ii) of the
CGMA entails that $\Omega$ is cyclic for $\Rs(A_{\delta})$, 
$0 < \delta < \delta_1$. And the Haag duality proven in Prop. 4.3.1 yields
$$\Rs(A_{\delta})' = 
(\underset{0 \leq \epsilon < \delta}\to{\cup} \Rs(W_{\epsilon}))' =
\underset{0 \leq \epsilon < \delta}\to{\cap} \Rs(W_{\epsilon})' =
\underset{0 \leq \epsilon < \delta}\to{\cap} \Rs(W_{\epsilon}') \quad ,$$
for which $\Omega$ is cyclic whenever $\delta$ is sufficiently small,
by hypothesis.
Hence, there exists a $\delta_1 > 0$ such that $\Omega$ is cyclic and
separating for $\Rs(A_{\delta})$, $0 < \delta < \delta_1$. Moreover, since
$W' \subset I_{\delta}'$, for all $\delta > 0$, $\Omega$ is also separating
for $\Rs(I_{\delta})$. Hence, the CGMA and the net continuity condition imply 
that the modular objects $J_{I_{\delta}}, \Delta_{I_{\delta}}$, resp. 
$J_{A_{\delta}}, \Delta_{A_{\delta}}$, corresponding to the pair 
$(\Rs(I_{\delta}), \Omega)$, resp. $(\Rs(A_{\delta}), \Omega)$, exist for
all $0 < \delta < \min\{\delta_0,\delta_1\}$. Below we shall tacitly take 
$0 < \delta < \min\{\delta_0,\delta_1\}$ without further comment. \par

\proclaim{Proposition 4.3.2} Assume the CGMA with the choices $\Ms = \RR^4$ 
and $\Ws$ as described, as well as the mentioned net continuity condition. Let 
$\{W_{\epsilon}\}_{\epsilon > 0}$ be a continuous net of wedges such that 
$W_{\epsilon} \rightarrow W$ as $\epsilon \rightarrow 0$.  
Then the net $\{J_{W_{\epsilon}}\}_{\epsilon > 0}$ converges strongly to 
$J_W$ as $\epsilon \rightarrow 0$. In addition, the net 
$\{\Delta_{W_{\epsilon}}^{it}\}_{\epsilon > 0}$ converges strongly to
$\Delta_W ^{it}$ as $\epsilon \rightarrow 0$.
\endproclaim

\demo{Proof} By Corollary A.2 of \cite{24}, which is based upon a result of
\cite{25}, it follows from the hypotheses that 
$\Delta_{I_{\delta}} \rightarrow \Delta_W$ and
$\Delta^{-1}_{A_{\delta}} \rightarrow \Delta^{-1}_W$ in the strong 
resolvent sense, and $J_{I_{\delta}} \rightarrow J_W$ and
$J_{A_{\delta}} \rightarrow J_W$ in the strong operator topology (note that 
$\Rs(W)' = (\underset{\delta > 0}\to{\cap} \Rs(A_{\delta}))' =
(\underset{\delta > 0}\to{\cup} \Rs(A_{\delta})')''$). On the other hand,
from equation (2.6) in \cite{31}, one has the inequality
$$ (\idty + \Delta_{A_{\delta}})^{-1} \leq 
(\idty + \Delta_{W_{\epsilon}})^{-1} \leq 
(\idty + \Delta_{I_{\delta}})^{-1} \quad , \tag{4.3.1} $$
for all $0 < \epsilon < \delta$. Employing this inequality, the 
polarization identity, and the stated strong resolvent convergence, it follows 
easily that $(\idty + \Delta_{W_{\epsilon}})^{-1}$ converges weakly to 
$(\idty + \Delta_{W})^{-1}$. By the positivity of the operators in (4.3.1) 
and the operator monotonicity of the operation of taking square roots, 
(4.3.1) also entails 

$$(\idty + \Delta_{A_{\delta}})^{-1/2} \leq 
(\idty + \Delta_{W_{\epsilon}})^{-1/2} \leq
(\idty + \Delta_{I_{\delta}})^{-1/2} \quad ,$$

\nind for all $0 < \epsilon < \delta$, so that by the same argument, also 
$(\idty + \Delta_{W_{\epsilon}})^{-1/2}$ converges weakly to 
$(\idty + \Delta_{W})^{-1/2}$. In order to make the following computations
somewhat more transparent, let 
$R_W \equiv (\idty + \Delta_{W})^{-1}$ and 
$R_{W_{\epsilon}} \equiv (\idty + \Delta_{W_{\epsilon}})^{-1}$. One observes 
then that for any vector $\Phi \in \Hs$ the expression

$$
\Vert (R_{W_{\epsilon}}^{1/2} - R_{W}^{1/2})\Phi\Vert^2 = 
\langle\Phi,(R_{W_{\epsilon}} - R_{W_{\epsilon}}^{1/2} R_{W}^{1/2} 
 - R_{W}^{1/2}R_{W_{\epsilon}}^{1/2}  + R_{W} )\Phi\rangle
$$

\nind must converge to zero as $\epsilon \rightarrow 0$. Since 
$R_{W_{\epsilon}}^{1/2}$ is uniformly bounded, $R_{W_{\epsilon}}$ converges 
also strongly to $R_W$. Standard arguments then yield the strong
convergence of $\{\Delta_{W_{\epsilon}}^{it}\}_{\epsilon > 0}$ to
$\Delta_W ^{it}$ as $\epsilon \rightarrow 0$. \par
     To proceed further, note that from the above it follows that
$\Delta_W^{1/2}$ is the strong graph limit of the net
$\{\Delta_{I_{\delta}}^{1/2}\}$. In particular, there exists a dense
subset $\Ks$ of $\Hs$ such that for each $\Phi \in \Ks$ there exists a
corresponding net $\{\Phi_{\delta}\}$ with 
$\Phi_{\delta} \in \Rs(I_{\delta})\Omega$ satisfying 
$\Phi_{\delta} \rightarrow \Phi$ and

$$\Delta_{I_{\delta}}^{1/2}\Phi_{\delta} \rightarrow \Delta_W^{1/2}\Phi 
\quad . $$

\nind Since the Tomita-Takesaki conjugations $S_{I_{\delta}}$ are 
restrictions of the corresponding conjugations $S_{W_{\epsilon}}$ to 
$\Rs(I_{\delta})\Omega$ (for all $0 < \epsilon < \delta$), one sees that this 
implies

$$J_{W_{\epsilon}}\Delta_{W_{\epsilon}}^{1/2}\Phi_{\delta} =
S_{W_{\epsilon}}\Phi_{\delta} = S_{I_{\delta}}\Phi_{\delta} 
= J_{I_{\delta}}\Delta_{I_{\delta}}^{1/2}\Phi_{\delta} \rightarrow
J_W \Delta_W^{1/2}\Phi \quad , $$

\nind for all $0 < \epsilon < \delta$, since $J_{I_{\delta}}$ converges 
strongly to $J_W$. But this convergence of 
$J_{W_{\epsilon}}\Delta_{W_{\epsilon}}^{1/2}\Phi_{\delta}$
entails the convergence of ($\delta \rightarrow 0$, $0 < \epsilon < \delta$) 

$$\align
\frac{1}{\idty + \Delta_{W_{\epsilon}}^{1/2}}J_{W_{\epsilon}}\Phi_{\delta}
&= \frac{1}{\idty + \Delta_{W_{\epsilon}}^{-1/2}} J_{W_{\epsilon}}
   \Delta_{W_{\epsilon}}^{1/2}\Phi_{\delta} =  
\frac{1}{\idty + \Delta_{W_{\epsilon}}^{-1/2}} J_{I_{\delta}}
   \Delta_{I_{\delta}}^{1/2}\Phi_{\delta} \\
&\rightarrow \frac{1}{\idty + \Delta_W^{-1/2}} J_W \Delta_W^{1/2}\Phi =  
\frac{1}{\idty + \Delta_W^{1/2}}J_{W}\Phi \quad . 
\endalign $$

\nind As the nets $\{\Phi_{\delta}\}_{\delta > 0}$ and 
$\{\frac{1}{\idty + \Delta_{W_{\epsilon}}^{1/2}}\}_{\epsilon > 0}$ converge
strongly, this proves the weak convergence 
$$\frac{1}{\idty + \Delta_{W}^{1/2}}J_{W_{\epsilon}}\Phi 
\rightarrow \frac{1}{\idty + \Delta_{W}^{1/2}}J_W \Phi \quad . $$
Hence, $J_{W_{\epsilon}}$ converges weakly (and thus also strongly, 
since the operators are antiunitary) to $J_W$. 
\hfill\qed\enddemo

     The preceding proposition establishes that to every  
continuous net of wedges is associated a strongly continuous net of modular
involutions. Using this fact and the explicit knowledge which Prop.\ 
4.2.10 furnishes about the geometric action of the generators of the group 
$\Gs$, we shall show that there exists a choice of $J(\Pid)$ which is strongly
continuous. In the following, $\Us(\Hs)$ denotes the group of unitary 
operators acting on the separable Hilbert space $\Hs$. 

\proclaim{Proposition 4.3.3} Assume the CGMA
with the choices $\Ms = \RR^4$ and $\Ws$ as described, along 
with the transitivity of the adjoint action of $\{ J_W \mid W \in \Ws \}$ on 
the net $\wrnet$, and the net continuity condition stated at the beginning 
of this section. Then there exists a strongly continuous projective 
representation $V(\Pid) \subset \Js$ of the group $\Pid$ which 
acts geometrically correctly upon the net $\wrnet$. 
\endproclaim

     As was already pointed out,
{\it any} of the projective representations $J(\Pid)$ furnished by 
Corollary 2.3 acts geometrically correctly upon the net $\wrnet$.
Here, we merely make a particular choice amongst these in order to explicitly
assure that the projective representation is continuous. The unitarity
of the representation is already guaranteed by Prop. 4.2.10. \par
     We shall prove Prop. 4.3.3 in a series of steps. To begin, we
shall define projective representations of certain subgroups of $\Pid$ and 
show that they actually yield continuous representations of their respective
subgroups.  \par
     Consider a wedge $W^{(0)} \in \Ws_0$ containing the origin of
$\RR^4$ in its edge and denote by $x^{(0)},y^{(0)}$, {\it etc.}, any 
translation in the two-dimensional subspace $\RR^2_{W^{(0)}}$ generated by the 
two lightlike directions fixing the boundaries of $W^{(0)}$. Denote by
$J_{z^{(0)}}$ the modular involution associated with
$(\Rs(W^{(0)}+z^{(0)}),\Omega)$. \par
     It follows from Props. 4.2.10 and 4.3.1 that
$$J_{z^{(0)}}\Rs(W)J_{z^{(0)}} = \Rs(\Lambda_{W^{(0)}}W + 2z^{(0)}) \quad ,$$
for all $W \in \Ws$, where $\Lambda_{W^{(0)}} \in \Ls_+$ is the reflection
which is equal to $-1$ on $\RR^2_{W^{(0)}}$ and equal to $1$ on the 
two-dimensional subspace of $\RR^4$ which forms the edge of $W^{(0)}$. 
(This relation was Assumption (1) in \cite{24}.) One therefore sees that
$$J_{x^{(0)}}J_{y^{(0)}}\Rs(W)J_{y^{(0)}}J_{x^{(0)}} = \Rs(W + 2x^{(0)} -
2y^{(0)}) \quad , $$
for any $W \in \Ws$, $W^{(0)} \in \Ws_0$, and $x^{(0)},y^{(0)}$ as 
described above. \par
     For $x^{(0)} \in \RR^2_{W^{(0)}} \subset \RR^4 \subset \Pid$, choose
$V_{W^{(0)}}(2x^{(0)}) \equiv J_{x^{(0)}}J_{W^{(0)}}$. Then Prop. 4.3.2 entails
immediately that $x^{(0)} \mapsto V_{W^{(0)}}(x^{(0)})$ is a strongly 
continuous family of unitary operators implementing the action of the subgroup
$\RR^2_{W^{(0)}}$ of the translation group on the net $\wrnet$. This is
true for any choice of $W^{(0)} \in \Ws_0$. \par
     Next, consider the wedges 
$W_i^{(0)} = \{ x \in \RR^4 \mid x_i > \vert x_0 \vert \}$, $i = 1,2,3$, and 
the corresponding projective representations $V_i(\RR^2_{W_i^{(0)}})$, 
$i = 1,2,3$. These unitary operators will be used to build the desired
representation of the translation group. We shall first show that they
coincide on the subgroup of time translations. To this end we make use of the 
fact that the rotations in the time-zero plane are induced by unitary 
operators in $\Js$, cf. Prop. 4.3.1. Hence, if $R$ is a rotation by $\pi/2$
about the 1-axis, we obtain from Lemma 2.1 (2), using the abbreviation 
$x_0 = (x_0,0,0,0)$, the equalities
$$J(R)V_1(x_0)J(R)^{-1} = J(R)J_{W_1^{(0)} + x_0}J_{W_1^{(0)}}J(R)^{-1} =
J_{RW_1^{(0)} + x_0}J_{RW_1^{(0)}} = V_1(x_0) \quad , $$
since $RW_1^{(0)} = W_1^{(0)}$. Here we have made use of the important fact,
a consequence of Prop. 4.3.1 and the uniqueness of modular objects, that the 
modular conjugations associated with wedges transform covariantly under the 
adjoint action of the (anti)unitary operators in $\Js$, {\it i.e.}
$$J(\lambda)J_W J(\lambda)^{-1} = J_{\lambda W} \quad , \tag{4.3.2} $$
for any choice of wedge $W\in\Ws$ and Poincar\'e transform $\lambda \in \Ps_+$.
Secondly, we know from Corollary 2.3 that $V_1(x_0) = Z(x_0)V_2(x_0)$,
where $Z(x_0)$ is an internal symmetry of the net $\wrnet$ in the center
of $\Js$. In the light of the equalities
$$J(R) V_2 (x_0) J(R)^{-1} = J(R) J_{W_2^{(0)} + x_0} J_{W_2^{(0)}} J(R)^{-1} =
J_{R(W_2^{(0)} + x_0)} J_{R W_2^{(0)}} = V_3 (x_0) \ \ , $$
using $R({W_2^{(0)}}+ x_0) = {W_3^{(0)} + x_0}$, we arrive at the relation  
$$ Z (x_0) V_2 (x_0) = V_1 (x_0) = J(R) V_1 (x_0) J(R)^{-1} 
= Z (x_0) V_3 (x_0) \quad . $$
Thus $V_2 (x_0) = V_3 (x_0)$, and in a similar way one proves 
$V_1 (x_0) = V_3 (x_0)$. We therefore write $V((x_0,0,0,0))$ for $V_i(x_0)$.
This technique of establishing the equality of unitary implementers will also 
be used in the subsequent arguments in order to solve the cohomological 
problems involved in the discussion of the projective representation. \par
     Now, for any $x = (x_0,x_1,x_2,x_3) \in \RR^4$, we define 
$$V(x) \equiv V((x_0,0,0,0))V_1((0,x_1,0,0))V_2((0,0,x_2,0))V_3((0,0,0,x_3)) 
\quad . $$
As in the proof of Prop. 2.2 in \cite{24}, one verifies that 
$x \mapsto V(x)$ is a projective unitary representation
of the translation subgroup of the Poincar\'e group acting geometrically 
correctly on the net $\wrnet$. In order to prove that it is actually a 
representation, we must show that the various factors in the definition
of $V$ commute. Let us consider, for example, the operator 
$V_1((0,x_1,0,0))$, which leaves $\Omega$ invariant and satisfies
$$V_1((0,x_1,0,0))\Rs(W_2^{(0)} + z^{(0)})V_1((0,x_1,0,0))^{-1} = 
\Rs(W_2^{(0)} + z^{(0)}) \quad , $$
for every $z^{(0)} \in \RR^2_{W_2^{(0)}}$. 
So it must commute with the modular involutions of the
coherent family 
$\{ \Rs(W_2^{(0)} + z^{(0)}) \mid z^{(0)} \in \RR^2_{W_2^{(0)}} \}$ and 
therefore also with $V((x_0,0,0,0)) = V_2((x_0,0,0,0))$ and $V_2((0,0,x_2,0))$.
Similarly, $V_1((0,x_1,0,0))$ commutes with 
$V_3((0,0,0,x_3))$, and by the same argument one can  establish the 
commutativity of the remaining unitaries. \par
       We next show that the unitaries in the definition of $V$ define 
continuous representations of the respective one--dimensional subgroups.       

\proclaim{Lemma 4.3.4} Under the assumptions of Prop. 4.3.3, for any
$i = 1,2,3$ and $x^{(0)} \in \RR^2(W_i^{(0)})$, the mapping
$\RR \ni t \mapsto V_i(tx^{(0)})$ is a strongly continuous homomorphism. 
\endproclaim

\demo{Proof} For convenience, set $J_t \equiv J_{W_i^{(0)} + tx^{(0)}}$
and $V(t) = J_{t/2} J_0$, $t \in \RR$. It follows from Prop. 4.3.2  
that $J_{t/2}$ and hence $V(t)$ is strongly continuous in $t$. Since 
$V(t) J_0 = J_0 V(t)^{-1}$ and similarly $V(t) J_{t/2} = J_{t/2} V(t)^{-1}$, 
one obtains with the help of relation (4.3.2), for $n \in \IN$,
$$ V(t)^{2n} J_0 = V(t)^{n} J_0 V(t)^{-n} = J_{nt} \quad , $$
and consequently one has
$$ V(t)^{2n} = V(t)^{2n} J_0^2 =  J_{nt} J_0 = V(2nt) \quad . $$
Similarly one finds 
$$ V(t)^{2n+1} = V(t)^{2n} J_{t/2} J_0 = V(t)^{n} J_{t/2} V(t)^{-n} J_0 = 
J_{(n+1/2)t} J_0 = V((2n+1)t) \quad . $$
{}From these relations one sees in particular that for $m_1,m_2\in \IN$
and $0 \neq n \in \ZZ$,
$$ V(m_1/n)V(m_2/n) = V(1/n)^{m_1} V(1/n)^{m_2} = V(1/n)^{m_1 + m_2} =
V((m_1+m_2)/n) \quad . $$ 
Since $m_1,m_2,n $ are arbitrary and $V(t)$ is continuous, the remaining 
portion of the assertion follows.
\hfill\qed\enddemo

     Combining this lemma with the preceding results, we have thus
established the fact that the unitary operators $V(x)$ introduced above 
define a continuous representation of the translations.   

\proclaim{Lemma 4.3.5} Under the assumptions of Prop. 4.3.3, there exists 
in $\Js$ a strongly continuous unitary representation $V(\RR^4)$ of the 
translation subgroup which acts geometrically correctly upon the net $\wrnet$.
\footnote{It is possible to show the existence of a continuous representation 
of the translation group without the assumption of the net continuity 
condition. This argument will be presented in a subsequent publication.} 
\endproclaim

     Let us now turn to the Lorentz transformations. As is well known, 
any Lorentz transformation $\Lambda \in \Lid$ can uniquely be decomposed 
in the chosen Lorentz system into a boost $B$ and a rotation $R$, 
$\Lambda =BR$, where $B=\sqrt{\Lambda \Lambda^T}$ and $R= B^{-1} \Lambda$. It 
is apparent that the factors appearing in this decomposition are continuous in 
$\Lambda$. We first define unitary operators corresponding to the boosts and 
rotations individually. \par 
     Given a nontrivial boost $B$ there exists a unique two-dimensional 
subspace $\RR^2_B$ in the time-zero plane $\{ x \in \RR^4 \mid x_0 = 0 \}$ of 
the chosen Lorentz system which is perpendicular to the boost direction and 
therefore pointwise invariant under the action of $B$. We pick an arbitrary 
unit vector $\vec{e} \in \RR^2_B$  and consider the corresponding wedge 
$W^{(0)}_{\vec{e}} = \{ x \in \RR^4 \mid \vec{x} \cdot \vec{e} > |x_0| \}$. 
An elementary computation using Prop. 4.2.10 shows that the Poincar\'e 
transformations associated with the corresponding modular conjugations satisfy 
$ g_{B W^{(0)}_{\vec{e}}} g_{W^{(0)}_{\vec{e}}} = B^2$. This leads us to
define 
$$V_{\vec{e}}(B) \equiv J_{B^{1/2} W^{(0)}_{\vec{e}}} J_{W^{(0)}_{\vec{e}}}
\quad , $$
where $B^{1/2}$ is the unique boost whose square is equal to $B$. If $B=1$, we 
set $V_{\vec{e}}(1) = \idty$. This definition is consistent since 
$J_{W^{(0)}_{\vec{e}}}^2 = \idty$ for any unit vector $\vec{e}$.\par
     In a similar manner we construct implementers of the rotations. Given
any proper rotation $R \neq 1$ there is a unique two-dimensional subspace 
$\RR^2_R$ which is perpendicular to the axis of revolution of $R$ and 
therefore stable under the action of this rotation. As in the case of the 
boosts, we consider for $\vec{e} \in \RR^2_R$ the corresponding wedge 
$W^{(0)}_{\vec{e}}$ and find 
$g_{R W^{(0)}_{\vec{e}}} g_{W^{(0)}_{\vec{e}}} = R^2$. Correspondingly, we set
$$V_{\vec{e}}(R) \equiv J_{R^{1/2} W^{(0)}_{\vec{e}}} J_{W^{(0)}_{\vec{e}}}
\quad , $$
where $R^{1/2}$ is defined as the rotation with the same axis of revolution as 
$R$ but with half the rotation angle. \par
     This definition requires a consistency check because of the 
nonuniqueness of the square root of rotations. So let $R_1, R_2$ be
two different square roots of $R$ (differing by a rotation by $\pi$). Then 
$R_2 W^{(0)}_{\vec{e}}=-R_1 W^{(0)}_{\vec{e}}$ and, consequently, 
$J_{R_2 W^{(0)}_{\vec{e}}} J_{W^{(0)}_{\vec{e}}} =
J_{- R_1 W^{(0)}_{\vec{e}}} J_{W^{(0)}_{\vec{e}}}$. Because of  
Haag duality, we have 
${\Cal R} (-W^{(0)}) = {\Cal R} ({W^{(0)}}\,{}' ) = {\Cal R} (W^{(0)})\,{}'$, 
for any wedge $W^{(0)}$, and consequently $J_{-W^{(0)}} = J_{W^{(0)}}$. Hence, 
we have the equality 
$J_{- R_1 W^{(0)}_{\vec{e}}} = J_{R_1 W^{(0)}_{\vec{e}}}$, 
proving the consistency of the definition of $V_{\vec{e}}(R) $. We shall show 
in the next lemma that the implementers of boosts and rotations defined above 
do not depend on the choice of the vector $\vec{e}$. \par

\proclaim{Lemma 4.3.6} Let $V_{\vec{e}}(B)$, $V_{\vec{e}}(R)$
be the unitary operators implementing the boost $B$ and rotation $R$,
respectively. These operators do not depend on the choice of the vector 
${\vec{e}}$ within the above-stated limitations. 
\endproclaim

\demo{Proof} Consider first the case of boosts. If $B=1$, there is nothing 
to prove. So let $B \neq 1$, let $\RR_B^2$ be the corresponding 
two-dimensional invariant subspace and let $B_1$ be any other boost which 
leaves this subspace pointwise invariant. As in the case of the 
translations discussed in Lemma 4.3.4, it follows from relation (4.3.2) that 
for any ${\vec{e}} \in \RR_B^2$ one has
$V_{\vec{e}}(B_1)^n = V_{\vec{e}}(B_1^n)$, for $n \in \IN$. \par
     Now let $R_\phi$ be a rotation by $\phi$ about the axis established by the
direction of the boost $B$ and let $J(R_\phi)$ be a corresponding implementer. 
Then one obtains from relation (4.3.2) 
$$ J(R_\phi) V_{\vec{e}}(B_1) J(R_\phi)^{-1} = 
J_{R_\phi B_1^{1/2} W^{(0)}_{\vec{e}}} J_{R_\phi W^{(0)}_{\vec{e}}} =
J_{B_1^{1/2} W^{(0)}_{R_\phi \vec{e}}} J_{W^{(0)}_{R_\phi \vec{e}}} =
V_{R_\phi \vec{e}}(B_1) \quad , $$
since $R_\phi$ and $B_1^{1/2}$ commute. On the other hand, according to 
Corollary 2.3, there exists some element $Z_\phi$ in the subgroup of internal 
symmetries ${\Cal Z}$ of $\Cal J$ such that 
$$ V_{R_\phi \vec{e}}(B_1) =  Z_\phi V_{\vec{e}}(B_1) \quad . $$
Setting $\phi = 2 m \pi / n$, for $m,n \in \IN$, one sees from the preceding
two relations that 
$$ V_{\vec{e}}(B_1) =  V_{R_{2 m \pi / n}^n \vec{e}}(B_1)  =
 J(R_{2 m \pi / n})^n V_{\vec{e}}(B_1) J(R_{2 m \pi / n})^{-n} = 
Z_{2 m \pi / n}^n V_{\vec{e}}(B_1) \quad , $$
and consequently $Z_{2 m \pi / n}^n = \idty$. Hence, 
$$ V_{R_{2 m \pi / n} \vec{e}}(B_1^n) = V_{R_{2 m \pi / n} \vec{e}}(B_1)^n 
=  Z_{2 m \pi / n}^n V_{\vec{e}}(B_1)^n = V_{\vec{e}}(B_1^n) \quad , $$
and setting $B_1 = B^{1/n}$ one obtains
$$ V_{R_{2m \pi / n} \vec{e}}(B) =  V_{\vec{e}}(B) \quad . $$
According to Prop.\ 4.3.2, the operator $J_{ R_\phi W^{(0)}}$ depends 
continuously on $\phi$ for any wedge $W^{(0)}$, and the same is thus
also true of $V_{R_\phi \vec{e}}(B)$. It therefore follows from the preceding 
relation that $V_{R_\phi \vec{e}}(B) = V_{\vec{e}}(B)$ for any rotation 
$R_\phi$, proving the assertion for the case of the boosts. \par
     For the rotations $R$, one proceeds in exactly the same way as above. The 
role of $R_\phi$ is here played by the rotations about the axis of revolution
fixed  by $R$.
\hfill\qed\enddemo

     In view of this result we may omit in the following the index ${\vec{e}}$
and set 
$$V(B) \equiv V_{\vec{e}}(B), \quad V(R) \equiv V_{\vec{e}}(R) \quad . $$
We next discuss the continuity properties of these operators with respect to 
the boosts and rotations. 

\proclaim{Lemma 4.3.7} The unitary operators $V(B)$ and $V(R)$ depend 
(strongly) continuously on the boosts $B$ and rotations $R$, respectively. 
\endproclaim

\demo{Proof} Let $B_n$ be a sequence of boosts which converges to $B$. If 
$B\neq 1$ it is clear that the distance between the unit disks in the 
corresponding invariant subspaces $\RR^2_{B_n}$ and $\RR^2_{B}$ converges to 
$0$. In particular, there exists a sequence of unit vectors 
$\vec{e}_n \in \RR^2_{B_n}$ which converges to some $\vec{e} \in \RR^2_{B}$ 
and consequently the sequence of wedges $B_n^{1/2} W^{(0)}_{\vec{e}_n}$ 
converges to $B^{1/2} W^{(0)}_{\vec{e}}$. Because of  the continuity of the 
modular operators $J_W$ with respect to $W$, established in Prop.\ 4.3.2,
one concludes that
$$ V(B_n) = J_{B_n^{1/2} W^{(0)}_{\vec{e}_n}} J_{W^{(0)}_{\vec{e}_n}} 
\longrightarrow J_{B^{1/2} W^{(0)}_{\vec{e}}} J_{W^{(0)}_{\vec{e}}} =
V(B) \quad .$$ 
If the sequence $B_n$ converges to $1$, the corresponding unit disks in 
$\RR^2_{B_n}$ need not converge. But, because of the compactness of the
unit ball in $\RR^3$, for any sequence of unit vectors 
$\vec{e}_n \in \RR^2_{B_n}$, there exists a subsequence $\vec{e}_{\sigma(n)}$ 
which converges to some unit vector $\vec{e}_{\sigma}$. Since 
$B_{\sigma(n)}^{1/2} \rightarrow 1$, the corresponding sequences of wedges 
$B_{\sigma(n)}^{1/2} W^{(0)}_{\vec{e}_{\sigma(n)}}$ and 
$W^{(0)}_{\vec{e}_{\sigma(n)}}$ converge to $W^{(0)}_{\vec{e}_{\sigma}}$, and 
consequently one has
$$ V(B_{\sigma(n)}) = J_{B_{\sigma(n)}^{1/2} 
W^{(0)}_{\vec{e}_{\sigma(n)}}} J_{W^{(0)}_{\vec{e}_{\sigma(n)}}} 
\rightarrow J_{W^{(0)}_{\vec{e}_\sigma}} J_{W^{(0)}_{\vec{e}\sigma}} =
\idty \quad . $$ 
Since the choice of the sequence $\vec{e}_n \in \RR^2_{B_n}$ was
arbitrary, the proof of the continuity of the boost operators is 
complete. The argument for the rotations is analogous; the only difference 
being that the boost direction must be replaced by the axis of revolution. 
\hfill\qed\enddemo

     We are now in the position to prove Proposition 4.3.3. Given an 
element $(\Lambda,x) \in \Pid$, we proceed to the unique and 
continuous decomposition $(\Lambda,x) = (1,x) (B,0) (R,0)$ and set 
$$ V((\Lambda,x)) \equiv V(x) V(B) V(R) \quad , $$
where the unitary operators corresponding to the translations, boosts
and rotations have been defined above. Since these operators depend 
continuously on their arguments, the assertion of Prop. 4.3.3 follows. As a 
matter of fact, we shall see that the unitary operators $V((\Lambda,x))$ 
actually define a true representation of $\Pid$. A first step in this 
direction is the following lemma. 

\proclaim{Lemma 4.3.8} Let $V(\cdot)$ be the continuous unitary 
projective representation of $\Pid$ introduced above. One has \par
   (1) $V(R) V(B) V(R)^{-1} = V(RBR^{-1})$ and 
$V(R) V(R_0) V(R)^{-1} = V(RR_0R^{-1})$, for all boosts $B$ and rotations 
$R,R_0$. \par 
   (2) $V(\cdot)$ defines a true representation of every continuous 
one-parameter subgroup of boosts or rotations.   
\endproclaim

\demo{Proof} The first statement in (1) follows from relation (4.3.2) and 
Lemma 4.3.6, which imply 
$$\align 
V(R) V(B) V(R)^{-1} &  = 
V(R) J_{B^{1/2} W^{(0)}_{\vec{e}}} J_{W^{(0)}_{\vec{e}}} V(R)^{-1} \\
& =  J_{R B^{1/2} R^{-1} W^{(0)}_{R \vec{e}}} J_{W^{(0)}_{R \vec{e}}}  = 
V(RBR^{-1}) \quad , 
\endalign $$
where the last equality follows from the fact that $RBR^{-1}$ is 
again a boost which leaves the subspace $R \, \RR^2_B$ pointwise invariant.  
The argument for the rotations is analogous.\par
     Now let $\{G(u) \mid u \in \RR\}$, be a continuous one-parameter group of 
boosts or rotations. As in the proof of Lemma 4.3.4, one shows by an 
elementary computation on the basis of relation (4.3.2) that 
$V(G(u))^n = V(G(u)^n)=V(G(nu))$. Consequently, one finds that, for 
$m_1,m_2 \in \IN $ and $0 \neq n \in \ZZ$,
$$\align  
V(G(m_1/n)) V(G(m_2/n)) & = V(G(1/n))^{m_1} V(G(1/n))^{m_2} \\
& =V(G(1/n))^{m_1+m_2} = V(G((m_1+m_2)/n)) \quad . 
\endalign $$
The stated assertion (2) thus follows once again from the continuity 
properties of $V(\cdot)$.  
\hfill\qed\enddemo

     Instead of proving by explicit but tedious computations that $V(\cdot)$ 
defines a true representation of $\Pid$, we prefer to give a more abstract 
argument based on cohomology theory. In the appendix it is shown that the 
existence of a continuous unitary projective representation $V(\Pid)$ with 
values in $\Cal J$ implies that there is a continuous unitary representation 
$U(\cdot)$ of the covering group $ISL(2,\CC)$ of $\Pid$. $U$ takes values in 
the closure $\overline{\Cal J}$ of $\Cal J$ in the weak operator topology. 
Moreover, there exists a mapping  $Z: \, ISL(2,\CC) \mapsto \overline{\Zs}$,
the closure of the internal symmetry group $\Zs$ in the 
center of $\overline{\Cal J}$, such that $U(A) = Z(A)V(\mu(A))$, for all
$A \in ISL(2,\CC)$, where $\mu: ISL(2,\CC) \mapsto \Pid$ is the
canonical covering homomorphism whose kernel is a subgroup of order 2, the
center of $ISL(2,\CC)$. \par  
     The preceding results enable us to show that $U(\cdot)$ acts trivially on 
the center of $ISL(2,\CC)$ and therefore defines a representation of 
$\Pid$. For let $A_1,A_2 \in ISL(2,\CC)$ be two elements corresponding to 
rotations by $\pi$ about two orthogonal axes, {\it i.e.} 
$\mu(A_i) = R_i(\pi)$, $i=1,2$. It then follows that 
$A_1 A_2 A_1^{-1} A_2^{-1} = C$, where $C$ is the nontrivial element in the
center of $ISL(2,\CC)$. Consequently, we have
$$ \align
U(C) & = U(A_1) U(A_2) U(A_1)^{-1} U(A_2)^{-1} \tag{4.3.3} \\ & =
Z(A_1) V(R_1 (\pi))  Z(A_2) V(R_2 (\pi)) V(R_1 (\pi))^{-1} Z(A_1)^{-1}
V(R_2 (\pi))^{-1} Z(A_2)^{-1} \\ & = 
V(R_1 (\pi)) V(R_2 (\pi)) V(R_1 (\pi))^{-1} V(R_2 (\pi))^{-1} \quad , 
\endalign $$
where we made use of the fact that the operators $Z(A_i), i=1,2, $
are elements of $\overline{\Zs}$ and therefore commute through the product and
cancel. Since $R_1 (\pi) R_2 (\pi) R_1 (\pi)^{-1} = R_2 (\pi)$, we see
from Lemma 4.3.8 (1) that
$$ V(R_1 (\pi)) V(R_2 (\pi)) V(R_1 (\pi))^{-1} = 
V(R_1 (\pi) R_2 (\pi) R_1 (\pi)^{-1}) = 
V(R_2 (\pi) ) \quad . $$
So we conclude that $U(C) = \idty$, as claimed. We can therefore set
$$ U(\mu(A)) \equiv U(A), \ \ A \in ISL(2,\CC) \quad . $$

    In the final step of our argument we make use of the fact that the 
Poincar\'e group is perfect. (Recall that a group is perfect if it is equal to 
its commutator subgroup -- see the appendix.) Given $\lambda_i \in  \Pid$, 
$i=1,2$, one can show in the same way as in relation (4.3.3) that
$$ U(\lambda_1 \lambda_2 \lambda_1^{-1} \lambda_2^{-1}) =
V(\lambda_1) V(\lambda_2) V(\lambda_1)^{-1} V(\lambda_2)^{-1} \quad . $$
Since the elements on the right hand side of this equation are contained in
$\Cal J$, we conclude that the representation $U(\cdot)$ also has values in 
$\Cal J$ (so  one does not need to proceed to the closure 
$\overline{\Cal J}$). It then follows from Prop.\ 4.3.1 that the unitary 
operators $U(\lambda), \lambda \in \Pid$,  act geometrically correctly on the
net $\wrnet$. \par
     Now, given a wedge $W$ and the corresponding modular conjugation 
$J_W$ and reflection $g_W \in {\Cal P}_+$ --  see Prop.\ 4.2.10 -- 
it follows from relation (4.3.2) that 
$ J_W V(\lambda) J_W = Z V(g_W \lambda g_W^{-1})$, where $Z \in {\Cal Z}$
is some internal symmetry. As these central elements drop out in group 
theoretic commutators of the operators $V(\lambda)$, we can compute the 
adjoint action of the modular conjugations $J_W$ on $U(\Pid)$ by making use of 
the relation  
$$\align  
J_W U(\lambda_1 \lambda_2 \lambda_1^{-1} \lambda_2^{-1})J_W &  = 
J_W V(\lambda_1) V(\lambda_2) V(\lambda_1)^{-1} V(\lambda_2)^{-1} J_W \\
& = V(g_W \lambda_1 g_W^{-1} ) V(g_W \lambda_2 g_W^{-1}) 
V(g_W \lambda_1 g_W^{-1})^{-1} V(g_W \lambda_2g_W^{-1} )^{-1} \\ 
& = U(g_W \lambda_1 \lambda_2 \lambda_1^{-1} \lambda_2^{-1} g_W^{-1}) \quad . 
\endalign $$ 
Since $\Pid$ is perfect, this shows that 
$$J_W U(\lambda) J_W = U(g_W \lambda g_W^{-1}) \, , \ \ \text{for } \ \lambda \in
\Pid \quad . \tag{4.3.4} $$ 
Hence the involution $J_W$ induces the outer automorphism corresponding to 
$g_W$ on $U(\Pid)$, so we may take $U(g_W) \equiv J_W$. The fact that 
$U(\lambda) \in {\Cal J}$ acts geometrically correctly on the net implies,
according to relation (4.3.2),
$$ U(\lambda) J_W  U(\lambda)^{-1} = J_{\lambda W} \, , 
\ \ \text{for } \ \lambda \in \Pid \quad . \tag{4.3.5}$$

Let $W_1,W_2 \in \Ws$ be arbitrary. There exists an element 
$\lambda \in \Pid$ such that $W_2 = \lambda W_1$. Hence the Poincar\'e
covariance of $\wrnet$ and condition (i) of the CGMA entail the relation
$g_{W_2} = \lambda g_{W_1} \lambda^{-1}$. Using (4.3.4) and (4.3.5), we 
therefore have the equalities
$$\align 
U(g_{W_1})U(g_{W_2}) &= J_{W_1}J_{W_2} = J_{W_1}J_{\lambda W_1} \\
  &= J_{W_1}U(\lambda)J_{W_1}U(\lambda)^{-1} \\
  &= U(g_{W_1}\lambda g_{W_1}^{-1})U(\lambda)^{-1} \\
  &= U(g_{W_1}\lambda g_{W_1}^{-1}\lambda^{-1}) \\
  &= U(g_{W_1}g_{W_2}) \quad , \tag{4.3.6}
\endalign $$
since $g_{W_1}^{-1} = g_{W_1}$. Hence, $U(\cdot)$ provides a representation
for all of $\Gs = \Ps_+$.  \par
     Since $\Cal J$ is generated by the conjugations 
$J_{\lambda W}, \lambda \in \Pid$, we conclude that 
${\Cal J} = U(\Pid) \cup  J_{W^{(0)}} U(\Pid) $, for any fixed wedge 
$W^{(0)} \in \Ws_0$. Moreover, ${\Cal J}^+ =  U(\Pid)$, where ${\Cal J}^+$ is 
the subgroup of unitary operators in $\Cal J$ which is generated by products 
of an even number of modular conjugations. As $U$ is a faithful representation 
of $\Pid$ -- cf.\ the standing assumptions in Chapter II -- and $\Pid$ has 
trivial center, the center of $\Cal J$ consists only of $\idty$. Hence the 
representation $U(\cdot)$ must coincide with $V(\cdot)$. This shows finally 
that $V(\cdot)$ defines a representation of $\Pid$, as claimed. We summarize 
these findings in the following theorem.

\proclaim{Theorem 4.3.9} Assume the CGMA
with the choices $\Ms = \RR^4$ and $\Ws$ the described set
of wedges. If $\Js$ acts transitively upon the set $\wrnet$ and the net 
continuity condition mentioned at the beginning of Section 4.3 holds, 
then there exists a strongly continuous (anti)unitary representation $U(\Ps_+)$
of the proper Poincar\'e group which acts geometrically correctly upon
the net $\wrnet$ and which satisfies $U(g_W) = J_W$, for every $W \in \Ws$. 
Moreover, $U(\Pid)$ equals the subgroup of $\Js$ consisting of all products of 
even numbers of $J_W$'s and $\Js = U(\Pid) \cup J_{W_R}U(\Pid)$. Furthermore,
$U(\cdot)$ coincides with the representation $V(\cdot)$, which has been 
explicitly constructed above.
\endproclaim

\bigpagebreak

\heading V. Geometric Action of Modular Groups and the Spectrum Condition 
\endheading

     A physically important property of a representation of the translation
group on $\RR^4$ is the spectrum condition, in other words, the condition that 
the generators of the given representation $U(\RR^4)$ have their joint 
spectrum $\sp(U)$ in the closed forward light cone $\overline{V_+}$ (for the
positive spectrum condition) or in the closed backward light cone
$\overline{V_-}$ (for the negative spectrum condition). In Section 5.1 we
examine how to incorporate the spectrum condition into our setting, using 
only the modular objects. We shall show that the (positive or negative)
spectrum condition holds whenever the group $\Js$ generated by the initial 
modular involutions contains also the initial modular groups. Some further 
consequences of the spectrum condition in our setting, such as the PCT and 
Spin \& Statistics Theorems, will also be discussed.  \par
     We then turn our attention to the possible geometric action of the
modular {\it unitaries}. In Section 5.2 we shall reconsider the condition of 
{\it modular covariance}, which has been extensively discussed in the 
literature \cite{22}\cite{36}\cite{35}\cite{26}. If $W_0 \in \Ws$ is a wedge, 
$\{\Delta_{W_0}^{it}\}_{t \in \RR}$ is the modular group corresponding
to $(\Rs(W_0),\Omega)$, and $\{ \lambda(t) \}_{t \in \RR}$ is the 
one-parameter subgroup of (suitably Poincar\'e-transformed) boosts leaving  
$W_0$ invariant, then modular covariance is said to hold if
$$\Delta_{W_0}^{it}\Rs(W)\Delta_{W_0}^{-it} = \Rs(\lambda(t)W) \quad , \quad 
\text{for all} \quad t \in \RR \quad , \quad  W \in \Ws \quad , $$
in other words, if the modular group associated to the
algebra for the wedge $W_0$ implements the mentioned boost subgroup. In fact,
the subgroup $\{ \lambda(t) \}_{t \in \RR}$ is usually more precisely 
specified: if $\ell_{\pm}$ are two positive lightlike translations such that
$W_0 \pm \ell_{\pm} \subset W_0$, one has in $\Pid$ the relation
$$\lambda(t)(1,\ell_{\pm})\lambda(t)^{-1} = (1,e^{\mp\alpha t}\ell_{\pm}) 
\quad , $$
with $\alpha = \pm 2 \pi$. The sign is a matter of convention fixing the 
direction of time. \par
     Bisognano and Wichmann \cite{9}\cite{10} (see also \cite{29}) have shown 
that modular covariance holds for nets associated to Wightman fields in a 
Poincar\'e covariant vacuum representation. We shall show that if the adjoint 
action of the modular groups corresponding to the wedge algebras leaves the 
set $\wrnet$ invariant, then modular covariance {\it follows} and either the 
positive or the negative spectrum condition holds. Moreover, under the same 
assumptions plus the locality of the net, the modular conjugations 
$\{ J_W \}_{W \in \Ws}$ will be seen to act geometrically as reflections about 
spacelike lines, {\it i.e.} as in Prop. 4.2.10. In Section 5.3 we shall 
present some examples of nets satisfying all assumptions made in our program 
through Chapter IV, but violating the condition of modular covariance. In one 
of these examples the spectrum condition is violated, in the other the positive
spectrum condition obtains. We then contrast the approaches to geometric 
modular action through the modular conjugations or through the modular 
groups in the light of the results of Section 5.2 and the 
mentioned examples. \par

\bigpagebreak
   {\bf 5.1. The Modular Spectrum Condition}
\bigpagebreak

     Let $V(\RR^4)$ be any representation of the translation group acting
covariantly on the net $\wrnet$ and satisfying the relativistic spectrum
condition with $\Omega$ as the ground state. Borchers \cite{12} has isolated a 
condition on the modular group $\Delta_{W_0}^{it}$ associated with the pair
$(\Rs(W_0),\Omega)$ which is intimately connected to the spectrum condition. 

\proclaim{Borchers' Relation} For every future-directed lightlike vector 
$\ell$ such that \newline
$W_0 + \ell \subset W_0$, there holds the relation
$$\Delta_{W_0}^{it}V(\ell)\Delta_{W_0}^{-it} = V(e^{-2\pi t}\ell) \quad ,
\quad \text{for all} \quad t \in \RR \quad . \tag{5.1.1}$$
\endproclaim

\nind Note that this is precisely the relationship which would result if 
$\Delta_{W_0}^{it}$ implemented the subgroup of boosts leaving the wedge
$W_0$ invariant. It has turned out that this condition is {\it equivalent} to 
the representation $V(\RR^4)$ satisfying the spectrum condition.\footnote{This
connection has also been shown to be useful in applications to quantum fields
defined on certain curved space-times associated with black holes \cite{61}.}
We cite the result as proven in \cite{24}; the appearance of our theorem was 
preceded by that of an analogous result proven under slightly more 
restrictive conditions by Wiesbrock \cite{68}. The proof of the deep result 
that the spectrum condition implies (5.1.1) is due to Borchers \cite{12}. For 
a recent, considerably simplified proof of Borchers' theorem, we recommend 
\cite{30} to the reader's attention.

\proclaim{Proposition 5.1.1} Let $V(\RR^4)$ be a strongly continuous unitary 
representation of the translation group on $\RR^4$ which acts geometrically 
correctly upon the net $\wrnet$ and leaves $\Omega$ invariant. Then $V(\RR^4)$ 
satisfies the (positive) relativistic spectrum condition, {\it i.e.} 
$\sp(V) \subset \overline{V}_+$, if and only if relation (5.1.1) holds for all 
wedges $W_0$, as described.
\endproclaim

     We intend to utilize this proposition in a discussion of the spectral 
properties of the representation $U(\RR^4)$ of the translation group obtained
in the previous chapter. Note that because we have a representation of $\Pid$ 
which acts geometrically correctly upon the net and which leaves the state 
invariant, if (5.1.1) holds (for $U(\cdot)$) for one such wedge $W_0$, it must 
hold for all such wedges. \par
     In our approach, employing the modular involutions to derive symmetry 
groups and their representations, the {\it only} role played by the modular 
groups $\Delta_W^{it}$ is to characterize algebraically the spectrum condition
as above. We next show that in our framework, the Borchers relation 
(5.1.1) (with $\pm 2 \pi$ in the exponent on the right-hand side instead of
$-2\pi$) already follows from the following assumption:

\proclaim{Modular Stability Condition} The modular unitaries are 
contained in the group generated by the modular involutions, {\it i.e.} 
$\Delta_W^{it} \in \Js$, for all $t \in \RR$ and $W \in \Ws$.
\endproclaim

     In the situation described by this condition, the group generated by the 
modular unitaries and the modular conjugations associated to the net 
$\wrnet$ by the vector $\Omega$ is minimal in a certain sense. The
name of this condition is motivated by the {\it use} we envisage for it. We 
shall prove in Theorem 5.1.2 that the CGMA and the modular stability condition 
imply the spectrum condition, {\it i.e.} physical stability, in the special 
case of Minkowski space. Since both conditions are well-defined for nets based 
on arbitrary space-times, the modular stability condition, in the context of 
the CGMA, could perhaps serve as a substitute for the spectrum condition on 
space-times with no timelike Killing vector. In fact, as discussed below in
Section 6.2, recent results \cite{19}\cite{17} in de Sitter space support
this picture. \par
     We remark that the Poincar\'e covariance we have established entails that 
$\Delta_W^{it} \in \Js$, for all $t \in \RR$ and {\it some} $W \in \Ws$, 
implies the modular stability condition. This follows from the 
transitive action of $\Pid$ upon the set $\Ws$ and the well-known fact that if 
$\Delta^{it}$ is the modular unitary for the pair $(\Ms,\Omega)$ and if the 
unitary $U$ leaves $\Omega$ invariant, then $U\Delta^{it}U^*$ is the modular 
unitary for the pair $(U\Ms U^*, \Omega)$. We can now show that, within our 
framework, the condition that the modular unitaries are contained in the group 
$\Js$ implies the spectrum condition, up to a sign. We shall see in the 
examples in Section 5.3 that each of the possible outcomes stated in this 
theorem can occur. \par   

\proclaim{Theorem 5.1.2} Assume the CGMA
with the choices $\Ms = \RR^4$ and $\Ws$ the collection of
wedgelike regions in $\RR^4$, the transitivity of the adjoint action of 
$\Js$ on the net $\wrnet$, and the net continuity condition mentioned at the 
beginning of Section 4.3. Let $U(\RR^4)$ be the representation of the
translation group obtained in Section 4.3. If $\Delta_W^{it} \in \Js$, for all 
$t \in \RR$ and some $W \in \Ws$, {\it i.e.} if the modular stability condition
obtains, then $\sp(U) \subset \overline{V_+}$ or 
$\sp(U) \subset \overline{V_-}$. Moreover, for every future-directed 
lightlike vector $\ell$ such that $W + \ell \subset W$, there holds the 
relation

$$\Delta_{W}^{it}U(\ell)\Delta_{W}^{-it} = U(e^{-\alpha t}\ell) \quad ,
\quad \text{for all} \quad t \in \RR \quad , $$

\nind where $\alpha = \pm 2\pi$.
\endproclaim

\demo{Proof} Recall that $\Js^+$ is the subgroup of $\Js$ consisting of all 
products of even numbers of elements of $\{ J_W \mid W \in \Ws \}$. Note that 
the relation $\Delta_W^{it/2}\Delta_W^{it/2} = \Delta_W^{it}$ and the 
assumption $\Delta_W^{it} \in \Js$, for all $t \in \RR$, imply that 
$\Delta_W^{it} \in \Js^+ = U(\Pid)$ (using Theorem 4.3.9).
Hence, for a fixed $W \in \Ws$ and each $t \in \RR$, there exists an element 
$(\Lambda_t,a_t) \in \Pid$ such that 
$$\Rs(W) = \Delta_W^{it}\Rs(W)\Delta_W^{-it} = 
U(\Lambda_t,a_t)\Rs(W)U(\Lambda_t,a_t)^{-1} = \Rs(\Lambda_t W + a_t) \quad . $$
Therefore, one must have $(\Lambda_t,a_t) \in \IPid(W)$, $t \in \RR$, and the 
group $\Gs_W = \{ (\Lambda_t,a_t) \mid t \in \RR \}$ constituted by these 
transformations must be a one-parameter subgroup of $\IPid(W)$ which is 
abelian, since the unitaries $\Delta^{it}_W$, $t \in \RR$, mutually commute. 
\par 
    Let $W_0 \in \Ws$ be any wedge. One observes that if $(\ell_+,\ell_-)$ is 
a pair of lightlike vectors such that $W_0 \pm \ell_{\pm} \subset W_0$,
the adjoint action of any element of $\ILid(W)$ transforms the Poincar\'e 
group element $(1,\ell_{\pm})$ to $(1,c\ell_{\pm})$, with $c > 0$. In 
particular, for each $t \in \RR$ there must exist an element $c_t^{\pm} > 0$ 
such that 
$$\Delta_{W_0}^{it}U(u\ell_{\pm})\Delta_{W_0}^{-it} = 
U(c_t^{\pm}u\ell_{\pm}) \quad , \tag{5.1.2} $$
for all $u \in \RR$. Thus one has
$$\align
U(c_{t+s}^{\pm}\ell_{\pm}) &= \Delta_{W_0}^{i(t+s)}U(\ell_{\pm})\Delta_{W_0}^{-i(t+s)}
  = \Delta_{W_0}^{is}\Delta_{W_0}^{it}U(\ell_{\pm})\Delta_{W_0}^{-it}\Delta_{W_0}^{-is} \\
&= U(c_s^{\pm}c_t^{\pm}\ell_{\pm}) \quad ,
\endalign $$
which implies that $c_s^{\pm}c_t^{\pm} = c_{t+s}^{\pm}$, since 
$U(\RR^4)$ acts geometrically correctly upon the net $\wrnet$ and since there 
exist wedges $W_1$ such that $W_1 + s\ell_{\pm} \neq W_1 + t\ell_{\pm}$ for 
$s \neq t$. {}From the left side of relation (5.1.2) one also sees that the map
$(t,u) \mapsto U(c_t^{\pm}u\ell_{\pm})$ is strongly continuous, uniformly on 
compact subsets of $\RR^2$. As shall be shown, this implies that $c^{\pm}_t$
is continuous in $t$. \par
     Assume that $c_t^{\pm}$ is discontinuous at $t = 0$. It then follows from 
the equation $c_s^{\pm}c_t^{\pm} = c_{t+s}^{\pm}$ that $c_t^{\pm}$ is
unbounded in any neighborhood of $t = 0$. Thus, for any $r \neq 0$, there exist
sequences $t_n \rightarrow 0$, $u_n \rightarrow 0$ such that
$u_n c_{t_n} \rightarrow r$. Therefore, equation (5.1.2) and the mentioned 
strong continuity entail the equality $\idty = U(r\ell_{\pm})$, which is a
contradiction. Thus, the function $t \mapsto c_t^{\pm}$ must be continuous at 
0. The relation $c_s^{\pm}c_t^{\pm} = c_{t+s}^{\pm}$ then implies that there
exist constants $\alpha_{\pm} \in \RR$ such that 
$c_t^{\pm} = e^{\alpha_{\pm}t}$. \par
    It is important to notice that $\alpha_{\pm} \neq 0$. If, for example, one 
had $c_t^+ = 1$ for all $t \in \RR$, then one would have 
$[\Delta_{W_0}^{it},U(\ell_+)] = 0$ and thus
$\Delta_{W_0 + \ell_+}^{it}=U(\ell_+)\Delta_{W_0}^{it}U(\ell_+)^{-1}=
\Delta_{W_0}^{it}$, for all $t \in \RR$. But since 
$\Rs(W_0 + \ell_+) \subsetneq \Rs(W_0)$, by the standing assumptions, this is 
in conflict with standard results in modular theory. For both algebras have 
$\Omega$ as a cyclic vector and the stability of the smaller algebra under the 
action of the modular group of the larger one would thus imply that these 
algebras must be equal (see \cite{18}). \par
    One can now apply the arguments of Prop. 2.3 in \cite{24}. There it was
shown that relation (5.1.2) implies that for any two vectors 
$\Phi,\Psi \in \Hs$ there exists a function $f(z)$ which
is continuous and bounded on the strip $0 \leq \text{Im}(z) \leq 1/2$,
analytic in the interior, satisfies the bound 
$\vert f(z) \vert \leq \Vert \Phi \Vert \Vert \Psi \Vert$, and
on the real axis has the boundary value

$$f(t) = \langle \Phi, U(-e^{\alpha_+ t}\ell_+)\Psi\rangle \quad . $$

\nind Since $\Phi$ and $\Psi$ are arbitrary, one may conclude that the
operator function $z \mapsto U(-e^{\alpha_+ z}\ell_+)$ is weakly continuous
on the strip $0 \leq \text{Im}(z) \leq 1/2$, analytic in the interior, and 
bounded in norm by 1. In particular, one has

$$\Vert U(-i \sin(u\alpha_+ )\ell_+) \Vert \leq 1 \quad , $$

\nind for $0 \leq u \leq 1/2$. Hence, it follows that either
$P\cdot\ell_+ \geq 0$, where $P$ is the generator of the strongly continuous
abelian unitary group $U(\RR^4)$, or $P\cdot\ell_+ \leq 0$. By Lorentz 
covariance, these relations hold for arbitrary lightlike vector $\ell_+$, 
hence the spectrum of $P$ must be contained either in the closed forward light 
cone or the closed backward light cone. The final assertion of the theorem 
then follows from Borchers' theorem \cite{12}.
\hfill\qed\enddemo

\nind This observation reinforces our belief that the modular involutions are 
of primary interest in this context. \par
     Theorem 5.1.2 seems to leave open the possibility that the modular 
group associated to the wedge algebra $\Rs(W)$ could conceivably act 
geometrically as some other subgroup of the invariance group $\IPid(W)$ of $W$
besides the boost subgroup. However, this is not the case, as we shall prove 
in the next section -- cf. Prop.\ 5.2.4 and Theorem 5.2.7. \par
     We wish to emphasize the point that Borchers' relation (5.1.1) is 
truly an {\it additional} assumption in our framework, as is the 
modular stability condition. In Section 5.3 we present a simple example of a
net satisfying our CGMA and all of the other assumptions made in this paper 
except the modular stability condition and (5.1.1). In this example 
the spectrum condition is therefore violated, and the action of the modular 
groups associated to wedge algebras does not coincide with the Lorentz boosts. 
 \par
     Next, we wish to make a few comments about the uniqueness of 
the representation of $\Pid$ which has been obtained above. There are
uniqueness results for representations of the translation subgroup satisfying
the spectrum condition in local quantum field theory - see \cite{24} and 
references cited there. For the case of nets $\wrnet$ based on wedges, the
assertion can be derived easily from Borchers' theorem. We state and
prove this fact for completeness.

\proclaim{Proposition 5.1.3} Let $V(\RR^4)$ be a continuous unitary 
representation of the translations on $\Hs$ which acts geometrically 
correctly on the net $\wrnet$, leaves $\Omega$ invariant, and satisfies the 
spectrum condition. Then there is no other representation on $\Hs$
with these properties.
\endproclaim

\demo{Proof} Let $W$ be any wedge and let $\ell$ be any positive lightlike 
vector such that $W + \ell \subset W$. Since $V(\cdot)$ acts geometrically
correctly on the net, one has 
$V(\ell)\Delta^{it}_W V(\ell)^{-1} = \Delta^{it}_{W + \ell}$, and because of
the hypothesized spectral properties of $V(\cdot)$, Borchers' relation
holds: $\Delta^{it}_W V(\ell)\Delta^{-it}_W = V(e^{-2\pi t}\ell)$. Combining
these two relations yields 
$V(\ell - e^{-2\pi t}\ell) = \Delta^{it}_{W+\ell}\Delta^{-it}_W$, and the
operators appearing on the right-hand side of this equation are fixed by
the net $\wrnet$ and the vector $\Omega$. Hence, $V(\cdot)$ is uniquely
determined by these data for all lightlike vectors; the group property then
yields the desired conclusion.
\hfill\qed\enddemo

     For the representation of the entire Poincar\'e group, the best result
seems to be that of \cite{22}, which asserts that if the distal 
split property holds, then the representation of $\Pid$ is also unique. 
(See also the results in the recent article by Borchers \cite{16}.)
In Section 5.3 we shall present an example of a well-behaved net
covariant under two distinct representations of the Poincar\'e group,
only one of which is selected by the CGMA. \par
     With the additional condition (5.1.1) yielding the spectrum condition, 
algebraic PCT and Spin \& Statistics theorems can be proven. A series of 
papers \cite{36}\cite{35}\cite{45} (see also \cite{26}) have 
demonstrated a purely algebraic version of the important relationship between 
spin and statistics, which was first pointed out by Fierz and Pauli and then 
proven rigorously in the context of Wightman quantum field theory by Burgoyne 
and L\"uders and Zumino (see \cite{60} for references). In the work 
\cite{36}\cite{35}\cite{26} the assumption of modular covariance was made, 
which, as we shall see in Section 5.3, does not necessarily hold in our 
more general setting. But if the conditions of Theorem 5.1.2 are satisfied,
then the results established above {\it do} imply the hypotheses made in the
approach by Kuckert \cite{45} in order to derive the PCT and Spin \& 
Statistics theorems. We shall not take further space to formulate the obvious 
theorem and refer the reader to \cite{45} for details.   \par

\bigpagebreak
   {\bf 5.2. Geometric Action of Modular Groups}
\bigpagebreak

     To obtain a deeper insight into the nature of the property of modular
covariance on the one hand and the relation between the geometric action of 
modular involutions and that of the modular groups on the other, we shall 
assume in this section that the modular groups have a geometric action similar 
to that which we have heretofore assumed for the modular involutions. In 
particular, we shall assume that the adjoint action of the modular groups of 
the wedge algebras leaves the set $\wrnet$ invariant. Throughout this section
we shall assume that $\Ms = \RR^4$ and $\Ws$ is the set of wedges, as 
previously described. \par

\proclaim{Condition of Geometric Action for the Modular Groups}
The Condition of Geometric Action for the modular groups is fulfilled if the 
net $\wrnet$ and vector $\Omega$ satisfy the first three conditions of the 
CGMA stated in Chapter III and the fourth condition is replaced by the
following requirement: For each $W_0 \in \Ws$, the adjoint action of 
$\{\Delta_{W_0}^{it}\}_{t \in \RR}$ leaves the set $\wrnet$ invariant, 
{\it i.e.} for any $W \in \Ws$ and any $t \in \RR$ there exists a wedge 
$W_t \in \Ws$ such that

$$\Delta_{W_0}^{it}\Rs(W)\Delta_{W_0}^{-it} = \Rs(W_t) \quad . $$
\endproclaim

     This {\it condition for the modular groups} will be called CMG for short.
We shall show that the analysis carried out in the preceding chapters in
the case of theories satisfying the CGMA can likewise be performed when one
takes the CMG as the starting point. \par
     We denote by $\Ks$ the unitary group generated by the set \newline
$\{ \Delta_W^{it} \mid t \in \RR, W \in \Ws \}$. As in  Chapter II one sees
that the CMG entails that each $\ad \Delta_W^{it}$ induces a bijection
$\upsilon_W (t)$ on the set $\Ws$ of wedges. The group generated by these 
bijections will be denoted by $\Us$. We state the following counterpart to
Lemma 2.1.

\proclaim{Lemma 5.2.1} The group $\Cal U$ defined above has the following 
properties.\par
   (1) For every $\upsilon \in {\Cal U}$ and $W \in {\Cal W}$, one has 
$\upsilon \upsilon_W (t) \upsilon^{-1} = \upsilon_{\upsilon (W)} (t)$,
$t \in \RR$. \par
   (2) If $\upsilon (W) = W$ for some $\upsilon \in {\Cal U}$ and 
$W \in {\Cal W}$, then $\upsilon \upsilon_W (t) = \upsilon_W (t)
\upsilon$, $t \in \RR$. \par
   (3) One has $\upsilon_W (t)(W) = W$, for all $W \in {\Cal W}$ and $t \in
\RR$. \par
   (4) If $W_1 \in \Ws$ and $\upsilon_W(t)(W_1) \subset W$, for all
$t \in \RR$, then $W_1 = W$.
\endproclaim

\demo{Proof} The first two statements can be established in the same way as
part (2) and (3) of Lemma 2.1. The third statement follows from the
fact that each algebra $\Rs(W)$ is stable under the adjoint action of the 
modular group $\{\Delta_{W}^{it}\mid t\in\RR\}$. Finally, the fourth
assertion is a consequence of the basic result from Tomita-Takesaki theory
that the only weakly closed subalgebra of a von Neumann algebra $\Rs$
which has $\Omega$ as a cyclic vector and  is stable under the action of the 
modular group of $(\Rs , \Omega)$ is $\Rs$ itself.
\hfill\qed\enddemo 

     Proposition 2.2 and Corollary 2.3, where $\Js$ is replaced by $\Ks$ and
$\Ts$ by $\Us$, also hold in the setting of the CMG, and it is still true
that $\Rs(W)$ is nonabelian for each $W \in \Ws$. Moreover, an analogue of
Proposition 3.1 obtains. We omit the straightforward proofs of these 
statements. For the set of wedgelike regions $\Ws$ in $\RR^4$, which we
consider here, the elements $\upsilon_W(t)$ of the transformation group
$\Us$ satisfy the conditions (A) and (B) in Section 4.1. We can thus apply
Theorem 4.1.15 to conclude the following result.
 
\proclaim{Lemma 5.2.2} Let the CMG hold as described. If 
$\{\Delta_{W_0}^{it}\}_{t \in \RR}$ is the modular group corresponding to an 
arbitrary wedge algebra  $\Rs({W_0})$ and the vector $\Omega$, then for 
each $t \in \RR$ there exists an element $L_{W_0}(t)$ of the extended 
(by the dilatations $\RR_+$) Poincar\'e group $\Ds\Ps$ such that
$$\ad \Delta_{W_0}^{it}(\Rs(W)) = \Rs(L_{W_0}(t) W) \quad , 
\quad \text{for all} \quad W \in \Ws \quad . $$

\endproclaim

    Because of the group law 
$\Delta_{W_0}^{is} \Delta_{W_0}^{it}  = \Delta_{W_0}^{i(s+t)}$ and the 
standing assumption that the relation between wedges and wedge algebras is a 
bijection, one has
$$ L_{W_0}(s) L_{W_0}(t) = L_{W_0}(s+t), \quad s,t \in \RR \quad , 
\tag{5.2.1}$$
for the corresponding transformations. In particular, 
$L_{W_0}(t) = L_{W_0}(t/2)^2$, so each $L_{W_0}(t)$ lies in the identity 
component $\Ds\Pid$ of the extended Poincar\'e group. We denote by $\Gs$ the 
subgroup of $\Ds\Pid$ generated by the set 
$\{ L_W(t) \mid t \in \RR, W \in \Ws \}$. In the next step of our analysis we 
shall determine this group. \par
     In order to abbreviate the argument, we shall make the additional 
simplifying assumption that $\Gs$ acts transitively on the set $\Ws$ of
wedges (which follows from the assumption that the adjoint action of $\Ks$
upon $\wrnet$ is transitive). However, this additional assumption can, in
fact, be {\it derived} from the CMG as it stands; we shall present the proof
in a subsequent publication. 

\proclaim{Lemma 5.2.3} If the CMG holds and $\Ks$ acts transitively upon 
$\wrnet$, then the  group $\Gs$ of transformations coincides with the proper 
orthochronous Poincar\'e group: $\Gs = \Pid$. 
\endproclaim

\demo{Proof} Note, to begin, that the elements of the commutator subgroup of 
$\Gs$ do not contain any nontrivial dilatations and therefore are contained in 
$\Pid$. Moreover, they act transitively on ${\Cal W}$, as can be seen as 
follows: Let $W_1$ be any wedge and let $L_W(t) \in \Gs$ be any transformation 
associated with some wedge $W$. As $\Gs$ is assumed to act transitively on 
$\Ws$, there exists, according to part (1) of Lemma 5.2.1, a transformation 
$L \in \Gs$ such that $L L_{W_1} (t) L^{-1} = L_W (t)$. On the other hand, 
according to part (3) of that lemma, one has the relation 
$L_{W_1} (s) W_1 = W_1$, for all $s \in \RR$, and consequently 
$$ L L_{W_1} (t) L^{-1} L_{W_1} (t)^{-1} \, W_1 = L_W (t) \, W_1 \quad . $$
Since the wedge $W_1$ and the transformation $L_W (t)$ were arbitrary,
the transitive action of the commutator subgroup follows. The first part of 
Prop. 4.2.9 then implies that this subgroup of $\Gs$ coincides with 
$\Pid$. \par
     Now let $W$ be any given wedge, let 
$L_W(t) = (\gamma_W(t), \Lambda_W(t), a_W(t)) \in \Gs$ be the
corresponding transformation on Minkowski space, where $\gamma_W(t) > 0$
is a dilatation, $\Lambda_W(t)$ a Lorentz transformation and $a_W(t)$ a 
translation, and let $(1,1,a) \in \Gs, a \in \RR^4$, be any other nontrivial 
translation which leaves $W$ invariant. Part (2) of Lemma 5.2.1 then implies 
that $L_W(t) (1,1,a)  L_W(t)^{-1} = (1,1,a)$, for all $t \in \RR$. On the 
other hand, one obtains by explicit computation 
$L_W(t) (1,1,a) L_W(t)^{-1} = (1, 1, \gamma_W(t) \Lambda_W(t) a)$. 
Hence $a$ is an eigenvector of $\Lambda_W(t)$ and thus would have to be 
lightlike if $\gamma_W(t) \neq 1$, in conflict with its choice. Therefore,
one has $\gamma_W(t)=1$ and $\Gs = \Pid$, as claimed. 
\hfill\qed\enddemo 

     In the next step we want to determine the geometric action of the 
transformations $L_W(t)$ associated with the modular groups. The preceding 
results suffice to show that these transformations are Lorentz boosts. More 
detailed information will be obtained by making use of the continuity and 
analyticity properties of the modular groups.   

\proclaim{Proposition 5.2.4} Given the CMG and the transitive action of $\Ks$
upon $\wrnet$, the transformations 
$L_{W_R}(t) \in \Gs$ associated with the standard wedge $W_R$ are, for all
$t \in \RR$, the boosts 
$$ L_{W_R} (t) = \left( \matrix B(t) & 0 \\
                                  0  & 1 \endmatrix \right) 
\ \text{with}  \ \   
B(t) = \left( \matrix \cosh \alpha t & \sinh \alpha t \\
                       \sinh \alpha t & \cosh \alpha t  \endmatrix
                                  \right) 
\tag{5.2.2}$$
and $\alpha \in \{ \pm 2 \pi  \}$. The form of $L_{W} (t)$ for arbitrary 
wedges $W$ is obtained from $L_{W_R} (t)$ by Poincar\'e transformations --
see the first part of Lemma 5.2.1.  
\endproclaim

\demo{Proof} According to the second part of Lemma 5.2.1 and Lemma 5.2.3, 
$L_{W_R}(t)$ commutes with all elements of the stability group of $W_R$ in 
$\Pid$. It thus must be a boost which leaves $W_R$ invariant and consequently 
has the block form given in (5.2.2). Moreover, because of relation (5.2.1), 
the matrix $B(t)$ has the form given in (5.2.2), where the argument 
$\alpha t$ of the hyperbolic functions could, however, be {\it a priori\/} any 
additive function (homomorphism) $\beta (t)$ on the reals. For the proof that 
$\beta (t)$ has the asserted form, it suffices to show that $\beta (t)$ is 
continuous - one then may apply standard results about continuous 
one-parameter subgroups of $GL(n,\CC)$ (see, {\it e.g.} Theorem 2.6 and 
Corollary 1.5 in \cite{38}). \par
    To this end one exploits the continuity properties of the group \newline
$\{\Delta_{W_R}^{it} \mid t \in \RR \}$. According to the information 
about the action of $L_{W_R} (t)$ accumulated up to this point, if $\ell$ is 
any positive lightlike vector such that $W_R + \ell \subset W_R$, one has
$$ \Delta_{W_R}^{it} \Rs (W_R + \ell) \Delta_{W_R}^{-it} 
= \Rs (W_R + e^{\beta(t)} \ell).$$
If $\beta (t)$ is discontinuous at $t=0$, one may assume without restriction 
(since $\beta(\cdot)$ is additive) that there exists a $\beta_0 > 0$ and 
a sequence $\{ t_n \}_{n \in \IN} \subset \RR$ such that 
$t_n \rightarrow 0$ and $\beta_{t_n} \geq \beta_0 > 0$.  By isotony and the 
preceding equality of algebras, one thus obtains 
$\Delta_{W_R}^{it_n} \Rs (W_R + \ell) \Delta_{W_R}^{-it_n} \subset 
\Rs (W_R + e^{\beta_0 } \ell) $. As $\Delta_{W_R}^{it}$ is continuous in
the strong operator topology and $\Rs (W_R + e^{\beta_0 }\ell)$
is weakly closed, one can proceed on the left-hand side of this inclusion to 
the limit, yielding 
$\Rs (W_R + \ell) \subset \Rs (W_R + e^{\beta_0 } \ell)$. Since also 
$\Rs (W_R + e^{\beta_0 } \ell) \subset \Rs (W_R + \ell)$, by isotony, one 
concludes that these two algebras are equal, in conflict with the CMG. So 
$\beta(\cdot)$ is continuous at $0$, and since it is a homomorphism it must be
continuous everywhere. This shows that for some constant $\alpha$,
$\beta (t) = \alpha t$, for all $t \in \RR$. \par
    In order to determine the value of this constant $\alpha$, one can rely 
on results of Wiesbrock \cite{71}\cite{72}, cf.\ also \cite{14}. If $\ell$ 
is a lightlike vector as above, the specific form of the action of 
$L_{W_R} (t)$, $t \in \RR$, on $\Rs (W_R + \ell)$ 
implies that $\big( \Rs (W_R + \ell) \subset \Rs(W_R), \Omega \big)$ is
a $\pm$-half-sided modular inclusion (where the $\pm $ depends on the sign of
$\alpha$). The claim $\alpha \in \{ \pm 2\pi \}$ then follows from the results 
in the quoted references. 
\hfill\qed\enddemo 

     We have therefore derived modular covariance from our {\it prima facie}
less restrictive Condition of Geometric Action for the modular groups. We next 
show that we have a strongly continuous unitary representation of $\Pid$ 
satisfying the spectrum condition with either negative or positive energy. 

\proclaim{Theorem 5.2.5} Assume that the CMG is satisfied and that the adjoint
action of $\Ks$ upon $\wrnet$ is transitive. Then there is a strongly 
continuous unitary representation $U(\cdot)$ of the covering group 
$ISL(2,\CC)$ of $\Pid$ which generates $\Ks$ and acts geometrically correctly 
on the net. If, in addition, the net $\wrnet$ satisfies locality, {\it i.e.\/} 
$\Rs(W) \subset \Rs(W')'$ for all $W \in \Ws$, then $U(\cdot)$ yields  
a strongly continuous unitary representation $U(\cdot)$ of $\Pid$ satisfying 
either the positive or negative spectrum condition, depending on the sign of 
$\alpha$ in Prop. 5.2.4.
\endproclaim

\demo{Proof} This may be proven analogously to the arguments of Section 4.3, 
but since Prop. 5.2.4 has already established that modular covariance holds, 
it suffices here simply to appeal to the results of \cite{22}\cite{36} - 
particularly Lemma 2.6 and Corollary 1.8 in \cite{22} and Prop. 2.8 in 
\cite{36}. In fact, the mentioned results of \cite{22} imply that $\Ks$ 
provides a strongly continuous unitary representation of the covering group 
$ISL(2,\CC)$.\footnote{Note that, given our hypotheses, the assumptions of 
locality and additivity in \cite{22} are not required for the cited results.}
For then the pair $(\Ks,\xi)$, where $\xi: \Ks \mapsto {\Cal U} \simeq \Pid$ 
is the canonical homomorphism, is what those authors call a central weak Lie 
extension of the group $\Pid$. With the additional assumption of locality, 
the results of \cite{36} imply that the projective representation obtained 
above is actually a strongly continuous representation of $\Pid$. The sign of 
$\alpha$ in Prop. 5.2.4 determines whether the inclusions 
$(\Rs(W_R + \ell) \subset \Rs(W_R) , \Omega)$, with $W_R,\ell$ as in the proof 
of Prop. 5.2.4, are all +-half-sided modular inclusions or $-$-half-sided 
modular inclusions. That, together with Poincar\'e covariance, then entails 
the spectrum condition with either positive or negative energy (see the 
argument of the proof of Theorem 5.1.2).
\hfill\qed\enddemo

     It is of particular interest to note that the weak geometric 
action of the modular groups we have been studying in this section also 
entails the corresponding geometric action of the modular involutions, if and 
only if the net $\wrnet$ is local.

\proclaim{Theorem 5.2.6} If the CMG is satisfied and the group $\Ks$ 
generated by the modular unitaries of all wedge algebras acts transitively
upon the net $\wrnet$, then $\Ks$ is equal to the 
group $\Js^+$ consisting of all products of even numbers of modular 
conjugations $\{ J_W \mid W \in \Ws \}$. The adjoint action of the 
modular conjugations in $\{ J_W \mid W \in \Ws \}$ leaves the net $\wrnet$
invariant (and so our CGMA holds) if and only if the net fulfills locality, 
{\it i.e.} $\Rs(W') \subset \Rs(W)'$, for all $W \in \Ws$. \par
     In that case, the modular conjugations $\{ J_W \mid W \in \Ws \}$ have 
the same geometric action upon the net as was found in Prop. 4.2.10 under 
different hypotheses. Furthermore, the net $\wrnet$ satisfies wedge duality 
and the modular conjugations yield a representation of the proper Poincar\'e 
group $\Ps_+$ which acts geometrically correctly upon the net.
\endproclaim

\noindent (A simple and well-known example of a net which complies with the 
CMG but where locality and hence also the CGMA fails is the net generated by a 
Fermi field \cite{10}. It satisfies a twisted form of locality, however.)

\demo{Proof} By the results of \cite{22} appealed to in the proof of 
Theorem 5.2.5, $\Ks$ is isomorphic to either $\Pid$ itself or to its covering 
group, $ISL(2,\CC)$, and by Theorem 5.2.5 one knows that $\Ks = U(ISL(2,\CC))$.
Under the stated hypothesis, the conclusion of Corollary 2.7 in \cite{36} 
still holds\footnote{In the proof of Prop. 2.6 in \cite{36}, which is 
appealed to in the argument for Corollary 2.7, one should replace 
$\Fs(W_1 \cap \Lambda_2(-t)W_1)$ by $\Rs(W_R) \cap \Rs(\Lambda_2(-t)W_R)$. 
Since $W_R \cap \Lambda_2(-t)W_R$ is not empty, assumption (ii) in our CGMA 
entails that $\Omega$ is cyclic and separating for 
$\Rs(W_R) \cap \Rs(\Lambda_2(-t)W_R)$. The rest of the argument proceeds as 
before.}, {\it i.e.} one has also here the relation for the modular 
conjugations and groups associated with the wedges $W_k^{(0)}, k = 1,2,3$, 
based on the time-zero plane, 

$$J_{W_R}\Delta^{it}_{W_k^{(0)}}J_{W_R} = \Delta^{-it}_{W_k^{(0)}} \quad
k = 2,3 \quad , $$

\nind where $W_R = W_1^{(0)}$ is the standard wedge and $J_{W_R}$ the 
corresponding modular involution. The corresponding relation for $k = 1$ is a 
basic result of Tomita-Takesaki theory. Furthermore, $J_{W_R}$ commutes with 
those elements of $\Ks$ which act upon the net $\wrnet$ as translations in the 
direction of the $2$- or $3$-axes, since their adjoint action leaves 
$\Rs(W_R)$ invariant and they leave $\Omega$ fixed. The adjoint action of 
$J_{W_R}$ on those elements of $\Ks$ which act upon $\wrnet$ as translations 
in the lightlike directions of $\ell_{\pm}$ fixed by $W_R$ inverts these 
elements, by \cite{71}. Let $\theta_1$ denote the element
$\diag (-1,-1,1,1) \in \Ps_+$. The above remarks imply the relations
$$J_{W_R}U(\mu^{-1}(\lambda))J_{W_R} = U(\mu^{-1}(\theta_1 \lambda \theta_1)) 
\quad , \tag{5.2.3} $$
for any $\lambda$ which is one of the translations or boosts just discussed, 
where $\mu$ is the canonical covering homomorphism from $ISL(2,\CC)$ onto 
$\Pid$. But since these boosts and these translations generate $\Pid$, it 
follows that (5.2.3) holds for any $\lambda \in \Pid$. Indeed, one has (5.2.3) 
for any wedge $W$, with $\theta_1$ replaced by the corresponding involution,
and it follows that $J_{W} \Ks J_{W} = \Ks$, for any $W \in \Ws$. \par
     Since the Poincar\'e group acts transitively on $\Ws$, for any pair of 
wedges $W_a, W_b$ there exists some Poincar\'e transformation 
$\lambda \in \Pid$ such that $\lambda W_a = W_b$. Consequently, one has 
$J_{W_b} = U(A(\lambda)) J_{W_a} U(A(\lambda))^{-1}$ for any 
$A(\lambda) \in ISL(2,\CC)$ with $\mu (A) = \lambda$, since $U(\cdot)$ acts 
geometrically correctly on the net and leaves $\Omega$ invariant. Hence, one 
has 
$$J_{W_a} J_{W_b} = \big( J_{W_a} U(A(\lambda)) J_{W_a} \big) 
U(A(\lambda))^{-1} \in {\Ks} \quad , $$
according to the preceding results, which shows that $\Js^+ \subset {\Ks}$. On 
the other hand, it follows from relation (5.2.3) that for $\lambda \in \Pid$
$$ J_{W_R} J_{\lambda^2 W_R}  = 
U(A(\theta_1 \lambda \theta_1)^{2} A(\lambda)^{-2}) \quad . $$
Hence the unitaries corresponding to the boosts in the $2$- and $3$-direction 
as well as to the lightlike translations in the direction of $\ell_{1\pm}$ are 
contained in ${\Cal J}^+$. Similarly, one can reproduce these
arguments with $W_R = W_1^{(0)}$ replaced by $W_2^{(0)}$ and $W_3^{(0)}$ to 
show that the unitaries corresponding to the boosts in the $1$-direction as 
well as the lightlike translations in the direction of $\ell_{2\pm}$ and
$\ell_{3\pm}$ are contained in $\Js^+$. Since these unitaries together 
generate $U(ISL(2,\CC))$, one concludes that $\Ks \subset {\Cal J}^+$, and 
therefore the two groups are equal.\par
     {}From the invariance of $\Rs(W_R)$ under the adjoint action of the 
unitaries implementing the stability group of $W_R$, it follows that also  
the algebra $\Rs(W_R)' = J_{W_R} \Rs(W_R) J_{W_R}$ is invariant under this
action. Hence, if $\Rs(W_R)\, '$ is a wedge algebra, then it must be equal to 
$\Rs(W_R ')$ -- it cannot coincide with $\Rs(W_R)$, since otherwise it would 
be abelian. Therefore, if the adjoint action of the elements of 
$\{ J_W \mid W \in \Ws \}$ leaves $\wrnet$ invariant, the net must satisfy 
wedge duality and hence locality. Conversely, if the net satisfies locality, 
then $\Rs(W') \subset \Rs(W)'$ is stable under the adjoint action of the 
modular group $\{ \Delta_W^{-it} \mid t \in \RR\}$, of $(\Rs(W)', \Omega)$ 
according to Prop.\ 5.2.4. Since $\Omega$ is cyclic and separating for both 
algebras, Tomita-Takesaki theory then entails the equality 
$\Rs(W') = \Rs(W)' = J_{W} \Rs(W) J_{W}$. But this implies that, for any 
$W_a, W_b \in \Ws$ with corresponding modular involutions $J_{W_a} ,J_{W_b}$, 
one has 
$$J_{W_a} \Rs(W_b) J_{W_a} = J_{W_a} J_{W_b} \Rs(W'_b) J_ {W_b} J_{W_a} 
\in \wrnet \quad , $$
since $J_{W_a} J_{W_b} \in \Js^+ = \Ks$ and $\wrnet$ is invariant under the
adjoint action of $\Ks$. The remaining assertions are therefore immediate
consequences of the results of Chapter IV.
\hfill\qed\enddemo   

     To close the circle of implications relating the geometric action of
the modular involutions to that of the modular groups, we conclude this 
section with the following result.

\proclaim{Theorem 5.2.7} Assume the CGMA, with the choices $\Ms = \RR^4$ and 
$\Ws$ the collection of wedgelike regions in $\RR^4$, and the transitivity of 
the adjoint action of $\Js$ on the net $\wrnet$. If $\Delta_W^{it} \in \Js$, 
for all $t \in \RR$ and some $W \in \Ws$, {\it i.e.} if the modular stability
condition obtains, and the adjoint action of $\Ks$ upon $\wrnet$ is 
transitive, then modular covariance is satisfied.
\endproclaim

\demo{Proof} Since, by hypothesis, the adjoint action of any element of $\Js$ 
leaves the set $\wrnet$ invariant and since their transitive action on
$\wrnet$ implies $\Delta_W^{it} \in \Js$, for all $t \in \RR$ and $W \in \Ws$, 
it is clear that the CMG is satisfied. Prop. 5.2.4 completes the proof.
\hfill\qed\enddemo

\nind Hence, we {\it derive} modular covariance from our CGMA, whenever 
the modular stability condition also holds. We remark once again that 
in a later publication we shall show that the additional assumption of the 
transitive action of $\Ks$ is superfluous. \par

\bigpagebreak
   {\bf 5.3. Modular Involutions Versus Modular Groups}
\bigpagebreak

     As explained in the introduction, there have been two distinctly different
approaches to the study of the geometric action of modular objects and its
consequences. In the one, initiated in \cite{24}, geometric action of the
modular involutions was assumed, whereas in the other, initiated in \cite{12},
the starting point was the geometric action of the modular groups. However, 
even within each of these approaches, differing forms of concrete action have 
been studied. In most of the papers concerned with the consequences of 
geometric action of the modular groups, the action was assumed in the form of 
modular covariance (see \cite{22}\cite{36}, among others). 
There are some variations of this condition in the literature 
\cite{35}\cite{26}, but they all have in common that from the outset 
one is given an action of the Lorentz group on the space-time. \par
     Certain exceptions are the papers by Kuckert \cite{46} and Trebels 
\cite{65}, where the geometric action was assumed in the guise of requiring 
the adjoint action of the modular groups (or the modular involutions) to leave 
the set of local algebras in Minkowski space invariant. However, in both
approaches the starting point is a vacuum representation of a net on Minkowski 
space which is covariant with respect to the translation group satisfying the 
spectrum condition. \par
     All of these approaches have in common that some {\it a priori} 
information about the geometric action of the modular groups or the
spacetime symmetry group is required. But, as we have shown in the above
analysis, this detailed information is {\it derived} if one starts from
our CGMA. We also wish to emphasize that the condition of modular covariance 
and Borchers' relation (5.1.1) are not implied in our framework. To 
illustrate these assertions, we present a simple example of a net satisfying 
our CGMA and all of the other assumptions made in this paper, except the 
modular stability condition. This example thus violates the spectrum 
condition and the modular groups associated to wedge algebras do not coincide 
with the representation of the Lorentz boosts, {\it i.e.} modular covariance 
{\it fails} in this example, though it is Poincar\'e covariant. Subsequently, 
we give another example violating modular covariance but satisfying the 
spectrum condition and all of our assumptions. It is therefore clear 
that the assumption of modular covariance is more restrictive than the CGMA, 
even when the spectrum condition is posited.  \par
     Turning to our first example, let $\cnet$ be the standard net of von 
Neumann algebras generated by a (hermitian, scalar, massive) free field 
on the Fock space $\Hs$. It is based on the set $\Cs$ of double cones in 
$\RR^4$ and covariant under the standard action $\alpha_{\lambda}$, 
$\lambda \in \Pid$, of the Poincar\'e group. Let $\Theta$ be the PCT-operator 
on $\Hs$ and $\theta$ be the corresponding reflection in Minkowski space. For 
each double cone $\Os$ define 
$\Bs(\Os) = \As(\theta\Os) = \Theta\As(\Os)\Theta$. 
Let $\hat{\As}(\Os) \equiv \As(\Os)\otimes\Bs(\Os)$ act on $\Hs\otimes\Hs$.
The net $\hnet$ is clearly local, since $\Theta$ is antiunitary and thus 
behaves properly under the taking of algebraic commutants. We observe that
$\hat{\alpha}_{\lambda} \equiv \alpha_{\lambda}\otimes\beta_{\lambda}$,
with $\beta_{\lambda} \equiv \alpha_{\theta\lambda\theta}$,
$\lambda\in\Ps^{\uparrow}_+$, defines an automorphic local action on $\hnet$, 
as can be seen as follows. With $\lambda \in \Ps^{\uparrow}_+$, one has
$$\align
\hat{\alpha}_{\lambda}(\hat{\As}(\Os)) &= 
\alpha_{\lambda}(\As(\Os))\otimes\beta_{\lambda}(\Bs(\Os)) =
\As(\lambda\Os)\otimes(\As((\theta\lambda\theta)\theta\Os)) \\
 &= \As(\lambda\Os)\otimes\As(\theta\lambda\Os) = \hat{\As}(\lambda\Os)
\quad .
\endalign $$
    With $U(\lambda)$ the unitary implementation of $\alpha_{\lambda}$ on 
$\Hs$, one easily checks that $V(\lambda) \equiv \Theta U(\lambda) \Theta$ 
implements the action of $\beta_{\lambda}$. Setting $U(x) = e^{ixP}$, where
$P$ is the generator of the translations satisfying the positive spectrum
condition, one has $V(x) = \Theta e^{ixP} \Theta = e^{-ixP}$. Hence 
$V(\lambda)$ satisfies the negative spectrum condition, but 
$\hat{U}(\lambda) \equiv U(\lambda)\otimes V(\lambda)$ violates both the
positive and the negative spectrum conditions. \par
     By the results of Bisognano and Wichmann \cite{9}, applicable to the 
free field, one knows that for the standard wedge $W_R$ the modular structure 
for the (weakly closed) wedge algebra $\As(W_R)$ and $\Omega$ is given by 
$J_{W_R} = \Theta_R = \Theta U_{\pi}$, where $U_{\pi}$ implements the 
rotation by $\pi$ about the $1$-axis, and 
$\Delta^{it}_{W_R} = U(\lambda_R(t))$, $t \in \RR$, where the
$\lambda_R(t)$ are the Lorentz boosts in the $1$-direction. The corresponding 
modular objects for 
$(\Bs(W_R),\Omega) = (\Theta \As(W_R) \Theta, \Omega) = (\As(W_R)',\Omega)$ 
are given by ${}_{\Bs}J_{W_R} = \Theta_R$ and
${}_{\Bs}\Delta^{it}_{W_R} = U(\lambda_R(t))^{-1} = U(\lambda_R(-t))$. It 
follows that the modular objects for 
$(\hat{\As}(W_R)=\As(W_R)\otimes\Bs(W_R),\Omega\otimes\Omega)$ are given by
$$\hat{J}_{W_R} =  \Theta_R \otimes\Theta_R  \quad , \quad
\hat{\Delta}^{it}_{W_R} = U(\lambda_R(t)) \otimes U(\lambda_R(-t)) \quad .$$
So, one has (with $\theta_R$ the transformation on Minkowski space 
corresponding to $\Theta_R$)
$$\align
\hat{J}_{W_R} \hat{\As}(\Os) \hat{J}_{W_R} &= 
\Theta_R \As(\Os) \Theta_R \otimes \Theta_R \Bs(\Os) \Theta_R =
\As(\theta_R \Os) \otimes \As(\theta_R \theta\Os) \\
  &= \As(\theta_R \Os) \otimes \As(\theta\theta_R \Os) = 
\hat{\As}(\theta_R \Os) \quad ,
\endalign $$
and the modular conjugation $\hat{J}_{W_R}$ acts geometrically correctly
on the net $\hnet$. By Poincar\'e covariance of the net, the same holds true 
for the modular involution $\hat{J}_W$, for any wedge $W$. \par 
     Turning to the modular groups, one sees
$$\align
\hat{\Delta}^{it}_{W_R} \hat{\As}(\Os) \hat{\Delta}^{-it}_{W_R}  &=
U(\lambda_R(t)) \As(\Os) U(\lambda_R(t))^{-1} \otimes 
U(\lambda_R(-t)) \Bs(\Os) U(\lambda_R(-t))^{-1} \\ 
&= \As(\lambda_R(t)\Os) \otimes \As(\lambda_R(-t)\theta\Os) =
\As(\lambda_R(t)\Os) \otimes \As(\theta\lambda_R(-t)\Os) \\
 &= \As(\lambda_R(t)\Os) \otimes \Bs(\lambda_R(-t)\Os) \neq 
\hat{\As}(\lambda_R(t)\Os) 
\quad .
\endalign $$
Hence, $\hat{\Delta}^{it}_{W_R}$ does not satisfy modular covariance. Note also
that the modular groups are {\it not} contained in $\Js = \hat{U}(\Ps_+)$, so
that the modular stability condition is violated, in accord with 
Theorem 5.2.7.  \par
     We mention as an aside that in \cite{36} Guido and Longo propose 
the split property, which yields the uniqueness of the representation of the 
Poincar\'e group, as a natural candidate for the hypothesis needed in order to 
conclude that the modular group of a wedge algebra satisfies modular 
covariance. However, in the preceding example, the split property holds, 
though modular covariance does not.  \par
    In our next example, we see that it is possible for all of our assumptions
to hold, as well as the positive spectrum condition, but for modular covariance
to be violated. For each $W \in \Ws$, the set of wedgelike regions in 
four-dimensional Minkowski space, we denote by $N(W)$ the 
unique wedge in the coherent family of wedges determined by $W$ which contains 
the origin in its edge. Once again taking $\cnet$ to be the usual net 
for the free field on four-dimensional Minkowski space, we 
consider the net $\wnet$ indexed by the wedgelike regions and define for this 
example $\hat{\As}(W) \equiv \As(W) \otimes \As(-N(W))$. (Note that
$-N(W) = N(W)'$.) This net is local, since $W_1 \subset W_2'$ entails 
$N(W_1) \subset N(W_2)'$ and since $\As(W)' = \As(W')$ (Haag duality). 
Moreover, for each $\Pid \ni \lambda = (\Lambda,a)$, we set 
$\hat{\alpha}_{\lambda} \equiv \alpha_{\lambda} \otimes \alpha_{(\Lambda,0)}$.
Hence, the translation subgroup acts trivially upon the second factor of each
local algebra. In this example, the unitary implementers of the action
$\hat{\alpha}_{\lambda}$ are given by 
$\hat{U}(\Lambda,a) = U(\Lambda,a) \otimes U(\Lambda,0)$, and the translation
subgroup is implemented by $\hat{U}(a) = U(a) \otimes \idty$. Thus, the 
positive spectrum condition holds (though the vacuum is infinitely
degenerate), whereas modular covariance is violated. In fact,
$\hat{\Delta}_{W_R}^{it} = U(\lambda_R(t)) \otimes U(\lambda_R(-t))$, since in 
the second factor of $\hat{\As}(W)$ there appears the algebra 
$\As(-N(W)) = \As(N(W)') = \As(N(W))'$. On the other hand, the modular 
conjugations corresponding to $(\hat{\As}(W),\Omega \otimes \Omega)$ are
given by $J_W \otimes J_{N(W)}$ and hence satisfy the CGMA and the assumption 
of transitive action on $\Ws$. \par
     It is of interest to note that this example also violates the condition 
of modular stability, $\hat{\Delta}_W^{it} \in \Js$, in spite of the 
validity of the spectrum condition. Furthermore, the local algebras associated 
with double cones $\Os$,
$$\hat{\As}(\Os) \equiv 
\underset{\Os \subset W \in \Ws}\to{\bigcap}
\hat{\As}(W) $$

\nind do not generate the wedge algebras. This resembles the situation
which one expects to meet for the bosonic part of the field algebra in 
theories with topological or gauge charges.\par
    We sketch a final illustrative example, which makes a number of points 
about the interrelationship of the CGMA, uniqueness of representation of
the Poincar\'e group, and some further properties of interest. Consider an
infinite component free hermitian Bose field with momentum space annihilation 
and creation operators satisfying the following canonical commutation relations
\cite{47}:

$$[a(\vec{p}\, ',q'), a^*(\vec{p},q)] = 2\omega_{\vec{p}} \,
\delta^{(3)}(\vec{p}-\vec{p}\, ')\delta^{(4)}(q-q') \quad , $$

\nind where $\vec{p},\vec{p}\, ' \in \RR^3$,  
$\omega_{\vec{p}} = \sqrt{\vec{p}\, ^2 + m^2}$, $m > 0$, and the variables
$q,q' \in \RR^4$ label the internal degrees of freedom. One unitary 
representation of the Poincar\'e group on the corresponding Fock space of this
field is determined by

$$U(\Lambda,x)a(\vec{p},q)U(\Lambda,x)^{-1} \equiv
e^{i\Lambda p \cdot x} a(\vec{\Lambda p},q) \quad , $$

\nind where $p = (\omega_{\vec{p}},\vec{p})$, while a second one is determined
by

$$\tilde{U}(\Lambda,x)a(\vec{p},q)\tilde{U}(\Lambda,x)^{-1} \equiv
e^{i\Lambda p \cdot x} a(\vec{\Lambda p},\Lambda q) \quad . $$

\nind It is evident that both representations satisfy the spectrum condition.
\par
     Let $\cnet$ be the net of von Neumann algebras generated by this free
field. Clearly, this net transforms covariantly under both $U(\Pid)$ and 
$\tilde{U}(\Pid)$. The work of Bisognano and Wichmann \cite{9} shows that, 
using the representation $U(\Pid)$, the net satisfies the special condition of 
duality, and hence it satisfies Haag duality for the wedge algebras, the CGMA, 
modular covariance and the modular stability condition, 
$\Ks \subset \Js$. The arguments of Bisognano and Wichmann break down for the 
representation $\tilde{U}(\Pid)$, because the extra action on the dummy 
variable would destroy the analytic continuation crucial to their arguments. 
\par
     Applying the CGMA to the net $\cnet$ in the Fock vacuum state would 
result in the construction of the representation $U(\Pid)$ and not the 
representation $\tilde{U}(\Pid)$. Note further that since both representations 
act geometrically correctly upon the net, we have

$$\tilde{U}(\lambda) = \tilde{Z}(\lambda)U(\lambda) \quad , $$

\nind for all $\lambda \in \Pid$, with coefficients $\tilde{Z}$ which
induce internal symmetries of the net and commute with $U(\Pid)$. But
they are not contained in $\Js$ and, for this reason, this example escapes 
the uniqueness statement in Theorem 4.3.9. On the other hand, the net $\cnet$ 
violates the distal split property and, for this reason, the example also 
escapes the uniqueness theorem of \cite{22}. \par
    As we have shown, the CMG implies both modular covariance and the CGMA for 
the involutions (the latter in the presence of locality). The results in 
Section 5.2 therefore generalize the results of both \cite{22} and \cite{46}. 
We have also seen that there exist Poincar\'e covariant 
nets of local algebras on Minkowski space which do not satisfy the condition 
of modular covariance but which satisfy all of our assumptions, with or 
without the additional condition of positive spectrum. \par
     Though the CGMA (in application to the special case of Minkowski space) 
is weaker than the condition of modular covariance, it nonetheless allows one 
to systematically establish the same results which were proven under the 
assumption of modular covariance in the literature. Moreover, since the modular
involutions depend only upon the characteristic cones of the pairs 
$(\As(W),\Omega)$, it would seem that they are more likely to encode some 
intrinsic information about the representation, as opposed to the modular
unitaries, which are strongly state-dependent. \par

\bigpagebreak

\heading VI. Geometric Modular Action and De Sitter Space \endheading

     As a further example of application of the program outlined in
Chapter III, we consider three-dimensional de Sitter space. The restriction
on the dimension is made for simplicity, as it will allow us to apply some
of the results obtained in the preceding analysis. \par
     It is well-known that three-dimensional de Sitter space $dS^3$ can
conveniently be embedded into the ambient four-dimensional Minkowski space
$\RR^4$. Choosing proper coordinates, it is described by

$$dS^3 \equiv \{ x \in \RR^4 \mid x_0^2 - x_1^2 - x_2^2 - x_3^2 = -1 \} \quad , $$

\nind with the induced metric and causal structure from Minkowski space. 
Accordingly, the restriction of the Lorentz group $\Ls$ in the ambient space
$\RR^4$ to $dS^3$ is the isometry group of this space, simply called here the
de Sitter group and commonly denoted by $O(1,3)$. As the elements of $\Ls$
are uniquely fixed by their action on $dS^3$, we will identify the de Sitter
group with $\Ls$ for later convenience. Similarly, the proper de Sitter
group and its identity component are identified with $\Ls_+$ and $\Lid$,
respectively.  \par
     In this chapter we shall assume the CGMA for a net $\wrnet$ on
$\Ms = dS^3$. Applying the reasoning advanced in Chapter III, one is presented 
once again with a unique minimal admissible family, namely
$\Ws \equiv \{ \tilde{W} \cap dS^3 \mid \tilde{W} \in \tilde{\Ws}_0 \}$, where
$\tilde{\Ws}_0$ is the family of wedgelike regions in the ambient 
four-dimensional Minkowski space $\RR^4$ containing the origin in their edges. 
Hence, we shall proceed with this choice of index set. Though there are 
clearly affinities between this setting and the Minkowski-space situation, 
there are nevertheless some nontrivial points to be worked out which do not 
automatically follow from the work in the previous chapters.  \par
     To begin, we shall prove in Section 6.1 a general Alexandrov-like result 
in $dS^3$ along the lines of Theorem 4.1.15. In view of the different geometric
structure of de Sitter space, the construction of the induced point 
transformations in $dS^3$ differs from the corresponding construction in 
Minkowski space. (An alternative construction made under stronger assumptions 
may be found in \cite{28}.) In Section 6.2 it will be shown that the CGMA, 
with an additional technical postulate, implies that 
the bijections on $\Ws$ induced by the adjoint action of the modular 
involutions $\{ J_W \mid W \in \Ws \}$ upon the net $\wrnet$ are obtained by 
elements of the de Sitter group. Then, under the assumption that the group
generated by
$\{ \ad J_W \mid W \in \Ws \}$ acts transitively upon the set $\wrnet$, it 
will be shown that the group thereby generated is $\Ls_+$, whenever one of the 
algebras $\Rs(W)$ is nonabelian. In contradistinction to the Minkowski space 
situation, all elements of the index set $\Ws$ are atoms; hence it is entirely
possible for the wedge algebras to be abelian here. If they are abelian, then
the group induced upon $dS^3$ is equal to the identity component
$\Lid$ of the de Sitter group. \par
   After this analysis, we shall proceed analogously to the development in 
Chapter IV to obtain a strongly continuous unitary representation of $\Ls_+$, 
resp. $\Lid$, which acts geometrically correctly upon the net $\wrnet$. \par 

\bigpagebreak
   {\bf 6.1. Wedge Transformations in de Sitter Space}
\bigpagebreak

     In this section, we shall work with bijections $\hal : \Ws \mapsto \Ws$ 
satisfying the condition
$$W_1 \cap W_2 = W_3 \cap W_4 \quad \Leftrightarrow \quad \hal(W_1) \cap \hal(W_2) =
\hal(W_3) \cap \hal(W_4) \quad , \tag{6.1.1}$$
for arbitrary pairs $W_1,W_2$ and $W_3,W_4$ in $\Ws$. In the next section, we 
shall provide assumptions on the net $\wrnet$ which entail condition (6.1.1).
\par
     We shall use constantly without further mention the elementary fact 
that $W \in \Ws$ determines uniquely a wedgelike region $\tilde{W}$ in $\RR^4$
such that $W = \tilde{W} \cap dS^3$ and {\it vice versa}. Hence we shall, 
where convenient for us, identify $W$ with $\tilde{W}$. It will be
clear from the context whether $W$ is regarded as a subset of $dS^3$ or of the
ambient space $\RR^4$. Adopting the notation of Chapter IV, we shall write
$W[\ell_1,\ell_2] \equiv \tilde{W}[\ell_1,\ell_2,0] \cap dS^3$, where
$\tilde{W}[\ell_1,\ell_2,0] \in \tilde{\Ws}_0$ is the wedge in the ambient
space fixed by the two positive lightlike vectors $\ell_1,\ell_2$ and the
translation $0$. For the analysis of condition (6.1.1) we must make some
elementary geometric points about pairs of wedges.

\proclaim{Definition} Let $W[\ell_1,\ell_2],W[\ell_3,\ell_4] \in \Ws$ be 
wedges. If the positive lightlike vectors $\ell_1,\ell_4$, respectively
$\ell_3,\ell_2$, are not parallel, then the pair of wedges 
$(W[\ell_1,\ell_4],W[\ell_3,\ell_2])$ will be called the pair of wedges dual 
to $(W[\ell_1,\ell_2],W[\ell_3,\ell_4])$ (or simply the dual pair).
\endproclaim

\nind If $(W_3,W_4)$ is the pair dual to $\pair$, then 
$W_1 \cap W_2 = W_3 \cap W_4$. If this intersection is nonempty, then
$\pair$ and $(W_3,W_4)$ are the only pairs in $\Ws$ with this intersection.
Hence, $\emptyset \neq W_1 \cap W_2 = W_3 \cap W_4$ implies that the 
(unordered) pairs $(W_1,W_2)$ and $(W_3,W_4)$ are either the same or dual
(for details, see \cite{28}). \par

     We immediately have the following counterpart to Lemma 4.1.7.

\proclaim{Lemma 6.1.1} Let $\ell_{1 \pm} = (1,\pm 1,0,0)$,
$\ell_{2 \pm} = (1,0,\pm 1,0)$ and $\ell = (1,a,b,c)$ with  
$a,b,c \in \RR$, $a^2 + b^2 + c^2 = 1$, $b \neq 1$. 
The wedges $W_1 = W[\ell_{1 \, +},\ell_{1 \, -}]$ and 
$W_2 = W[\ell_{2 \, +},\ell]$ have empty intersection if and
only if $0 < a \leq 1$, $0 \leq b < 1$ and $c = 0$. The statement is still true
if $W_1$ is replaced by $W_1'$ and the condition $0 < a \leq 1$ is replaced by
$-1 \leq a < 0$, or also if $\ell_{2+}$ is replaced by $\ell_{2-}$ and
$0 \leq b < 1$ by $-1 < b \leq 0$.
\endproclaim

     This result will be used in the proof of the next lemma.

\proclaim{Lemma 6.1.2} Let $\hal : \Ws \mapsto \Ws$ be a bijection satisfying
(6.1.1) and let $\ell_0$ be a fixed future-directed lightlike vector. Then
$\hal$ maps collections of wedges 
$\{ W[\ell_0,\ell] \mid \ell \, \text{lightlike}, \ell\cdot\ell_0 > 0 \}$ 
and 
$\{ W[\ell,\ell_0] \mid \ell \, \text{lightlike}, \ell\cdot\ell_0 > 0 \}$ 
onto sets of the same form.\footnote{Note that these collections of wedges are 
such that any two elements form a self-dual pair of wedges with nonempty 
intersection.} Furthermore,
$$W_1 \cap W_2 = \emptyset \quad \Leftrightarrow \quad \hal(W_1) \cap \hal(W_2) = \emptyset \quad , \tag{6.1.2}$$
for any $W_1,W_2 \in \Ws$, and
$$\hal(W') = \hal(W)' \quad , \quad \text{for any} \quad W \in \Ws \quad . 
\tag{6.1.3}$$
Therefore, if $W_1 \cap W_2 \neq \emptyset$ and the pair $(W_3,W_4)$ is 
dual to $(W_1,W_2)$, then $(\hal(W_1),\hal(W_2))$ is dual to 
$(\hal(W_3),\hal(W_4))$.
\endproclaim

     Henceforth, we shall abbreviate 
$\{ W[\ell_0,\ell] \mid \ell \ \text{lightlike}, \ \ell \cdot \ell_0 > 0\}$ 
by \newline
$\{ W[\ell_0,\ell] \mid \ell \}$, {\it etc.}

\demo{Proof} Let $W_1 \cap W_2 = \emptyset$, with $W_1, W_2 \in \Ws$. There 
clearly exist infinitely many distinct pairs of disjoint wedges in $\Ws$. 
Let $(W_1,W_2)$, $(W_3,W_4)$ and $(W_5,W_6)$ be any three of them. Then 
(6.1.1) implies

$$\hal(W_1) \cap \hal(W_2) = \hal(W_3) \cap \hal(W_4) = 
\hal(W_5) \cap \hal(W_6) \quad . $$

\nind If this intersection is nonempty, \! 
$(\hal(W_1),\!\hal(W_2))$, \! $(\hal(W_3),\!\hal(W_4))$ and \!
$(\hal(W_5),\!\hal(W_6))$
are distinct (since $\hal$ is a bijection), mutually dual pairs of wedges,
which is impossible. Hence, the assertion (6.1.2) is proven. The final 
assertion of the lemma follows at once. \par  
    Let $\ell_1,\ell_2,\ell_3,\ell_4$ be given. There exist corresponding 
lightlike vectors $\ell_1',\ell_2',\ell_3',\ell_4'$ such that

$$\hal(W[\ell_1,\ell_2]) = W[\ell_1',\ell_2'] \quad \text{and} \quad 
\hal(W[\ell_3,\ell_4]) = W[\ell_3',\ell_4'] \quad . $$

\nind Now $(W[\ell_1,\ell_2],W[\ell_3,\ell_4])$ equals its dual pair if and
only if $\ell_1$ is parallel to $\ell_3$ or $\ell_2$ is parallel to $\ell_4$, 
and since self-dual pairs of 
nondisjoint wedges are mapped by $\hal$ to self-dual pairs of nondisjoint 
wedges, this is equivalent to $\ell_1'$ is parallel to $\ell_3'$ or $\ell_2'$
is parallel to $\ell_4'$, respectively. Thus, all pairs of images of the 
wedges $W[\ell_0,\ell]$, $\ell_0$ fixed but $\ell$ arbitrary, are self-dual. 
Therefore, one has 
$$\{ \hal(W[\ell_0,\ell]) \mid \ell \} \subset 
\{ W[\ell'_0,\ell] \mid \ell \} $$
or
$$\{ \hal(W[\ell_0,\ell]) \mid \ell \} \subset 
\{ W[\ell,\ell'_0] \mid \ell \} $$
for a suitable $\ell'_0$. Since the same statement holds for $\hal^{-1}$, the
equality of these sets follows. \par
     It remains to prove (6.1.3). To this end assume 
$W[\ell_3,\ell_4] = W[\ell_1,\ell_2]'$, {\it i.e.} $\ell_3$ is parallel to 
$\ell_2$ and $\ell_4$ is parallel to $\ell_1$. The collection 
$\{ \hal(W[\ell_1,\ell]) \mid \ell \}$, 
which contains the wedge $W[\ell_1',\ell_2']$, coincides with either 
$\{ W[\ell_1',\ell] \mid \ell \}$ or $\{ W[\ell,\ell_2'] \mid \ell \}$. But 
by relation (6.1.1) and a straightforward application of Lemma 6.1.1, each 
element of $\{ \hal(W[\ell_1,\ell]) \mid \ell \}$ is 
disjoint from $\hal(W[\ell_1,\ell_2]') = W[\ell_3',\ell_4']$. So $\ell_1'$ is 
a positive multiple of $\ell_4'$ in the first case (otherwise, one would have
$W[\ell_1',\ell_4'] \in \{ \hal(W[\ell_1,\ell]) \mid \ell \}$ and
$W[\ell_1',\ell_4'] \cap W[\ell_3',\ell_4'] = \emptyset$, in contradiction to 
Lemma 6.1.1); in the second case one concludes that $\ell_2'$ is a positive 
multiple of $\ell_3'$. On the other hand, by considering the collection 
$\{ \hal(W[\ell,\ell_2]) \mid \ell \}$ instead, one can see that $\ell_2'$ is 
a positive multiple of $\ell_3'$, resp. that $\ell_4'$ is a positive multiple 
of $\ell_1'$.
\hfill\qed\enddemo

     We next show that $\hal$ induces a map on the set of characteristic 
planes in the ambient space $\RR^4$, as in Section 4.1. We use notation 
established there and recall that we identify $\Ws$ with $\tilde{\Ws}_0$.

\proclaim{Corollary 6.1.3} Let $\hal : \Ws \mapsto \Ws$ be a bijection 
satisfying (6.1.1). Then $\hal$ induces a bijection of characteristic planes,
which we shall also denote by $\hal$, such that $\hal(H_0[\ell_1])$ and
$\hal(H_0[\ell_2])$ are the characteristic planes determined by 
$\hal(W[\ell_1,\ell_2])$ (with $H_0[\ell_1] \neq H_0[\ell_2]$). 
\endproclaim

\demo{Proof} According to Lemma 6.1.2, one has for fixed $\ell_0$ either
$$\{ \hal(W[\ell_0,\ell]) \mid \ell \} = 
\{ W[\ell'_0,\ell] \mid \ell \} \quad \text{or} \quad
\{ \hal(W[\ell_0,\ell]) \mid \ell \} =
\{ W[\ell,\ell'_0] \mid \ell \} \quad , $$
for a suitable $\ell'_0$. Set $\hal(H_0[\ell_0]) = H_0[\ell'_0]$; the claim
then follows easily.
\hfill\qed\enddemo

     By considering disjoint pairs of wedges instead of maximal pairs 
of wedges, one can follow the argument of Lemma 4.1.11 to prove the following.

\proclaim{Lemma 6.1.4} Let $\hal : \Ws \mapsto \Ws$ be a bijection 
satisfying (6.1.1). If $\ell_1,\ell_2,\ell_3,\ell_4$ are linearly dependent
future-directed lightlike vectors such that any two of them are linearly
independent, then

$$\underset{i=1}\to{\overset{4}\to{\cap}} \hal(H_0[\ell_i]) =
\underset{i \neq k}\to{\cap} \hal(H_0[\ell_i]) \quad \text{for} \quad
k = 1,2,3,4 \quad . $$

\endproclaim  

     This leads to an induced map on spacelike lines through the origin.

\proclaim{Lemma 6.1.5} Let $\hal : \Ws \mapsto \Ws$ be a bijection 
satisfying (6.1.1), and let $x \in \RR^4$ be spacelike. Then the intersection

$$\underset{\{\ell \mid x \in H_0[\ell]\}}\to{\cap} \hal(H_0[\ell])$$

\nind is one-dimensional and spacelike. Hence, $\hal$ induces a bijection 

$$\RR x \mapsto 
\underset{\{\ell \mid x \in H_0[\ell]\}}\to{\cap} \hal(H_0[\ell])$$

\nind on the set of spacelike one-dimensional subspaces of $\RR^4$. This map 
will again be denoted by $\hal$.
\endproclaim

\demo{Proof} Let $\ell_1,\ell_2,\ell_3,\ell_4$ be pairwise
linearly independent lightlike vectors such that $x \in H_0[\ell_i]$, for 
$i = 1,2,3,4$. Then this quadruple of vectors is linearly dependent and
consequently, by Lemma 6.1.4, one has 

$$\underset{\{\ell\mid x \in H_0[\ell]\}}\to{\cap} \hal(H_0[\ell]) = 
\underset{i=1}\to{\overset{3}\to{\cap}} \hal(H_0[\ell_i]) \quad . $$

\hfill\qed\enddemo 

     We shall need the following geometric result about wedges.

\proclaim{Lemma 6.1.6} Let $x \in \RR^4$ be spacelike and let
$\ell_k = (1,a_k,b_k,c_k)$, where $a_k,b_k,c_k \in \RR$ satisfy 
$a_k^2 + b_k^2 +c_k^2 = 1$, $k = 1,2$. Set $W_0 = W[\ell_1,\ell_2]$. Then 
$\RR x \cap W_0 \neq \emptyset$ if and only if $W_0 \cap W \neq \emptyset$, 
for all $W \in \Ws$ whose edge contains $\RR x$. For $x = (0,0,1,0)$, this 
is also equivalent to the statement that $b_1b_2 < 0$. Moreover, 
$(0,0,1,0) \in W_0$ implies $b_1 > 0$ and
$-(0,0,1,0) \in W_0$ implies $b_1 < 0$ (when $b_1b_2 < 0$).
\endproclaim

\demo{Proof} One may assume without loss of generality that $x = (0,0,1,0)$. 
Since $W_0$ is open, it is trivial that $\RR x \cap W_0 \neq \emptyset$ 
implies $W_0 \cap W \neq \emptyset$, for all $W \in \Ws$ whose edge contains 
$x$. \par
     For the converse, it will first be shown that 
$W_0 \cap W \neq \emptyset$, for all $W \in \Ws$ whose edge contains $x$,
implies $b_1b_2 < 0$. The case $b_1 = b_2 = 0$ is excluded, since it would 
imply that $W_0$ is invariant under the translations $\RR x$ and consequently
also $W_0 '$ would be so invariant. Hence, it would follow that 
$W_0 \cap W_0 ' \neq \emptyset$,
which is a contradiction. By considering $W_0 '$ instead of $W_0$ if
$b_1 = 0$, one may assume that $b_1 \neq 0$. By applying suitable
Lorentz transformations leaving $(0,0,1,0)$ invariant, one may further assume
that $a_1 = c_1 = 0$ and, after applying a suitable rotation, $a_2 > 0$
and $c_2 = 0$. Lemma 6.1.1 entails that if $b_1 > 0$, $a_2 > 0$ and 
$b_2 \geq 0$, or $b_1 < 0$, $a_2 > 0$ and $b_2 \leq 0$, then one 
has $W_0 \cap W[\ell_{1+},\ell_{1-}] = \emptyset$, where $\ell_{1\pm}$ are as
in the lemma. Since the wedge $W[\ell_{1+},\ell_{1-}]$ contains the line
$\RR x$ in its edge, this is a contradiction. Hence, there holds 
$b_1 b_2 < 0$. \par
     Proceeding further, it may still be assumed that $a_1 = c_1 = c_2 = 0$.
The remaining assertion of the lemma follows for $b_1 > 0$, $b_2 < 0$ (and
similarly for $b_1 < 0$, $b_2 > 0$), if one notices that the vector
$$\align
(0,0,(&1-b_2)^2,0) \\
&= -b_2(1-b_2)(1,0,1,0) - (1-b_2)(1,a_2,b_2,0) + a_2(a_2,1-b_2,a_2,0)  
\endalign $$
is an element of $W_0$.
\hfill\qed\enddemo

     This enables us to prove this final preparatory lemma.

\proclaim{Lemma 6.1.7}Let $\hal : \Ws \mapsto \Ws$ be a bijection 
satisfying (6.1.1), and let $x \in \RR^4$ be spacelike. If $x \in W_1 \cap W_2$
for $W_1,W_2 \in \Ws$, then
$$\emptyset \neq \hal(\RR x) \cap \hal(W_1) = \hal(\RR x) \cap \hal(W_2)
\quad . $$
\endproclaim

\demo{Proof} It has been seen that $\hal$ maps the set of wedges in
$\Ws$ whose edges contain the line $\RR x$ onto the set of wedges in
$\Ws$ whose edges contain the line $\hal(\RR x)$. Lemmas 6.1.5 and 6.1.6 
entail that both $\hal(\RR x) \cap \hal(W_1)$ and $\hal(\RR x) \cap \hal(W_2)$
are nonempty. There exist lightlike vectors $\ell_1,\ell_2,\ell_3,\ell_4$ such 
that $\hal(W_1) = W[\ell_1,\ell_2]$ and $\hal(W_2) = W[\ell_3,\ell_4]$. 
It is not possible for both the vectors $\ell_1,\ell_4$ and the vectors 
$\ell_2,\ell_3$ to be parallel, for otherwise one would have 
$W[\ell_3,\ell_4] = W[\ell_1,\ell_2]'$, which implies that 
$W_1 \cap W_2 = \emptyset$. Assuming that $\ell_1$ and $\ell_4$ are not 
parallel, Lemmas 6.1.2 and 6.1.6 then yield
$$\emptyset \neq \hal(\RR x) \cap W[\ell_1,\ell_2] = 
\hal(\RR x) \cap W[\ell_1,\ell_4] =
\hal(\RR x) \cap W[\ell_3,\ell_4] \quad , $$
and a similar argument can be applied if $\ell_2$ and $\ell_3$ are not
parallel. 
\hfill\qed\enddemo

     We have seen above that every point $x$ in the three-dimensional de Sitter
space can be identified with a spacelike $x \in \RR^4$ with $x \cdot x = -1$. 
By Lemma 6.1.7, this then determines the nonempty intersection
$$\hal(\RR x) \cap \hal(W_0) = 
\underset{x \in W}\to{\underset{W \in \Ws}\to{\cap}} (\hal(\RR x) 
\cap \hal(W)) = 
\hal(\RR x) \cap ( \underset{x \in W}\to{\underset{W \in
    \Ws}\to{\cap}} \hal(W)) \quad , $$

\nind where $W_0 \in \Ws$ contains $x$. Since there exists a point $y \neq 0$ 
in this intersection, and $\hal(W_0) \in \Ws$, while $\hal(\RR x)$ is a 
spacelike line, the intersection $\hal(\RR x) \cap \hal(W_0)$ must contain the 
ray $\RR_+ y$. Hence, there exists a unique point, call it $\delta(x)$, such 
that $\delta(x) \in \hal(\RR x) \cap \hal(W_0)$ and 
$\delta(x) \cdot \delta(x) = -1$. It thus represents a point in 
three-dimensional de Sitter space. We have therefore proven the following
result.

\proclaim{Proposition 6.1.8} Let $\hal : \Ws \mapsto \Ws$ be a bijection 
satisfying (6.1.1). Then there exists a bijection $\delta : dS^3 \mapsto dS^3$
such that 

$$\hal(W) = \{ \delta(x) \mid x \in W \} \quad , $$

\nind for all $W \in \Ws$.
\endproclaim

     The following Alexandrov-like theorem has been established for the case
of de Sitter space by Lester \cite{48}:

\proclaim{Lemma 6.1.9} If $\phi : dS^3 \mapsto dS^3$
is a bijection such that lightlike separated points are mapped to lightlike
separated points, then there exists a Lorentz transformation $\Lambda$
of the ambient Minkowski space $\RR^4$ such that $\phi(x) = \Lambda x$, for 
all $x \in dS^3$.
\endproclaim 

We may therefore proceed to obtain the following extension of 
Lester's theorem. Details may be found in Section 1.5.2 of \cite{28}.

\proclaim{Theorem 6.1.10} Let $\hal : \Ws \mapsto \Ws$ be a bijection 
satisfying (6.1.1), and let \newline
$\delta : dS^3 \mapsto dS^3$ be the associated 
bijection. Then there exists a Lorentz transformation $\Lambda$ of the 
ambient Minkowski space $\RR^4$ such that $\delta(x) = \Lambda x$, for all 
$x \in dS^3$, and $\hal(W) = \Lambda W$, for all $W \in \Ws$.
\endproclaim \newpage

\bigpagebreak
   {\bf 6.2. Geometric Modular Action in de Sitter Space and the de Sitter 
Group}
\bigpagebreak

     We now turn to the discussion of nets on de Sitter space satisfying the
Condition of Geometric Modular Action given in Chapter III with the choices 
$\Ms = dS^3$ and the collection of wedges $\Ws$ specified in the previous 
section. In order to simplify the discussion, we work with the following
somewhat more restrictive version of the CGMA.

\proclaim{Strong CGMA} A theory complies with the strong form of the CGMA
if the net $\wrnet$ satisfies \par
   (i) $W \mapsto \Rs(W)$ is an order-preserving bijection, \par
   (ii) $\Omega$ is cyclic and separating for $\Rs(W_1) \cap \Rs(W_2)$ if and 
only if $W_1 \cap W_2 \neq \emptyset$, for $W_1,W_2 \in \Ws$, \par
   (iii) for any $W_0,W_1,W_2 \in \Ws$ with $W_1 \cap W_2 \neq \emptyset$, 
there holds
$$\Rs(W_1) \cap \Rs(W_2) \subset \Rs(W_0) \quad \text{if and only if} \quad
W_1 \cap W_2 \subset W_0 \quad , \tag{6.2.1} $$
and \par
   (iv) for each $W \in \Ws$, the adjoint action of $J_W$ leaves the set
$\wrnet$ invariant.
\endproclaim

     The first and fourth conditions are the same as in Chapter III and entail
the existence of an involution $\hal_W : \Ws \mapsto \Ws$ for each $W \in \Ws$ 
satisfying (3.1) and (3.2). The second condition is a strengthened version of 
the previous conditions (ii) and (iii). It directly implies relation (3.3).\par
     The third condition is an additional natural assumption which has no
counterpart in the original CGMA. We note that the restriction to intersecting 
pairs of wedges is motivated by a curious fact pointed out to us by E.H. 
Wichmann. Already for the standard net of von Neumann algebras of the free 
field, there exists a counterexample to relation (6.2.1) if $W_1,W_2$ are 
unrestricted wedges \cite{67}. However, it has been shown that in a net 
satisfying the usual axioms as well as the condition of additivity of wedge 
algebras, the relation (6.2.1) holds for pairs satisfying 
$W'_1 \cap W'_2 \neq \emptyset$ \cite{63}\cite{64}. But for
wedges $W_1,W_2 \in \Ws_0$, it is easy to see that
$W'_1 \cap W'_2 \neq \emptyset$ if and only if $W_1 \cap W_2 \neq \emptyset$.

\proclaim{Lemma 6.2.1} Let the strong CGMA with the choices $\Ms = dS^3$ and 
the set of wedges $\Ws$ in $dS^3$ hold. Then for each $W \in \Ws$ the 
associated involution $\hal_W : \Ws \mapsto \Ws$ satisfies (6.1.1).
\endproclaim

\demo{Proof} As already pointed out, the strong CGMA entails relation (3.3).
Therefore, in order to prove (6.1.1), it suffices to show that 
$W_1 \cap W_2 = W_3 \cap W_4 \neq \emptyset$ implies 
$\hal_W(W_1) \cap \hal_W(W_2) = \hal_W(W_3) \cap \hal_W(W_4)$. But
$W_1 \cap W_2 = W_3 \cap W_4$ implies
$\Rs(W_1) \cap \Rs(W_2) \subset \Rs(W_3)$, which itself entails
$\Rs(\hal_W(W_1)) \cap \Rs(\hal_W(W_2)) \subset \Rs(\hal_W(W_3))$. In the light
of (3.3), one concludes that also 
$\hal_W(W_1) \cap \hal_W(W_2) \neq \emptyset$, so by (6.2.1) one finds 
$\hal_W(W_1) \cap \hal_W(W_2) \subset \hal_W(W_3)$. By proving three similar 
inclusions, it follows that 
$\hal_W(W_1) \cap \hal_W(W_2) = \hal_W(W_3) \cap \hal_W(W_4)$.
\hfill\qed\enddemo

     Given the hypotheses of Lemma 6.2.1, we conclude from Theorem 6.1.10
that $\Ts$ is isomorphic to a subgroup $\Gs$ of the de Sitter group
$\Ls$. Since one has for $W_1, W_2 \in \Ws$ the fact that the inclusion 
$W_1 \subset W_2$ entails the equality $W_1 = W_2$, the index set $\Ws$ 
considered in this chapter consists exclusively of atoms,
{\it i.e.} we cannot conclude from the argument of Chapter II that the algebras
$\Rs(W)$ are nonabelian. Indeed, we shall see that this is quite possible.
Note that none of the arguments in Section 4.2 relied
upon the nonabelianness of the algebras $\Rs(W)$. Hence, with the additional
assumption that the adjoint action of $\Js$ upon the net $\wrnet$ is 
transitive, we conclude that the entire identity component of $\Lid$ of
$\Ls$ is contained in $\Gs$ (Prop. 4.2.2).

\proclaim{Lemma 6.2.2} Let the strong CGMA with the choices $\Ms = dS^3$ and 
the set of wedges $\Ws$ in $dS^3$ hold. Moreover, let the adjoint action of 
$\Js$ upon the set $\wrnet$ be transitive. Then either the algebra $\Rs(W)$ is 
nonabelian for every $W \in \Ws$ and the geometric action of the modular 
involutions is precisely that found in Prop. 4.2.10, or all these wedge 
algebras are abelian and the geometric action of the modular involutions is 
that found in Prop. 4.2.10 times the reflection about the origin, $\theta$.
\endproclaim

\demo{Proof} Because of the transitive action of $\Gs$ upon $\Ws$, it suffices 
to make the argument for the standard wedge $W_R$ and the corresponding 
involution $g_{W_R} \in \Gs$. Since $\Lid \subset \Gs$, one sees from Lemma 2.1
that $g_{W_R}$ commutes with the elements of the subgroup $\ILid(W_R)$ of 
$\Lid$ leaving $W_R$ invariant. But $W_R$ and $W_R '$ are the only wedges 
which are stable under the action of $\ILid(W_R)$, so it follows that either 
$g_{W_R} W_R = W' _R$ or $g_{W_R} W_R = W_R$. \par
     In both cases one can proceed in a manner similar to the proof of Prop.
4.2.10. Making use of the fact that $g_{W_R}$ is an involution which commutes 
with $\ILid(W_R)$, it is not hard to show that $g_{W_R}$ has the block form

$$ g_{W_R} = \left( \matrix X & 0 \\ 0 & Y \endmatrix \right) \quad , $$

\nind where $X,Y = \pm 1$. In the first case, $g_{W_R} W_R = W' _R$,
one clearly has $X = -1$. If also $Y = -1$, then $g_{W_R}$ commutes with all
elements of $\Lid \subset \Gs$, which would be in conflict with the transitive
action of $\Js$ upon $\wrnet$. Hence $Y = 1$ and $g_{W_R}$ has the form 
given in Prop. 4.2.10. The second case, $g_{W_R} W_R = W_R$, can be
treated in the same manner. \par
     Finally, the relation 
$\Rs(W_R)' = J_{W_R} \Rs(W_R) J_{W_R} = \Rs(g_{W_R}W_R)$ shows that if
$g_{W_R} W_R = W_R$ then $\Rs(W_R)$ is abelian. Conversely, if
$\Rs(W_R)$ is abelian  (and hence maximally abelian by the cyclicity of 
$\Omega$), then one has $\Rs(W_R) = \Rs(W_R)'$, and the above relation 
together with the first part of the CGMA implies $g_{W_R}W_R = W_R$.
\hfill\qed\enddemo

     It is now clear how to modify the arguments of Section 4.2 to obtain the
following result. 

\proclaim{Corollary 6.2.3} Let the conditions of Lemma 6.2.2 be satisfied.
If $\Rs(W)$ is nonabelian, for some $W \in \Ws$, then $\Gs$ coincides with
$\Ls_+$. On the other hand, if $\Rs(W)$ is abelian, for some $W \in \Ws$, then 
$\Gs$ coincides with $\Lid$.
\endproclaim

     The assumptions of Lemma 6.2.2 also directly yield an obvious 
counterpart to Proposition 4.3.1. The net continuity condition introduced
in Section 4.3 and the arguments presented there again entail that there 
exists a strongly continuous projective representation of $\Ls_+$ in the 
nonabelian case (which is of primary interest here). Moreover, the reasoning 
in Section 4.3 implies that this gives a true representation 
$U(\Ls_+)$ of the proper de Sitter group. We summarize in the following 
theorem.

\proclaim{Theorem 6.2.4} Let the strong CGMA with the choices $\Ms = dS^3$ and 
wedges $\Ws$ in $dS^3$ hold, and let the adjoint action of $\Js$ upon the set 
$\wrnet$ be transitive. If $\Rs(W)$ is nonabelian, for some $W \in \Ws$, then 
there exists a strongly continuous unitary representation of the proper de 
Sitter group $\Ls_+$ which acts geometrically correctly upon the net $\wrnet$. 
Moreover, the net satisfies Haag duality and is local. 
\endproclaim

     In light of the fact that the restricted Lorentz group $\Lid$ is also
isomorphic to the group of motions of Lobaschewskian space, which can be 
modelled on a surface of transitivity of $\Lid$ in $\RR^4$ (see, {\it e.g.},
\cite{34}), it is likely that the preceding arguments can be employed 
to handle that space-time, as well. \par
     To demonstrate that this theorem is not vacuous, we recall an example due 
to Fredenhagen \cite{33}. Consider once again the net from Section 5.3 
associated with the free scalar field on $\RR^4$. We define for each region 
$W \in \Ws$ a corresponding algebra $\Rs(W) \equiv \As(\tilde{W})$, where 
$\tilde{W}$ is the wedge fixed by $W$ in the ambient space $\RR^4$ and 
$\As(\tilde{W})$ the corresponding algebra generated by the free field.
The results of Bisognano and Wichmann \cite{9} and Thomas and Wichmann
\cite{64} entail that this net is covariant under the de Sitter group, and the 
assumptions in Theorem 6.2.4 are satisfied by this net in the vacuum state. 
\par
     Moreover, recent results in \cite{17} and \cite{19} concerning
quantum field theory on de Sitter space-time are fully consistent with our 
findings, even though the starting point is quite different. These authors
assume the existence of a preferred (vacuum-like) state vector $\Omega$
which is invariant under the de Sitter group $\Lid$ and satisfies a stability
condition which can be expressed in terms of certain analyticity properties
of the corresponding correlation functions. With this input they are then able 
to prove a Bisognano-Wichmann type theorem. In fact, they establish the 
Reeh-Schlieder property of $\Omega$ for wedge algebras (so the modular objects 
exist in their setting), and they also show that the modular conjugations
associated with these algebras and $\Omega$ induce the geometric action upon
the net found in the analysis presented here. Moreover, the modular groups
comply with our proposal for a modular stability condition. These facts 
support our view of the relevance of our selection criterion for vacuum-like 
states in theories on curved space-times. \par

\bigpagebreak

\heading VII. Summary and Further Remarks  \endheading

     As this paper is lengthy and involves many steps, it is perhaps not
amiss to provide a final summary here. First of all, we showed that our
Condition of Geometric Modular Action, CGMA, in the abstract form of the 
Standing Assumptions, yielded special Coxeter groups $\Ts$ of automorphisms on 
the index set $(I,\leq)$ of the net $\inet$ and provided them with projective 
representations having coefficients in an abelian group $\Zs$ of internal net 
symmetries. Some general properties of these groups, following from the 
modular theory, and a discussion of the finite case were given. \par
     In Chapter III it was explained how, starting from a smooth
manifold $\Ms$ and with a target space-time $(\Ms,g)$ in mind, one would
go about identifying the index set $\Ws$ before testing states on the net
$\wrnet$ for the CGMA. The resultant program using the CGMA for the
determination of much of the geometrical structure of the space-time was
then described. \par
     This program was then exemplified in application to the four-dimensional
Minkowski space as target space. This involved a series of results of
quite distinct natures. To begin, we showed that bijective inclusion-preserving
mappings on the set of wedges which satisfy one additional condition are 
implemented by elements of the extended Poincar\'e group, thus extending
the Alexandrov-type theorems for Minkowski space. Then, it was shown that
subgroups of the Poincar\'e group which act transitively upon the set of
wedges must contain the identity component $\Pid$ of the Poincar\'e group.
These results enabled us to show that the CGMA, applied to nets indexed by
wedges in $\RR^4$ and supplemented by the transitivity condition, implied that
the induced isometry group $\Gs$ is equal to the proper Poincar\'e group,
and that the implementers for the generating involutions have exactly the
geometric action found by Bisognano and Wichmann in their setting. \par
     This explicit knowledge of the geometric nature of the adjoint action
of the modular involutions $J_W$ upon the net, along with the additional
structure accompanying the modular theory, was used to construct a
continuous projective representation of $\Pid$, under the assumption of the
net continuity condition. Using Moore's Borel measurable cohomology theory, we 
showed that this projective representation of $\Pid$ lifted to a true 
representation of its universal covering group. The explicit geometric 
properties of the modular involutions already alluded to were then employed to 
prove that this representation of the covering group restricted to a strongly 
continuous unitary representation of $\Pid$ and actually coincided with the
constructed projective representation. In other words, the projective 
representation constructed in Section 4.3 is actually a true representation.
\par
     In Section 5.1, we showed that if the modular unitaries are all contained
in the group $\Js$ generated by the modular involutions, {\it i.e.}
if the modular stability condition holds, then the spectrum
condition must hold. This is a purely algebraic stability condition which
can be sensibly stated on any space-time. We next investigated the 
geometric action of the modular unitaries in detail. It was proven that,
if the Condition of Geometric Action for modular groups is satisfied, then 
both modular covariance and the modular stability condition, 
$\Ks \subset \Js$, hold and, if the net is local, the group $\Ks$ yields a 
strongly continuous unitary representation of $\Pid$ satisfying the spectrum 
condition. Moreover, under the same assumptions, the CGMA holds if and only if 
the net is local. Furthermore, if the CGMA and the modular stability 
condition are satisfied, then again modular covariance follows. In Section 5.3 
a number of examples were given which make clear that modular covariance is, 
in fact, strictly stronger than the CGMA. \par
     Finally, in Chapter VI we discussed the case of de Sitter space. In 
spite of its different geometric structure, results similar to the case of 
Minkowski space were recovered.  \par
     Among other space-times, we expect our approach to function with little 
change in such examples as the (static) Robertson-Walker space-times. It is
an interesting problem whether also in these cases the maps induced upon the 
index sets of the corresponding nets of algebras are implemented by point
transformations. In this regard, it is relevant to note that Alexandrov-type 
theorems are available for many of the classical Lorentzian space-times
(see \cite{8}). But even if not every element of the group $\Ts$ of 
transformations is implemented by a point transformation on the space-time 
(and we have already presented such an example in Section 4.1), we still 
anticipate that the CGMA could usefully select physically interesting states. 
Whatever the group of transformations which results, we would propose it as 
the symmetry group of the theory. \par
     We complete our comments in this final chapter by returning briefly to
the conceptually interesting question of whether one can derive the space-time
itself from our initial algebraic data. In this paper we began with a
particular smooth manifold $\Ms$ and saw how the CGMA, for a certain 
choice of index set which was determined by the target space-time $(\Ms,g)$, 
enabled us to derive a metric-characterizing isometry subgroup. But is it 
possible to do without these initial data? \par
     We shall sketch here our program for meeting this question. We have
shown that the abstract version of the CGMA in the form of the Standing 
Assumptions leads to a certain Coxeter group $\Ts$ of automorphisms on the
index set $(I,\leq)$ of the net $\irnet$. There exists in the mathematical
literature a branch of geometry known as {\it absolute geometry}, whose
point of departure is precisely an abstract group $\Ts$ generated by 
involutions and whose aim is to investigate which algebraic relations in the 
group $\Ts$ entail the existence of a space-time $(\Ms,g)$ such that the group 
$\Ts$ can be realized as a metric-characterizing subgroup of the isometry group
of $(\Ms,g)$. This has been carried out for all planar geometries 
\cite{5}\cite{75} and for three-dimensional Euclidean space \cite{1}. \par
     So a first step in an attempt to characterize Minkowski space
entirely in terms of the data $(\irnet,\Omega)$ would be to find the 
algebraic relations in the group $\Ts$ which would enable one to derive
in this manner four-dimensional Minkowski space. This has been accomplished 
in one form \cite{44}, but the particular geometric significance 
of the initial algebraic data in our setting entails that a different set
of algebraic axioms be determined \cite{66}. The second
step would be to determine which additional structure on $I$, or 
equivalently, which relations among the algebras in the net $\irnet$, imply
{\it via} modular theory the requisite relations among the generating
involutions $J_i$ (equivalently, $\tau_i$) found in the first step.
In application to Minkowski space, this would give an intrinsic 
characterization of ``wedge algebras'' (equivalently ``wedges'').  \par
    The results in this paper demonstrate that the CGMA is sufficiently strong 
to select physically interesting states and to actually determine 
metric-characterizing isometry groups in the examples of Minkowski and 
de Sitter space-times. We hope that the suggestive results and interesting
perspectives of the present analysis will draw attention to the various 
mathematical problems opened up by our program. \par

\bigpagebreak

\heading Appendix: Cohomology and the Poincar\'e Group
\endheading

     In this appendix we shall prove the technical cohomological result
used in the main text to the effect that the continuous projective 
representation $V(\Pid)$ constructed in Section 4.3 can be lifted to a true 
representation of the covering group. We include this appendix since we
have not found in the literature the results in the form we need. Assume that 
$G \ni g \mapsto V(g) \in \Us(\Hs)$ is a continuous projective representation 
of a semisimple Lie group by unitary operators on a separable Hilbert space
$\Hs$, which has coefficients in a closed\footnote{It is no loss of generality
to take $\Zs$ closed. Though the subgroup $\Zs \subset \Us(\Hs)$ in the 
main text is not {\it a priori} closed, closing it in the weak operator
topology still yields a trivial $\Pid$-module, as used in this appendix. 
However, the restriction that $\Zs$ be closed offers a technical problem in
the main text which is dealt with there.}
subgroup $\Zs \subset \Us(\Hs)$ left 
pointwise fixed by the adjoint action of the elements of 
$\{ V(g) \mid g \in G \}$. Note that the group $\Us(\Hs)$ of unitary operators 
acting on the separable Hilbert space $\Hs$ is, when provided with the strong 
(or weak) operator topology, a complete, metrizable, second countable 
topological group (cf. p. 33 in \cite{27} and references cited there). It 
therefore follows that also $\Zs$ is a complete, metrizable, second countable 
topological group; hence, it is a polonais (polish) group. In particular, 
$\Zs$ is a trivial $G$-module. The first main theorem we want to prove is the 
following. (A related theorem with different assumptions and proof may 
be found in \cite{22}.) \par

\proclaim{Theorem A.1} Let $G$, $V(G)$ and $\Zs$ be as described 
above.\footnote{In fact, the arguments presented below are valid for a larger
class of groups $G$ than semisimple Lie groups, but we shall not tax the
reader's patience here with this generalization.}
Then there exists a strongly continuous unitary representation of the covering 
group $E$ of the group $G$. 
\endproclaim

     The proof of this theorem will proceed in several steps, which we present
in separate lemmata for the sake of clarity. For the reader's convenience, we 
shall present some background information about the two-dimensional cohomology 
of groups, which can be found in textbooks on the subject (see, {\it e.g.} 
\cite{20}). Since we are interested in the continuity of the representations, 
we shall need to work in the category of topological groups but find ourselves 
obliged to use the Borel cohomology on locally compact groups initiated by 
Mackey \cite{49} and fully defined and extended by Moore \cite{51}-\cite{55}, 
since the computational situation for continuous cohomologies seems to be 
exceedingly complicated. Fortunately, it can be shown that this will be 
sufficient for our purposes. For an overview of the various cohomologies for 
topological groups, see the review by Stasheff \cite{59}. \par
     Let $G$ be a group and $G' \equiv [G,G]$ denote its derived subgroup, 
{\it i.e.} the group generated by the set $\{ghg^{-1}h^{-1} \mid g,h \in G \}$ 
of commutators in $G$. If $G' = G$, the group $G$ is said to be perfect, and 
any connected semisimple Lie group has this property.\footnote{Indeed,
Moore \cite{53} suggests the property 
$G \! = \! G'$ as the algebraic analogue of 
connectedness.} In particular, the group of interest to us in this paper, 
the proper orthochronous Poincar\'e group $\Ps_+^{\uparrow}$, is a perfect 
group. \par
     Let $G$ be a group and $A$ be an abelian group. A central extension of 
$G$ by $A$ is a triple $(\tilde{G},\phi,\iota)$ with
$\tilde{G}$ a group, $\iota$ an injective homomorphism from $A$ to 
$\tilde{G}$ satisfying $\iota(A) \subset \text{center}(\tilde{G})$ and $\phi$ 
a homomorphism from $\tilde{G}$ onto $G$ satisfying 
$\text{kernel}(\phi) = \iota(A)$. In other words, the sequence

$$\{ 1 \} \longrightarrow A \overset{\iota}\to{\longrightarrow} \tilde{G} 
\overset{\phi}\to{\longrightarrow} G \longrightarrow \{ 1 \} \tag{A.1}$$

\nind is exact, with $\{ 1 \}$ denoting the trivial group. Such a central 
extension is said to be equivalent to the central extension

$$\{ 1 \} \longrightarrow A \overset{\iota'}\to{\longrightarrow} \tilde{G}' 
\overset{\phi'}\to{\longrightarrow} G \longrightarrow \{ 1 \}$$

\nind if there exists an isomorphism $\rho: \tilde{G} \mapsto \tilde{G}'$ such 
that the diagram

$$\CD
\{ 1 \}   @>>>  A   @>\iota>>  \tilde{G}  @>\phi>>  G  @>>> \{ 1 \}  \\
@.        @VV\text{id.}V   @VV\rho V   @VV\text{id.}V  @.  \\
\{ 1 \}   @>>>  A   @>\iota ' >>  \tilde{G}'  @>\phi ' >>  G  @>>> \{ 1 \}
\endCD$$

\nind is commutative. The direct product $G \times A$ is an example of a 
central extension with the inclusion $a \mapsto (1,a)$ and the projection 
$(g,a) \mapsto g$, where $g \in G$ and $a \in A$. If the groups involved are 
topological groups and one wishes to keep track of continuity, as we do in 
this paper, then in the above the homomorphism $\iota$ is required also to be 
a homeomorphism onto a closed subgroup of $\tilde{G}$, $\phi$ must also be 
continuous and open (so that $\tilde{G}/\iota(A) \simeq 
\tilde{G}/\text{kernel}(\phi) \simeq G$), and $\rho$ must be an isomorphism in 
the category of topological groups. \par
    If $E$ is a topological group such that $[E,E]$ is dense in $E$ and 
$p: E \mapsto G$ is a surjective continuous homomorphism, following Moore, we 
shall say that the pair $(E,p)$ is a cover of $G$ if the kernel of $p$ is 
contained in the center of $E$. Then $E$ is an extension of $G$ by the trivial 
$G$-module $\text{kernel}(p)$ (and, of course, $[G,G]$ is necessarily dense 
in $G$). Moore showed that if $G$ is locally compact and separable, then $G$ 
has at most one simply connected covering group (in this sense) up to 
isomorphism of topological group extensions (see Lemma 2.2 in \cite{53}).
Moreover, if $G$ is perfect, then there does exist such a (unique) simply 
connected covering group (called the universal covering group) $E$, which 
turns out to be perfect and a Lie group itself (Theorem 2.2 in \cite{53}
and Theorem 10 in \cite{55}).
What will be important for our arguments below is that if $G$ is a 
semisimple Lie group, then this universal covering group {\it coincides} with 
the standard, topologically defined, universal covering group (cf. p. 49 in
\cite{55}). A central extension $(U,\nu,\jmath)$ is called universal if 
for every central extension $(\tilde{G},\phi,\iota)$ of $G$ by $A$, there 
exists a (continuous, open) homomorphism $h$ from $U$ to $\tilde{G}$ such that 
$\phi \circ h = \nu$. If such a universal central extension exists, then it 
is unique up to isomorphism over $G$. And it is known (cf. Theorem 5.7 in 
\cite{50}) that a group $G$ admits a universal central extension if and 
only if $G$ is perfect. {}From the remarks above, it is now clear that for 
semisimple Lie groups, the (standard) universal covering group coincides with 
the universal covering group in the sense of Moore, which coincides with the 
universal central extension. \par
     Given a central extension (A.1) of $G$ by $A$,\footnote{For the purposes 
of his cohomology theory, in \cite{51}\cite{52} Moore took $A$ to be an
abelian, locally compact and second countable topological group. However, in
\cite{54} he extended his results to include second countable, Hausdorff 
polonais groups $A$. We may, therefore, take $A = \Zs$ below.} assume that
$\sigma : G \mapsto \tilde{G}$ is a section with $\sigma(1) = 1$, in other 
words it is a (Borel measurable) set map such that $\phi(\sigma(g)) = g$ for 
all $g \in G$. The function 
$\gamma(\sigma) = \gamma : G \times G \mapsto \tilde{G}$ defined 
by $\gamma(g,h) \equiv \sigma(g)\sigma(h)\sigma(gh)^{-1}$ is a measure of the 
amount $\sigma$ diverges from a homomorphism, and, of course, the 
associativity in $\tilde{G}$ implies that $\gamma$ is a 2-cocycle. Note that 
because $\phi(\gamma(g,h)) = 1$, $\gamma$ actually takes values in the 
subgroup $A$. Let $Z^2(G,A)$ denote the set of all such (Borel measurable)
2-cocycles (which turns out to be an abelian group). Let $B^2(G,A)$ denote the 
$A$-valued coboundaries, {\it i.e.} the subgroup of $Z^2(G,A)$ consisting of 
functions $\gamma : G \times G \mapsto A$ for which there exists a (Borel 
measurable) $\beta : G \mapsto A$ such that 
$\gamma(g,h) = \beta(g)\beta(h)\beta(gh)^{-1}$ for all $g,h \in G$. The 
quotient group $Z^2(G,A)/B^2(G,A)$ is precisely the second cohomology group 
$H^2(G,A)$. One therefore sees that if $H^2(G,A) = \{ 1 \}$, then every 
($A$-valued) projective representation $\sigma$ of $G$ in $\tilde{G}$ 
determines a 2-cocycle $\gamma$ which is actually a 2-coboundary. Thus, by 
defining $\tilde{\sigma} \equiv \beta(g)^{-1}\sigma(g)$, a straightforward 
calculation shows that $\tilde{\sigma}: G \mapsto \tilde{G}$ is a (Borel 
measurable) homomorphism \footnote{The passage from Borel measurable to 
continuous will be addressed separately below.}, {\it i.e.} a representation, 
as desired. And if $H^2(G,A)$ is nontrivial, then it is possible to start with 
a section $\sigma$ for which there exists no $\beta$ for which 
$\beta^{-1}\sigma$ yields a homomorphism. In this case, the question would 
have to be settled for a given section individually.  \par
     In the setting of relevance to this paper, $E \ni e \mapsto V(p(e))$ is a 
continuous projective representation of $E$ with coefficients in $\Zs$. We 
prove the relevant cohomological result for the covering group $E$. \par

\proclaim{Lemma A.2} Let $G$ be a connected semisimple Lie group and $E$ be its
universal covering group. Then the second cohomology group $H^2(E,\Zs)$ is 
trivial. 
\endproclaim

\demo{Proof} In the proof of Theorem 9 in \cite{55}, it is shown that
for a perfect, almost connected group $G$, the second cohomology group
$H^2(E,S^1)$ in Moore's Borel measurable cohomology theory is trivial, where
$S^1$ is the circle group. This result is thus applicable to the situation
described by the hypothesis. Moreover, since $G$, and hence $E$, is perfect,
it follows that also the first cohomology group $H^1(E,S^1)$ is trivial
(see p. 48 in \cite{55}). Thus Prop. 4 in \cite{55} may be 
applied, yielding $H^2(E,A)$ is trivial for {\it any} unitary trivial
$G$-module $A$, and, in particular, for $A = \Zs$.
\hfill\qed\enddemo

     Hence, there exists a function $Z : E \mapsto \Zs$ such that
$U(e) \equiv Z(e)V(p(e))$, $e \in E$, is a true representation of $E$. One 
does indeed obtain a (unitary) representation of the group $E$. But in 
Moore's cohomology, the cochains are only Borel measurable on the group; 
in other words, although the original section $\sigma$ is continuous, the 
function $\beta$ may only be Borel measurable, so that 
$\tilde{\sigma} \equiv \beta^{-1}\sigma$, {\it i.e.} $U$, may be only Borel 
measurable. However, the following result, attributed to Mackey in \cite{78}, 
closes this gap. \par

\proclaim{Lemma A.3} If $H_1$ is a locally compact second countable group,
$H_2$ is any second countable topological group, and $h:H_1 \mapsto H_2$ is a 
Borel measurable homomorphism, then $h$ is continuous.
\endproclaim

\demo{Proof} This is Theorem B.3 in \cite{78}.
\hfill\qed\enddemo

\nind Hence, by taking $H_1 = E$ and $H_2 = \Us(\Hs)$, it follows that
$E \ni e \mapsto U(e)$ is, in fact, a strongly continuous
unitary representation of $E$, completing the proof of Theorem A.1. \par
     In the more structured setting of the main text of this paper, $G$
is the Poincar\'e group $\Pid$. There we get by an application of the 
preceding results:

\proclaim{Corollary A.4} Let $V(\cdot)$ be the continuous unitary projective
representation of $\Ps_+^{\uparrow}$ with values in ${\Cal J}$ which
has been constructed in Section IV.3, let  $\overline{\Cal J}$ be the 
closure of ${\Cal J}$ in the weak operator topology and let 
$\overline{\Zs}$ be the center of $\overline{\Cal J}$. There exists a strongly 
continuous unitary representation $U(\cdot)$ of the covering group 
$ISL(2,\CC)$ of the Poincar\'e group $\Ps_+^{\uparrow}$ with values in 
$\overline{\Cal J}$ and a mapping  $Z: \, ISL(2,\CC) \mapsto \overline{\Zs}$ 
with $U(A) = Z(A)V(\mu(A)), A \in ISL(2,\CC)$. Here, 
$\mu: ISL(2,\CC) \mapsto \Pid$ is the canonical covering homomorphism whose 
kernel is a subgroup of order 2, the center of $ISL(2,\CC)$. 
\endproclaim

\bigpagebreak

{\eightpoint
\nind {\bf Acknowledgments}: As this paper has been simmering for many years,
the authors have reason to thank many persons and institutions. DB thanks the 
Institute for Fundamental Theory at the University of Florida for an 
invitation in 1993, where this work was begun. He also acknowledges financial 
support from the Deutsche Forschungsgemeinschaft. SJS wishes to 
thank the Second Institute for Theoretical Physics at the University of 
Hamburg and DESY for invitations in the summers of 1993-95, as well as the 
University of Florida for travel support, which made the continuation of 
this collaboration possible. Part of this work was completed while SJS was the 
Gauss Professor at the University of G\"ottingen in 1994. For that opportunity 
SJS wishes to thank Prof. H.-J. Borchers and the Akademie der Wissenschaften 
zu G\"ottingen. Further progress was made while DB was a guest of the 
Department of Physics of the University of California at Berkeley in 1997, and
he gratefully acknowledges the hospitality of E.H. Wichmann as well as a travel
grant from the Alexander von Humboldt Foundation. Finally, DB and SJS express 
their gratitude to Prof. J. Yngvason and the Erwin Schr\"odinger International 
Institute for Mathematical Physics for providing the circumstances permitting 
the completion of this paper. All authors are grateful for useful comments by 
Profs. H.-J. Borchers, P. Ehrlich, C. Stark, H. V\"olklein and E.H. Wichmann,
which helped bring this long-standing project to a successful end.}         

\bigpagebreak

\heading References  \endheading

\roster
\eightpoint
\item{}J. Ahrens, Begr\"undung der absoluten Geometrie des Raumes aus dem
Spiegelungsbegriff, {\sl Math. Zeitschr., \bf 71}, 154-185 (1959).
\item{}A.D. Alexandrov, On Lorentz transformations, {\sl Uspehi Mat. Nauk.,
\bf 5}, 187 (1950).
\item{}A.D. Alexandrov, Mappings of spaces with families of cones and 
space-time transformations, {\sl Annali di Mat. Pura Appl., \bf 103}, 229-257 
(1975).
\item{}H. Araki, Symmetries in theory of local observables and the
choice of the net of local algebras, {\it Rev. Math. Phys., \bf Special 
Issue}, 1-14 (1992).
\item{}F. Bachmann, {\it Aufbau der Geometrie aus dem Spiegelungsbegriff},
second edition, Berlin, New York, Springer-Verlag, 1973.
\item{}U. Bannier, Intrinsic algebraic characterization of space-time 
structure, {\sl Int. J. Theor. Phys., \bf 33}, 1797-1809 (1994). 
\item{}U. Bannier, R. Haag and K. Fredenhagen, Structural definition of
space-time in quantum field theory, unpublished preprint, 1989.
\item{}W. Benz, {\it Real Geometries}, Mannheim, Leipzig, Vienna and Z\"urich,
B.I. Wissenschaftsverlag, 1994.
\item{}J. Bisognano and E.H. Wichmann, On the duality condition for a 
hermitian scalar field, {\sl J. Math. Phys., \bf 16}, 985-1007 (1975).
\item{}J. Bisognano and E.H. Wichmann, On the duality condition for quantum 
fields, {\sl J. Math. Phys., \bf 17}, 303-321 (1976).
\item{}H.-J. Borchers and G.C. Hegerfeldt, The structure of space-time 
transformations, {\sl Commun. Math. Phys., \bf 28}, 259-266 (1972).
\item{}H.-J. Borchers, The CPT-theorem in two-dimensional theories of
local observables, {\sl Commun. Math. Phys.  \bf 143}, 315-332 (1992).
\item{}H.-J. Borchers, On modular inclusion and spectrum condition, {\sl
Lett. Math. Phys., \bf 27}, 311-324 (1993).
\item{}H.-J. Borchers, On the use of modular groups in quantum field theory, 
{\sl Ann. Inst. Henri Poincar\'e, \bf 63}, 331-382 (1995).
\item{}H.-J. Borchers, Half-sided modular inclusion and the construction of
the Poincar\'e group, {\sl Commun. Math. Phys., \bf 179}, 703-723 (1996).
\item{}H.-J. Borchers, On Poincar\'e transformations and the modular group of 
the algebra associated with a wedge, preprint. 
\item{}H.-J. Borchers and D. Buchholz, Global properties of vacuum states
in de Sitter space, ESI-preprint and gr-qc/9803036.
\item{}O. Bratteli and D.W. Robinson, {\it Operator Algebras and
Quantum Statistical Mechanics I}, Berlin, Heidelberg, New York:
Springer-Verlag, 1979.
\item{}J. Bros, H. Epstein and U. Moschella, Analyticity properties and
thermal effects for general quantum field theory on de Sitter space-time,
preprint.
\item{}K.S. Brown, {\it Cohomology of Groups}, New York, Heidelberg and 
Berlin: Springer-Verlag, 1982.
\item{}R. Brunetti, D. Guido and R. Longo, Modular structure and duality in
conformal quantum field theory, {\sl Commun. Math. Phys., \bf 156}, 201-219
(1993).
\item{}R. Brunetti, D. Guido and R. Longo, Group cohomology, modular theory
and space-time symmetries, {\sl Rev. Math. Phys., \bf 7}, 57-71 (1995).
\item{}D. Buchholz, On the structure of local quantum fields with non-trivial
interactions, in: {\it Proceedings of the International Conference on Operator 
Algebras}, Leipzig, Teubner Verlagsgesellschaft, 1978.
\item{}D. Buchholz and S.J. Summers, An algebraic characterization of vacuum 
states in Minkowski space, {\sl Commun. Math. Phys., \bf 155}, 449-458 (1993).
\item{}C. D'Antoni, S. Doplicher, K. Fredenhagen and R. Longo, Convergence of
local charges and continuity properties of $W^*$-inclusions, {\sl Commun.
Math. Phys., \bf 110}, 325-348 (1987).
\item{}D.R. Davidson, Modular covariance and the algebraic PCT/Spin-Statistics
theorem, preprint. 
\item{}J. Dixmier, {\it Von Neumann Algebras}, Amsterdam, New York and Oxford: 
North-Holland, 1981.
\item{}O. Dreyer, {\it Das Prinzip der geometrischen Wirkung im de 
Sitter-Raum}, Diplomarbeit, University of Hamburg, 1996.
\item{}W. Driessler, S.J. Summers and E.H. Wichmann, On the connection
between quantum fields and von Neumann algebras of local operators, {\sl
Commun. Math. Phys. \bf 105}, 49-84 (1986).
\item{}M. Florig, On Borchers' theorem, preprint. 
\item{}K. Fredenhagen, On the modular structure of local algebras of
observables, {\sl Commun. Math. Phys., \bf 97}, 79-89 (1985). 
\item{}K. Fredenhagen, Global observables in local quantum physics, in:
{\it Quantum and Non-Commutative Analysis}, Amsterdam: Kluwer Academic
Publishers, 1993.
\item{}K. Fredenhagen, Quantum field theories on nontrivial spacetimes, in:
{\it Mathematical Physics Towards the 21st Century}, ed. by R. Sen and A. 
Gersten, Beer-Sheva: Ben-Gurion University Negev Press, 1993.
\item{}I.M. Gel'fand, R.A. Minlos and Z.Ya. Shapiro, {\it Representations of
the Rotation and Lorentz Groups and Their Applications}, New York: The 
MacMillan Company, 1963.
\item{}D. Guido, Modular covariance, PCT, Spin and Statistics, {\sl Ann.
Inst. Henri Poincar\'e, \bf 63}, 383-398 (1995).
\item{}D. Guido and R. Longo, An algebraic spin and statistics theorem, I,
{\sl Commun. Math. Phys., \bf 172}, 517-533 (1995).
\item{}W. Hein, {\it Struktur- und Darstellungstheorie der Klassischen 
Gruppen}, Berlin, Heidelberg, New York: Springer-Verlag, 1990.
\item{}S. Helgason, {\it Differential Geometry and Symmetric Spaces},
New York and London: Academic Press, 1962.
\item{}P.D. Hislop and R. Longo, Modular structure of the local algebras 
associated with the free massless scalar field theory, {\sl Commun. Math. 
Phys., \bf 84}, 71-85 (1982).
\item{}B. Huppert, {\it Endliche Gruppen I}, Berlin, Heidelberg, New York:
Springer-Verlag, 1983.
\item{}B.S. Kay, The double-wedge algebra for quantum fields on Schwarzschild
and Minkowski spacetimes, {\sl Commun. Math. Phys., \bf 100}, 57-81 (1985).
\item{}B.S. Kay and R.M. Wald, Theorems on the uniqueness and thermal
properties of stationary, nonsingular, quasifree states on space-times with
a bifurcate Killing horizon, {\sl Phys. Rep., \bf 207}, 49-136 (1991).
\item{}M. Keyl, Causal spaces, causal complements and their relations to
quantum field theory, {\sl Rev. Math. Phys., \bf 8}, 229-270 (1996). 
\item{}B. Klotzek and R. Ottenberg, Pseudoeuklidische R\"aume im Aufbau der
Geometrie aus dem Spiegelungsbegriff, {\sl Zeitschr. f. math. Logik und
Grundlagen d. Math., \bf 26}, 145-164 (1980).
\item{}B. Kuckert, A new approach to spin \& statistics, {\sl Lett. Math.
Phys., \bf 35}, 319-331 (1995). 
\item{}B. Kuckert, Borchers' commutation relations and modular symmetries,
{\sl Lett. Math. Phys., \bf 41}, 307-320 (1997).
\item{}L.J. Landau, Asymptotic locality and the structure of local internal
symmetries, {\sl Commun. Math. Phys., \bf 17}, 156-176 (1970).
\item{}J.A. Lester, Separation-preserving transformations of De Sitter
spacetime, {\sl Abh. Math. Sem. Univ. Hamburg, \bf 53}, 217-224 (1983).
\item{}G. Mackey, Les ensembles Bor\'eliens et les extensions des groupes, 
{\sl J. Math. Pures Appl., \bf 36}, 171-178 (1957).
\item{}J. Milnor, {\it Introduction to Algebraic K-Theory}, Annals of 
Mathematics Studies, \# 72, Princeton: Princeton University Press, 1971.
\item{}C.C. Moore, Extensions and low dimensional cohomology theory of 
locally compact groups, I, {\sl Trans. Amer. Math. Soc., \bf 113}, 40-63 
(1964).
\item{}C.C. Moore, Extensions and low dimensional cohomology theory of 
locally compact groups, II, {\sl Trans. Amer. Math. Soc., \bf 113}, 64-86 
(1964).
\item{}C.C. Moore, Group extensions of $p$-adic and adelic linear groups, 
{\sl Publ. Math. I.H.E.S., \# 35}, 157-222 (1968). 
\item{}C.C. Moore, Group extensions and cohomology for locally compact 
groups, III, {\sl Trans. Amer. Math. Soc., \bf 221}, 1-33 (1976).
\item{}C.C. Moore, Group extensions and cohomology for locally compact 
groups, IV, {\sl Trans. Amer. Math. Soc., \bf 221}, 35-58 (1976).
\item{}M.-J. Radzikowski, {\it The Hadamard Condition and Kay's Conjecture
in (Axiomatic) Quantum Field Theory on Curved Space-Times}, Ph.D. Dissertation,
Princeton University, 1992.
\item{}J.E. Roberts and G. Roepstorff, Some basic concepts of algebraic
quantum theory, {\sl Commun. Math. Phys., \bf 11}, 321-338 (1969).
\item{}B. Schroer, Recent developments of algebraic methods in quantum field 
theory, {\sl Int. J. Modern Phys., \bf B6}, 2041-2059 (1992).
\item{}J.D. Stasheff, Continuous cohomology of groups and classifying spaces,
 {\sl Bull. Amer. Math. Soc., \bf 84}, 513-530 (1978).
\item{}R.F. Streater and A.S. Wightman, {\it PCT, Spin and Statistics, and
All That}, Reading, Mass.: Benjamin, 1964.
\item{}S.J. Summers and R. Verch, Modular inclusion, the Hawking temperature
and quantum field theory in curved space-time, {\sl Lett. Math. Phys., \bf
37}, 145-158 (1996). 
\item{}S.J. Summers, Geometric modular action and transformation groups, {\sl
Ann. Inst. Henri Poincar\'e, \bf 64}, 409-432 (1996).
\item{}L.J. Thomas, {\it About the Geometry of Minkowski Spacetime and Systems
of Local Algebras in Quantum Field Theory}, Ph.D. Dissertation, University of
California, Berkeley, 1989.
\item{}L.J. Thomas and E.H. Wichmann, Standard forms of local nets in quantum 
field theory, to appear in {\sl J. Math. Phys.}.
\item{}S. Trebels, {\it \"Uber die geometrische Wirkung modularer 
Automorphismen }, Ph.D. Dissertation, University of G\"ottingen, 1997.
\item{}R. White, manuscript in preparation.
\item{}E.H. Wichmann, private communication. 
\item{}H.-W. Wiesbrock, A comment on a recent work of Borchers,
{\it Lett. Math. Phys., \bf 25}, 157-159 (1992).
\item{}H.-W. Wiesbrock, Symmetries and half-sided modular inclusions of von 
Neumann algebras, {\sl Lett. Math. Phys., \bf 28}, 107-114 (1993).
\item{}H.-W. Wiesbrock, Conformal quantum field theory and half-sided modular
inclusions of von Neumann algebras, {\sl Commun. Math. Phys., \bf 158}, 
537-543 (1993).
\item{}H.-W. Wiesbrock, Half-sided modular inclusions of von Neumann algebras,
{\sl Commun. Math. Phys., \bf 157}, 83-92 (1993) ({\it Errata}:
{\sl Commun. Math. Phys., \bf 184}, 683-685 (1997)).
\item{}H.-W. Wiesbrock, A note on strongly additive conformal field theory
and half-sided modular conformal standard inclusions, {\sl Lett. Math. Phys.,
\bf 31}, 303-307 (1994).
\item{}H.-W. Wiesbrock, Symmetries and modular intersections of von
Neumann algebras, {\sl Lett. Math. Phys., \bf 39}, 203-212 (1997).
\item{}H.-W. Wiesbrock, Modular intersections of von Neumann algebras in
quantum field theory, preprint.
\item{}H. Wolf, Minkowskische und absolute Geometrie, I, {\sl Math. Annalen,
\bf 171}, 144-164 (1967); Minkowskische und absolute Geometrie, II, {\sl
ibid.}, 165-193 (1967).
\item{}M. Wollenberg, On the relation between a conformal structure in
spacetime and nets of local algebras of observables, {\sl Lett. Math. Phys.,
\bf 31}, 195-203 (1994). 
\item{}E.C. Zeeman, Causality implies the Lorentz group, {\sl J. Math. Phys., 
\bf 5}, 490-493 (1964).
\item{}R.J. Zimmer, {\it Ergodic Theory and Semisimple Groups}, Boston, Basel 
and Stuttgart: Birkh\"auser, 1984.
\endroster

\enddocument